\documentstyle[12pt]{article}

\newcommand{\mysection}{\setcounter{equation}{0}\section}

\begin{document}
\vskip 0.2cm
\hfill{YITP-SB-03-02}
\vskip 0.2cm
\hfill{INLO-PUB-01/03}\\[0.5cm]
\vskip 0.2cm
\centerline{\large\bf {NNLO corrections to the total cross section}}
\centerline{\large\bf {for Higgs boson production in hadron-hadron
collisions}}
\vskip 0.4cm
\centerline {\sc V. Ravindran}
\centerline{\it Harish-Chandra Research Institute,}
\centerline{\it Chhatnag Road, Jhunsi,}
\centerline{\it Allahabad, 211019, India.}
\vskip 0.2cm
\centerline {\sc J. Smith 
\footnote{partially supported
by the National Science Foundation grant PHY-0098527.}
}
\centerline{\it C.N. Yang Institute for Theoretical Physics,}
\centerline{\it State University of New York at Stony Brook,
New York 11794-3840, USA.}
\vskip 0.2cm
\centerline {\sc W.L. van Neerven}
\centerline{\it Instituut-Lorentz}
\centerline{\it University of Leiden,}
\centerline{\it PO Box 9506, 2300 RA Leiden,}
\centerline{\it The Netherlands.}
\vskip 0.2cm
\centerline{February 2003}
\vskip 0.2cm
\centerline{\bf Abstract}
\vskip 0.3cm
We present the next-to-next-to-leading order (NNLO) corrections to
the total cross section for (pseudo-) scalar Higgs boson production
using an alternative method than those used in previous calculations.
All QCD partonic subprocesses have been included and
the computation is carried out in the effective Lagrangian approach
which emerges from the standard model by taking the limit
$m_t \rightarrow \infty$ where $m_t$ denotes the mass of the top quark. 
Our results agree with those published earlier in the literature. 
We estimate the theoretical uncertainties by comparing the $K$-factors  
and the variation with respect to the mass factorization/renormalization
scales with the results obtained by lower order calculations. We also 
investigate the dependence of the cross section on several parton
density sets provided by different groups. Further we study which part
of the coefficient functions dominates the cross section.
This is of interest for the resummation of large
corrections which occur near the boundary of phase space. 
It turns out that depending on the definition of the total cross section
the latter is dominated by the 
the soft-plus-virtual gluon corrections represented by 
$\delta(1-x)$ and $(\ln^i(1-x)/(1-x))_+$ terms.

\vfill

\mysection{Introduction}
\newcommand{\be}{\begin{eqnarray}}
\newcommand{\ee}{\end{eqnarray}}
The Higgs boson, which is the corner stone of the standard model, is the 
only particle which has not been discovered yet. Its discovery or its
absence will shed light on the mechanism how particles acquire mass
as well as answer questions about supersymmetric extensions of the standard
model or about compositeness of the existing particles and the Higgs boson. 
The LEP experiments \cite{lep} give a lower mass limit of about 
$m_H\sim 114~{\rm GeV/c^2}$ and fits to the data using precision calculations
in the electro-weak sector of the standard model indicate an upper limit 
$m_H<200~{\rm GeV/c^2}$ with $95~\%$ confidence level. 
After the end of the LEP program the search for the Higgs will be
continued at hadron colliders in particular at the TEVATRON and the LHC.

In this paper we concentrate on Higgs production
where the lowest order reaction proceeds via the gluon-gluon fusion
mechanism. In the standard model the Higgs boson couples to the gluons
via heavy quark loops. 
Since the coupling of the scalar Higgs boson ${\rm H}$ to a fermion loop 
is proportional to the mass of the fermion (for a review see \cite{ghkd}), 
the top-quark loop is the most important.  The latter 
contribution is also dominant for the pseudo-scalar Higgs boson ${\rm A}$ 
provided $\tan \beta$ is small where $\beta$ denotes the mixing angle in 
the Two-Higgs-Doublet model (2HDM).
In lowest order (LO) the gluon-gluon fusion process $g + g \rightarrow {\rm B}$
with ${\rm B}={\rm H},{\rm A}$, represented by the top-quark triangle graph, 
was computed in \cite{wil}. The two-to-two body tree graphs, given by
gluon bremsstrahlung $g + g \rightarrow g + {\rm B}$, 
$g + q(\bar q) \rightarrow q(\bar q) + {\rm B}$ and
$q+\bar q \rightarrow g + {\rm B}$ were computed  
for ${\rm B}={\rm H}$ in \cite{ehsb} and for ${\rm B}={\rm A}$ 
in \cite{gsz}. From these reactions one can derive 
the transverse momentum ($p_T$) and rapidity ($y$) distributions of the
(pseudo-) scalar Higgs boson. The total integrated
cross section, which also involves the computation of the QCD corrections 
to the top-quark loop, has been calculated in \cite{gsz}.
This calculation is rather cumbersome since it involves the computation
of two-loop triangular graphs with massive quarks. Furthermore also the 
two-to-three parton reactions have been computed in \cite{dkosz} using 
helicity methods. From the experience gained from the next-to-leading (NLO) 
corrections presented in \cite{gsz} it 
is clear that it will be very difficult to obtain the exact NLO corrections
to one-particle inclusive distributions as well as the NNLO corrections to
the total cross section.

Fortunately one can simplify the calculations if one takes the
large top-quark mass limit $m_t \rightarrow \infty$. 
In this case the Feynman graphs are obtained from an effective Lagrangian 
describing the direct coupling of the (pseudo-) scalar Higgs boson
to the gluons. The LO and NLO contributions to the total cross section
in this approximation were computed in \cite{dawson}. A thorough analysis
\cite{gsz}, \cite{kls} reveals that the error 
introduced by taking the 
$m_t\rightarrow \infty$ limit is less than about $5\%$ provided $m_H\le 2~m_t$. 
The two-to-three body processes were computed with the effective 
Lagrangian approach for the scalar and pseudo-scalar Higgs bosons 
in \cite{kdr} and \cite{kade} respectively using helicity methods. 
The one-loop corrections
to the two-to-two body reactions above were computed for the scalar
Higgs boson in \cite{schmidt} and the pseudo-scalar Higgs boson in \cite{fism}. 
These matrix elements were used to compute the transverse 
momentum and rapidity distributions of the scalar Higgs boson up to NLO in
\cite{fgk}, \cite{rasm1} and the pseudo-scalar Higgs boson in \cite{fism}.
The effective Lagrangian method was also applied to obtain the NNLO total 
cross section for scalar Higgs production by the calculation of the two-loop 
corrections to the 
Higgs-gluon-gluon vertex in \cite{harland}, the soft-plus-virtual gluon 
corrections in \cite{cafl1}, \cite{haki1} and the computation of the 
two-to-three 
body processes in \cite{haki2}, \cite{anme1}. These calculations were 
repeated for pseudo-scalar Higgs production in \cite{haki3}, \cite{anme2}.
 
In this paper we recalculate the NNLO corrections to the total cross section
for (pseudo-) scalar Higgs production which have been computed
recently in \cite{haki2}-\cite{anme2} and we found complete agreement
with their analytic results when the number of colours is set $N=3$. 
However our method differs
from the one presented in \cite{haki2}, \cite{haki3} and the approach
followed in \cite{anme1}, \cite{anme2}. The authors in the latter
references compute the total
cross section using the Cutkosky \cite{cut} technique where one- and 
two-loop Feynman integrals are cut in certain ways. 
These Feynman integrals can be computed using various techniques (for more
details see the references in \cite{anme1}).
Furthermore this method been also applied to compute
rapidity distributions in NLO  which is presented in \cite{anme3}.
The authors
in \cite{haki2} choose a more conventional method which was already used in 
\cite{mmn1}, \cite{mmn2}, \cite{ham} to compute the coefficient
functions for the Drell-Yan process. However instead of an exact
computation of the $2\rightarrow 2$ and $2\rightarrow 3$ body phase space 
integrals they expand them around $x=1$. Here $x=m^2/s$ 
where $m$ and $\sqrt{s}$ denote the (pseudo-) scalar Higgs mass and the
partonic centre-of-mass energy respectively. Since the coefficient functions can
be expressed into a finite number of Polylogarithms of the types 
${\rm Li}_n(x)$, ${\rm S}_{n,p}(x)$ (see e.g. \cite{lewin}) and logarithms of 
the types $\ln^i x~\ln^j(1-x)$, which are all multiplied by polynomials in $x$,
one can expand them in the limit 
$x\rightarrow 1$ and match them with the expressions coming from the
phase space integrals. In this way the coefficients of the above functions
are determined. However as is shown in \cite{ham} this procedure is 
not necessary since one can obtain the Polylogarithms in a direct way
by expanding the hypergeometric functions \cite{abst} in $\varepsilon=n-4$ 
where $n$ is the number of dimensions characteristic of $n$-dimensional
regularization. The point is that there is no essential difference
between Higgs production in the effective Lagrangian approach and the 
Drell-Yan production mechanisms except that the matrix elements in the former 
case contain higher powers of the invariants in the numerators. In
\cite{ham} a program was made using the algebraic manipulation program
FORM \cite{form} to evaluate the phase space integrals
analytically. For this calculation we have extended the program so that
it can accommodate integrals having higher powers of invariants
in the numerator. Our approach can be also used
for differential distributions in particular for jet production (see e.g.
\cite{glov}). Here one cannot avoid multi-particle phase space integrals.
Of course two-to-four body processes are even more cumbersome to deal
with but one can at least try to compute the pole terms $(1/\varepsilon)^k$,
which arise from infrared and collinear divergences, analytically or  
numerically (for numerical methods see \cite{hein}). Another
difference with the previous results is that we present the radiative
parts of the coefficient
functions for general colour factors of the local gauge group SU(N).
The latter is important if one wants to resum the dominant contributions.
(see \cite{kls}, \cite{kist}).

Our paper will be organized as follows. In Section 2 we give definitions
and discuss the regularization method to compute the partonic
cross sections in the effective Lagrangian approach. In Section 3 we present
the calculation of the coefficient functions. In particular we discuss
the various frames in which the three-particle phase space integrals
are computed. In Section 4 we present the total cross sections for
scalar Higgs production at the LHC and the TEVATRON. 
The long expressions for the coefficient functions expressed in the
various colour factors are presented in Appendix A (scalar Higgs)
and Appendix B (pseudo-scalar Higgs).

\mysection{Application of the effective Lagrangian approach to Higgs production}
In the large top-quark mass limit the Feynman rules (see e.g. \cite{kdr})
for scalar Higgs  production (${\rm H}$) can be derived from the following 
effective Lagrangian density
\begin{eqnarray}
\label{eqn2.1}
{\cal L}^{\rm H}_{eff}=G_{\rm H}\,\Phi^{\rm H}(x)\,O(x) \quad 
\mbox{with} \quad O(x)=-\frac{1}{4}\,G_{\mu\nu}^a(x)\,G^{a,\mu\nu}(x)\,,
\end{eqnarray}
whereas pseudo-scalar Higgs (${\rm A}$) production is obtained from
\begin{eqnarray}
\label{eqn2.2}
&&{\cal L}_{eff}^{\rm A}=\Phi^{\rm A}(x)\Bigg [G_{\rm A}\,O_1(x)+
\tilde G_{\rm A}\,O_2(x)\Bigg ] \quad \mbox{with} \quad
\nonumber\\[2ex]
&&O_1(x)=-\frac{1}{8}\,\epsilon_{\mu\nu\lambda\sigma}\,G_a^{\mu\nu}\,
G_a^{\lambda\sigma}(x) \,,
\nonumber\\[2ex]
&&O_2(x) =-\frac{1}{2}\,\partial^{\mu}\,\sum_{i=1}^{n_f}
\bar q_i(x)\,\gamma_{\mu}\,\gamma_5\,q_i(x)\,,
\end{eqnarray}
where $\Phi^{\rm H}(x)$ and  $\Phi^{\rm A}(x)$ represent the scalar and 
pseudo-scalar fields respectively and $n_f$ denotes the number of light 
flavours.
Furthermore the gluon field strength is given by $G_a^{\mu\nu}$ and the 
quark field is denoted by $q_i$.
The factors multiplying the operators are chosen in such a way that the 
vertices are normalised to the effective coupling constants $G_{\rm H}$, 
$G_{\rm A}$ and $\tilde G_{\rm A}$. The latter are determined by the 
top-quark triangular loop graph, including all QCD corrections, taken in the
limit $m_t\rightarrow \infty$ which describes the decay process 
${\rm B} \rightarrow g + g$ with ${\rm B}={\rm H},{\rm A}$ namely
\begin{eqnarray}
\label{eqn2.3}
G_{\rm B}&=&-2^{5/4}\,a_s(\mu_r^2)\,G_F^{1/2}\,
\tau_{\rm B}\,F_{\rm B}(\tau_{\rm B})\,{\cal C}_{\rm B}
\left (a_s(\mu_r^2),\frac{\mu_r^2}{m_t^2}\right )\,,
\nonumber\\[2ex]
\tilde G_{\rm A}&=&-\Bigg [a_s(\mu_r^2)\,C_F\,\left (\frac{3}{2}-3\,
\ln \frac{\mu_r^2}{m_t^2}\right )+\cdots \Bigg ]\,G_{\rm A}\,,
\end{eqnarray}
and $a_s(\mu_r^2)$ is defined by
\begin{eqnarray}
\label{eqn2.4}
a_s(\mu_r^2)=\frac{\alpha_s(\mu_r^2)}{4\pi}\,,
\end{eqnarray}
where $\alpha_s(\mu_r^2)$ is the running coupling constant and $\mu_r$ denotes 
the renormalization scale. Further $G_F$ represents the Fermi constant and the
functions $F_{\rm B}$ are given by
\begin{eqnarray}
\label{eqn2.5}
&& F_{\rm H}(\tau)=1+(1-\tau)\,f(\tau)\,, \qquad  F_{\rm A}(\tau)=f(\tau)\,\cot
\beta\,,
\nonumber\\[2ex]
&& \tau=\frac{4\,m_t^2}{m^2} \,, 
\nonumber\\[2ex]
&&f(\tau)=\arcsin^2 \frac{1}{\sqrt\tau}\,, \quad \mbox{for} \quad \tau \ge 1\,,
\nonumber\\[2ex]
&& f(\tau)=-\frac{1}{4}\left ( \ln \frac{1-\sqrt{1-\tau}}{1+\sqrt{1-\tau}}
+\pi\,i\right )^2 \quad \mbox{for} \quad \tau < 1\,,
\end{eqnarray}
where $\cot \beta$ denotes the mixing angle in the Two-Higgs-Doublet Model.
Further $m$ and $m_t$ denote the masses of the (pseudo-) scalar Higgs
boson and the top quark respectively. In the large $m_t$-limit
we have
\begin{eqnarray}
\label{eqn2.6}
 \mathop{\mbox{lim}}\limits_{\vphantom{\frac{A}{A}} \tau \rightarrow \infty}
F_{\rm H}(\tau)=\frac{2}{3\,\tau}\,, \qquad
 \mathop{\mbox{lim}}\limits_{\vphantom{\frac{A}{A}} \tau \rightarrow \infty}
F_{\rm A}(\tau)=\frac{1}{\tau}\,\cot \beta\,. 
\end{eqnarray}
The coefficient functions ${\cal C}_{\rm B}$ originate from the corrections 
to the top-quark triangular graph provided one takes the 
limit $m_t\rightarrow \infty$.  We have presented the Born level
couplings $G_{\rm B}$ in Eq.(\ref{eqn2.3}) for general $m_t$ for on-shell
gluons only in order to keep some part of the top-quark mass 
dependence. This is an approximation because the gluons which couple to the
H and A bosons via the top-quark loop in partonic
subprocesses are very often virtual. The virtual-gluon momentum dependence is 
neither described by $F_{\rm B}(\tau)$ nor by ${\cal C}_{\rm B}$. However
for total cross sections the main contribution comes from the region
where the gluons are almost on-shell so that this approximation is
better than it is for differential cross sections with 
large transverse momentum. The coefficient functions are computed
up to order $\alpha_s^2$ in \cite{kls}, \cite{cks} for the H 
and in \cite{cksb} for the A. They read as follows
\begin{eqnarray}
\label{eqn2.7}
&&{\cal C}_{\rm H}\left (a_s(\mu_r^2),\frac{\mu_r^2}{m_t^2}\right )=
1+a_s^{(5)}(\mu_r^2)\,\Bigg [5\,C_A-3\,C_F\Bigg ] +
\left (a_s^{(5)}(\mu_r^2)\right )^2\,\Bigg [
\frac{27}{2}\,C_F^2
\nonumber\\[2ex]
&&-\frac{100}{3}\,C_A\,C_F+\frac{1063}{36}\,C_A^2-\frac{4}{3}\,
C_F\,T_f-\frac{5}{6}\,C_A\,T_f +\Big (7\,C_A^2
\nonumber\\[2ex]
&& -11\,C_A\,C_F\Big )\,\ln \frac{\mu_r^2}{m_t^2}
+ n_f\,T_f\,\left (-5\,C_F-\frac{47}{9}\,C_A
+8\,C_F\,\ln \frac{\mu_r^2}{m_t^2} \right ) \Bigg ]\,,
\\[2ex]
\label{eqn2.8}
&&{\cal C}_{\rm A}\left (a_s(\mu_r^2),\frac{\mu_r^2}{m_t^2}\right )=1\,,
\end{eqnarray}
where $a_s^{(5)}$ is presented in the five-flavour number scheme. Notice that
the coefficient function in Eq. (2.7) is derived for general colour factors 
of the group SU(N) from Eq. (6) in \cite{cks}. These factors are given by
\begin{eqnarray}
\label{eqn2.9}
C_A=N\quad\,,\quad C_F=\frac{N^2-1}{2N}\quad\,,\quad T_f=\frac{1}{2}\,.
\end{eqnarray}
Notice that $T_f$ is also incorporated into $G_{\rm B}$ in Eq. (\ref{eqn2.3}) 
where it is set to the value $T_f=1/2$.

Using the effective Lagrangian approach we will calculate the 
total cross section of the reaction
\begin{eqnarray}
\label{eqn2.10}
H_1(P_1)+H_2(P_2)\rightarrow {\rm B}(-p_5)+'X'\,,
\end{eqnarray}
where $H_1$ and $H_2$ denote the incoming hadrons and $X$ represents an 
inclusive hadronic state. The total cross section is given by
\begin{eqnarray}
\label{eqn2.11}
&&\sigma_{\rm tot}=\frac{\pi\,G_{\rm B}^2}{8\,(N^2-1)}\,\sum_{a,b=q,\bar q,g}\,
\int_x^1 dx_1\, \int_{x/x_1}^1dx_2\,f_a(x_1,\mu^2)\,f_b(x_2,\mu^2)\,
\nonumber\\[2ex] && \qquad\qquad\times
\Delta_{ab,{\rm B}}\left ( \frac{x}{x_1\,x_2},\frac{m^2}{\mu^2} \right ) \,,
\quad {\rm B}={\rm H},{\rm A}\,,
\nonumber\\[2ex]
&&\mbox{with}\quad x=\frac{m^2}{S} \quad\,,\quad S=(P_1+P_2)^2\quad
\,,\quad p_5^2=m^2\,,
\end{eqnarray}
where the factor $1/(N^2-1)$ originates from the colour average
in the case of the local gauge group SU(N). Further
we have assumed that the (pseudo-) scalar Higgs boson is mainly
produced on-shell i.e. $p_5^2\sim m^2$. The parton densities denoted by
$f_a(y,\mu^2)$ ($a,b=q,\bar q,g$)
depend on the mass factorization/renormalization scale $\mu$. 
The same scales also enter the coefficient functions $\Delta_{ab,{\rm B}}$ 
which are derived from the partonic cross sections
\begin{eqnarray}
\label{eqn2.12}
&&\sigma_{ab,{\rm B}}\left (z,\frac{m^2}{\mu^2}\right )=\frac{\pi}{s}\,
\int \frac{d^np_5}{(2\pi)^n}\,\delta^+(p_5^2-m^2)\,
T_{ab,{\rm B}}(p_5,p_1,p_2)\,,
\end{eqnarray}
where the incoming parton momenta, denoted by $p_1$ and $p_2$, are related
to the hadron momenta by
\begin{eqnarray}
\label{eqn2.13}
&&p_1=x_1\,P_1\quad\,,\quad p_2=x_2\,P_2 \quad\,,\quad
\nonumber\\[2ex]
&&s=(p_1+p_2)^2 \quad,\quad \Longrightarrow \qquad  s=x_1\,x_2\,S\,,
\qquad z=\frac{m^2}{s}\,.
\end{eqnarray}
The amplitude $T_{ab,{\rm B}}$ can be written as
\begin{eqnarray}
\label{eqn2.14}
T_{ab,{\rm H}}(p_5,p_1,p_2)&=&G_{\rm H}^2\int d^4y\,e^{i\,p_5\cdot y}
\, \langle a,b|O(y)\,O(0)|a,b\rangle \,,
\\[2ex]
\label{eqn2.15}
T_{ab,{\rm A}}(p_5,p_1,p_2)&=&\int d^4y\,e^{i\,p_5\cdot y}\,
\langle a,b|\Big (G_{\rm A}\,O_1(y)+\tilde G_{\rm A}\,O_2(y)\Big )
\nonumber\\[2ex]
&&\times\Big (G_{\rm A}\,O_1(0)+\tilde G_{\rm A}\,O_2(0)\Big )
|a,b\rangle \,.
\end{eqnarray}
The expressions above for the amplitude $T_{ab}$
are similar to those given for the Drell-Yan process except that the
conserved electroweak currents are replaced by the operators $O$ and $O_1,O_2$.
The latter are not conserved so that they acquire additional renormalization
constants defined by
\begin{eqnarray}
\label{eqn2.16}
O(y)=Z_O\,\hat O(y) \quad\,,\quad O_i(y)=Z_{ij}\, \hat O_j(y)\,.
\end{eqnarray}
where the hat indicates that the quantities under consideration are 
unrenormalized.
Insertion of the above equations into Eqs. (\ref{eqn2.14}),(\ref{eqn2.15})
leads to the renormalized expressions
\begin{eqnarray}
\label{eqn2.17}
&&T_{ab,{\rm H}}(p_5,p_1,p_2)=\,G_{\rm H}^2\,Z_O^2\,\int d^4y\,
e^{i\,p_5\cdot y} \, \langle a,b|\hat O(y)\,\hat O(0)|a,b\rangle\,,
\\[2ex]
\label{eqn2.18}
&&T_{ab,{\rm A}}(p_5,p_1,p_2)=\int d^4y\,e^{i\,p_5\cdot y}\,
\Bigg [\Bigg \{G_{\rm A}^2\,Z_{11}^2+\tilde G_{\rm A}^2\,Z_{21}^2+
2\,G_{\rm A}\,\tilde G_{\rm A}\,Z_{11}\,Z_{21}\Bigg \}
\nonumber\\[2ex]
&&\times \langle a,b|\hat O_1(y)\,\hat O_1(0)|a,b\rangle+\Bigg \{ G_{\rm A}^2
\,Z_{12}^2 +\tilde G_{\rm A}^2\,Z_{22}^2
+2\,G_{\rm A}\,\tilde G_{\rm A}\,Z_{12}\,Z_{22}\Bigg \}
\nonumber\\[2ex]
&&\times \langle a,b|\hat O_2(y)\,\hat O_2(0)|a,b\rangle
+\Bigg \{ G_{\rm A}^2\,Z_{11}\,Z_{12}
+\tilde G_{\rm A}^2\,Z_{22}\,Z_{21}+G_{\rm A}\,\tilde G_{\rm A}
\nonumber\\[2ex]
&&\times \Big (Z_{11}\,Z_{22}+Z_{12}\,Z_{21}\Big )\Bigg \}\,
\langle a,b|\hat O_1(y)\,\hat O_2(0) +\hat O_2(y)\,\hat O_1(0)|a,b\rangle 
\Bigg ]\,.
\end{eqnarray}
The operator renormalization constants depend on the regularization scheme 
in particular on the 
prescription for the $\gamma_5$-matrix and the Levi-Civita tensor in 
Eq. (\ref{eqn2.2}). The computation of $T_{ab,{\rm H}}$ will be carried 
out by choosing n-dimensional regularization where in the case of
$T_{ab,{\rm A}}$ we adopt the HVBM prescription
\cite{hv}, \cite{brma} for the $\gamma_5$-matrix. For this choice
the contraction of the Levi-Civita tensors proceeds as
\begin{eqnarray}
\label{eqn2.19}
\epsilon_{\mu_1\nu_1\lambda_1\sigma_1}\,\epsilon^{\mu_2\nu_2\lambda_2\sigma_2}=
\large{\left |
\begin{array}{cccc}
\delta_{\mu_1}^{\mu_2}&\delta_{\mu_1}^{\nu_2}&\delta_{\mu_1}^{\lambda_2}&
\delta_{\mu_1}^{\sigma_2}\\
\delta_{\nu_1}^{\mu_2}&\delta_{\nu_1}^{\nu_2}&\delta_{\nu_1}^{\lambda_2}&
\delta_{\nu_1}^{\sigma_2}\\
\delta_{\lambda_1}^{\mu_2}&\delta_{\lambda_1}^{\nu_2}&
\delta_{\lambda_1}^{\lambda_2}&\delta_{\lambda_1}^{\sigma_2}\\
\delta_{\sigma_1}^{\mu_2}&\delta_{\sigma_1}^{\nu_2}&
\delta_{\sigma_1}^{\lambda_2}&\delta_{\sigma_1}^{\sigma_2}
\end{array}
\right |}\,,
\end{eqnarray}
where all Lorentz indices are taken to be n-dimensional. To facilitate the
calculation one can replace $\gamma_{\mu}\,\gamma_5$ in Eq. (\ref{eqn2.2})
by (see \cite{akde})
\begin{eqnarray}
\label{eqn2.20}
\gamma_{\mu}\,\gamma_5=\frac{i}{6}\,\epsilon_{\mu\rho\sigma\tau}\,
\gamma^{\rho}\, \gamma^{\sigma}\,\gamma^{\tau}\,,
\end{eqnarray}
which is equivalent to the HVBM scheme. Choosing the 
${\overline{\rm MS}}$ subtraction scheme the renormalization constant 
corresponding to the operator $O$ becomes \cite{klzu}
\begin{eqnarray}
\label{eqn2.21}
Z_O&=&1+a_s(\mu_r^2)\,S_{\varepsilon}\,\frac{2}{\varepsilon}\,\beta_0+
a_s^2(\mu_r^2)\,S_{\varepsilon}^2\,\Bigg [\frac{4}{\varepsilon^2}\,\beta_0^2
+\frac{2}{\varepsilon}\,\beta_1\Bigg ]+\cdots
\end{eqnarray}
where $S_{\varepsilon}$ denotes the spherical factor characteristic
of n-dimensional regularization. It is defined by
\begin{eqnarray}
\label{eqn2.22}
S_{\varepsilon}=\exp\left \{\frac{\varepsilon}{2}\Big [\gamma_E-\ln 4\pi\Big ]
\right \}\,.
\end{eqnarray}
The lowest order coefficients $\beta_0$ and $\beta_1$ originate from the
beta-function given by
\begin{eqnarray}
\label{eqn2.23}
&&\beta(\alpha_s)=-a_s^2(\mu_r^2)\,\beta_0-a_s^3(\mu_r^2)\,\beta_1+\cdots\,,
\nonumber\\[2ex]
&&\beta_0=\frac{11}{3}\,C_A-\frac{4}{3}\,n_f\,T_f \,, \quad
\beta_1=\frac{34}{3}\,C_A^2- 4\,n_f\,T_f\,C_F-\frac{20}{3}\,n_f\,T_f\,C_A\,.
\nonumber\\[2ex]
\end{eqnarray}
The operator renormalization constants corresponding to $O_1$ and $O_2$
are computed in \cite{larin1} and they read
\begin{eqnarray}
\label{eqn2.24}
Z_{11}=Z_{\alpha_s}=1+a_s(\mu_r^2)\,S_{\varepsilon}\,\frac{2}{\varepsilon}\,
\beta_0+ a_s^2(\mu_r^2)\,S_{\varepsilon}^2\,\Bigg [\frac{4}{\varepsilon^2}\,
\beta_0^2
+\frac{1}{\varepsilon}\,\beta_1\Bigg ]+\cdots
\end{eqnarray}
where $Z_{\alpha_s}$ denotes the coupling constant renormalization factor
defined by
\begin{eqnarray}
\label{eqn2.25}
\hat a_s =Z_{\alpha_s}\,a_s(\mu_r^2)\,.
\end{eqnarray}
The remaining constants are
\begin{eqnarray}
\label{eqn2.26}
Z_{21}&=&0\,,
\\[2ex]
\label{eqn2.27}
Z_{12}&=&a_s(\mu_r^2)\,S_{\varepsilon}\,\frac{1}{\varepsilon}\,
\Big [-6\,C_F\Big ]\,,
\\[2ex]
\label{eqn2.28}
Z_{22}&=&Z^s_{\rm MS}\,Z_5^s\,,
\end{eqnarray}
where $Z^s_{\rm MS}$ and $Z_5^s$ are the constants characteristic of the 
HVBM scheme. They are given by
\begin{eqnarray}
\label{eqn2.29}
Z^s_{\rm MS}&=&1+a_s^2(\mu_r^2)\,S_{\varepsilon}\,\frac{1}{\varepsilon}\,
\left [-\frac{44}{3}\,C_A\,C_F-n_f\,T_f\,C_F\,\frac{20}{3}\right ]\,,
\nonumber\\[2ex]
Z_5^s&=&1+a_s(\mu_r^2)\,\Big [-4\,C_F\Big ]+a_s^2(\mu_r^2)\,\left [22\,C_F^2
-\frac{107}{9}\,C_A\,C_F+\frac{31}{9}\,n_f\,T_f\,C_F\right ]\,.
\nonumber\\[2ex]
\end{eqnarray}
The latter renormalization constant is determined in such a way that
the Adler-Bell-Jackiw anomaly \cite{abj}
\begin{eqnarray}
\label{eqn2.30}
O_2(y)=4\,a_s(\mu_r^2)\,n_f\,T_f\,O_1(y)\,,
\end{eqnarray}
is preserved in all orders in perturbation theory according to the 
Adler-Bardeen theorem \cite{adba}.

\mysection{Computation of the partonic cross section up to NNLO} 
In this section we will give a short outline of the computation of
the partonic cross sections $\sigma_{ab,{\rm B}}$ in Eq. 
(\ref{eqn2.12})
up to next-to-next-leading order (NNLO). Insertion of a complete set of 
intermediate states and using translational invariance in Eq. (\ref{eqn2.13})
we obtain for scalar Higgs production
\begin{eqnarray}
\label{eqn3.1}
&&\sigma_{ab,{\rm H}}=K_{ab}\,\frac{\pi\,G_{\rm H}^2}{s}\,Z_O^2\,
\int d^np_5\,\delta^+(p_5^2-m^2)
\sum_{m=3}^{\infty}\,\prod_{i=3,i\not =5}^m\int\frac{d^np_i}{(2\pi)^{n-1}}\,
\delta^+(p_i^2)
\nonumber\\[2ex]
&&\qquad\qquad\times
\delta^{(n)}(\sum_{j=1}^m\,p_j)\, |M_{ab\rightarrow X~{\rm H}}|^2\,,
\nonumber\\[2ex]
&& M_{ab\rightarrow X~{\rm H}}=\langle p_1,p_2|\hat O(0)| X,p_5\rangle  
\quad \mbox{with} \quad | X,p_5\rangle=|p_3,p_4,p_6 \cdots p_m,p_5\rangle\,,
\nonumber\\[2ex]
\end{eqnarray}
where $K_{ab}$ represents the spin and colour average over the initial states.
Further one can write down a similar expression for the pseudo-scalar Higgs 
(see Eq. (\ref{eqn2.15})). Up to NNLO we have to compute the following 
subprocesses.
On the Born level we have the reaction
\begin{eqnarray}
\label{eqn3.2}
g+g \rightarrow {\rm B}\,.
\end{eqnarray}
In NLO we have in addition to the one-loop virtual corrections to the above
reaction the following two-to-two body processes
\begin{eqnarray}
\label{eqn3.3}
g+g \rightarrow {\rm B} + g \quad\,,\quad
g+q(\bar q) \rightarrow {\rm B} + q(\bar q)
\quad\,,\quad q+ \bar q \rightarrow {\rm B} + g\,.
\end{eqnarray}
In NNLO we receive contributions from the two-loop virtual corrections to
the Born process in Eq. (\ref{eqn3.2}) and the one-loop corrections to
the reactions in  Eq. (\ref{eqn3.3}). To these contribution one has to add
the results obtained from the following two-to-three body reactions
\begin{eqnarray}
\label{eqn3.4}
&&g+g \rightarrow {\rm B} + g + g \quad\,,\quad
 g+g \rightarrow {\rm B} + q_i + \bar q_i\,,
\\[2ex]
\label{eqn3.5}
&&g+q(\bar q) \rightarrow {\rm B} + q(\bar q) + g\,,
\\[2ex]
\label{eqn3.6}
&& q+ \bar q \rightarrow {\rm B} + g + g \quad\,,\quad
q+ \bar q \rightarrow {\rm B} + q_i + \bar q_i\,,
\\[2ex]
\label{eqn3.7}
&&q_1+ q_2 \rightarrow {\rm B} + q_1+ q_2 \quad\,,\quad 
q_1+ \bar q_2 \rightarrow {\rm B} + q_1+ \bar q_2\,, 
\\[2ex]
\label{eqn3.8}
&&q+ q \rightarrow {\rm B} + q+ q\,.
\end{eqnarray} 
In the case of pseudo-scalar Higgs production one also has to add the
contributions due to interference terms coming from the operators $\hat O_1$ 
and $\hat O_2$ in Eq. (2.18). Up to order $\alpha_s^2$ we have to compute the 
following expression for the reactions in (\ref{eqn3.4})
\begin{eqnarray}
\label{eqn3.9}
&&\sigma_{gg,{\rm A}}=K_{ab}\,\frac{\pi}{s}\,\int d^np_5\,
\delta^+(p_5^2-m^2)
\Bigg [ \Bigg \{G_{\rm A}^2\,Z_{11}^2\,\sum_{m=3}^4\,\prod_{i=3}^m\int 
\frac{d^np_i}
{(2\pi)^{n-1}}\,\delta^+(p_i^2)
\nonumber\\[2ex]
&& \times \delta^{(n)}(\sum_{j=1}^m\,p_j)\,
|\langle p_1,p_2|\hat O_1|X,p_5\rangle |^2\Bigg \}
+\Bigg \{G_{\rm A}\,\tilde G_{\rm A}\,\delta^{(n)}(p_1+p_2+p_5)
\nonumber\\[2ex]
&& \times\Big ( \langle p_1,p_2|\hat O_1|p_5\rangle\,\langle p_5|\hat O_2|
p_1,p_2\rangle +\langle p_1,p_2|\hat O_2|p_5\rangle \,
\langle p_5|\hat O_1|p_1,p_2\rangle \Big )\Bigg \}\Bigg ]\,.
\end{eqnarray}
The Feynman graphs for $\langle p_1,p_2|\hat O_1|p_5\rangle$ and
$\langle p_1,p_2|\hat O_2|p_5\rangle$ can be found in Fig. 1a and Fig. 2a of 
\cite{haki3}
respectively. From the above equation one infers that the interference term
contributes as a delta-function $\delta(1-z)$ to $\sigma_{gg,{\rm A}}$. 
Furthermore since $\tilde G_{\rm A}$ is proportional to $G_{\rm A}$ 
(see Eq. (\ref{eqn2.3})) one can extract the latter constant as an overall
factor from the above equation. For the reactions in Eq. (\ref{eqn3.5})
we have to compute
\begin{eqnarray}
\label{eqn3.10}
&&\sigma_{gq,{\rm A}}=K_{ab}\,\frac{\pi\,G_{\rm A}^2}{s}\,\int d^np_5\,
\delta^+(p_5^2-m^2)\Bigg [ \Bigg \{Z_{11}^2\,\sum_{m=3}^4\, \prod_{i=3}^m\int 
\frac{d^np_i}{(2\pi)^{n-1}}\,
\delta^+(p_i^2)
\nonumber\\[2ex]
&&\times \delta^{(n)}(\sum_{j=1}^m\,p_i)\, 
|\langle p_1,p_2|\hat O_1|X,p_5\rangle |^2\Bigg \}
+\Bigg \{Z_{12}\,\int \frac{d^np_3}{(2\pi)^{n-1}}\,\delta^+(p_3^2)\,
\nonumber\\[2ex]
&&\times\delta^{(n)}(p_1+p_2+p_3+p_5)\,\Big (\langle p_1,p_2|\hat O_1|p_5,p_3 
\rangle\, \langle p_5,p_3|\hat O_2|p_1,p_2\rangle 
\nonumber\\[2ex]
&&+\langle p_1,p_2|\hat O_2|p_5,p_3\rangle \,
\langle p_5,p_3|\hat O_1|p_1,p_2\rangle \Big )\Bigg \}\Bigg ]\,.
\end{eqnarray}
The Feynman graphs for $\langle p_1,p_2|\hat O_1|p_3,p_5\rangle$ and 
$\langle p_1,p_2|\hat O_2|p_3,p_5\rangle$ can be found in Fig. 1b and Fig. 2b 
of \cite {haki3} respectively. In the HVBM-scheme the latter matrix element is 
proportional to $\varepsilon=n-4$. However this contribution does not vanish in
the limit $\varepsilon\rightarrow 0$ because of the single pole term present
in $Z_{12}$ in Eq. (\ref{eqn2.27}). The same expression as in 
Eq. (\ref{eqn3.10}) also exists for $\sigma_{q\bar q,{\rm A}}$ originating 
from the first reaction in Eq. (\ref{eqn3.6}). 
In this case the Feynman graphs corresponding to 
$\langle p_1,p_2|\hat O_1|p_3,p_5\rangle$ and $\langle p_1,p_2|\hat O_2|
p_3,p_5\rangle$ are shown in Fig. 1d and Fig. 2d of \cite {haki3} respectively. 
Notice that these
contributions also survive in the calculation of differential distributions,
see \cite{fism}.

The computation of the phase space integrals proceeds as follows. First we take
over the expressions from the one- and two-loop corrections to the Born 
reaction in Eq.(\ref{eqn3.2}) which are computed in \cite{dawson} 
(one-loop) and \cite{harland} (two-loop).
Unfortunately the two-loop result is not presented for arbitrary colour factors
so that we cannot distinguish between $n_f\,C_A\,T_f$ and $n_f\,C_F\,T_f$.
Hence we have shuffled all contributions proportional to $n_f$ into
the term proportional to $n_f\,C_A\,T_f$. To compute the two-to-two body 
processes including the virtual corrections we have chosen the 
centre-of-mass frame of the incoming partons given by
\begin{eqnarray}
\label{eqn3.11}
p_1&=&\frac{1}{2}\,\sqrt{P_{12}}\,(1,0,\cdots,0,1)\,,
\nonumber\\[2ex]
p_2&=&\frac{1}{2}\,\sqrt{P_{12}}\,(1,0,\cdots,0,-1)\,,
\nonumber\\[2ex]
-p_3&=&\frac{P_{12}-m^2}{2\sqrt{P_{12}}}\,(1,0,\cdots,-\sin 
\theta,-\cos \theta)\,,
\nonumber\\[2ex]
-p_5&=&\frac{1}{2\sqrt{P_{12}}}\,(P_{12}+m^2,0,\cdots,
(P_{12}-m^2) \,\sin \theta,(P_{12}-m^2)\,\cos \theta)\,,
\nonumber\\[2ex]
\end{eqnarray}
where we have introduced the invariants
\begin{eqnarray}
\label{eqn3.12}
P_{ij}=(p_i+p_j)^2 \quad\,,\quad P_{12}=s\,.
\end{eqnarray}
In this frame the two-to-two body phase space integral becomes
\begin{eqnarray}
\label{eqn3.13}
\sigma_{ab\rightarrow c~{\rm H}}&=&K_{ab}\,Z_O^2\,\frac{\pi\,G_{\rm H}^2}
{P_{12}}\, \frac{2^{3-n}}{(4\pi)^{n/2}}\,
\frac{1}{\Gamma(n/2-1)}\,\left (\frac{P_{12}}{\mu^2}\right )^{1-n/2}
\left (\frac{P_{12}-m^2}{\mu^2}\right )^{n-3}\,
\nonumber\\[2ex]
&&\times \int_0^{\pi}d\theta\,
\sin^{n-3}\theta\,|M_{ab\rightarrow c~{\rm B}}|^2\,.
\end{eqnarray}
Here $\mu$ indicates the dimension of the strong coupling constant 
$g \rightarrow g\,\mu^{(4-n)/2}$ which is characteristic of n-dimensional
regularization.
In order to make all ratios dimensionless we have already included the
factor $\mu^{(4-n)/2}$ in the phase space integrals. Notice that in principle
the scale $\mu$ has nothing to do with the factorization or 
renormalization scale.
However for convenience one puts the latter scales equal to $\mu$.
The evaluation of the integral in Eq. (\ref{eqn3.13}) is rather easy even when
$|M_{ab\rightarrow c~{\rm B}}|$ contains virtual contributions. Since the
integrals can be evaluated analytically we have made a routine using 
the algebraic manipulation program FORM which provides us with the results.
A similar routine is made for the two-to-three body phase space integrals
but here the computation is not so easy unless one chooses a suitable frame.
Since we integrate over the total phase space the integrals are Lorentz
invariant and therefore frame independent. The matrix elements of the partonic
reactions can be partial fractioned in such a way that maximally two
factors $P_{ij}$ in Eq. (\ref{eqn3.12}) depend on the 
polar angle $\theta_1$ and 
the azimuthal angle $\theta_2$. Furthermore one factor only depends on
$\theta_1$ whereas the other one depends both on $\theta_1$ and $\theta_2$.
Therefore the following combinations show up in the matrix elements
\begin{eqnarray}
\label{eqn3.14}
&&P_{ij}^k\,P_{mn}^l \,, P_{ij}^k\,P_{m5}^l \,, P_{i5}^k\,P_{m5}^l
\quad\,,\quad 4\ge k\ge -2 \,, 4\ge l\ge -2\,, 
\nonumber\\[2ex]
&&p_i^2=p_j^2=p_m^2=p_n^2=0 \quad\,,\quad p_5^2=m^2\,.
\end{eqnarray}
For the first combination it is easy to perform the angular integration
since all momenta represent massless particles and one obtains a hypergeometric 
function (see e.g. Eq. (4.19) in \cite{rasm1}).
The angular integral of the second combination is more difficult to compute
because one particle is massive and the result is an one dimensional integral 
over a hypergeometric function
which however can be expanded around $\varepsilon$. Examples of these types
of integrals can be found in Appendix C of \cite{bkns}. The last combination is 
very difficult
to compute in n dimensions because both factors contain the massive particle 
indicated by $p_5$. Therefore one has to avoid this combination at any cost.
This is possible if one chooses the following three frames. The first one is
the centre-of-mass frame of the incoming partons. The momenta are
\begin{eqnarray}
\label{eqn3.15}
p_1&=&\frac{1}{2}\,\sqrt{P_{12}}\,(1,0,\cdots ,0,0,1)\,,
\nonumber\\[2ex]
p_2&=&\frac{1}{2}\,\sqrt{P_{12}}\,(1,0,\cdots ,0,0,-1)\,,
\nonumber\\[2ex]
-p_3&=&\frac{P_{12}-P_{45}}{2\,\sqrt{P_{12}}}\,(1,0,\cdots ,0,\sin \theta_1,
\cos \theta_1)\,,
\nonumber\\[2ex]
-p_4&=&\frac{P_{12}-P_{35}}{2\,\sqrt{P_{12}}}\,(1,0,\cdots ,\sin \psi\,\sin 
\theta_2,\cos \psi\, \sin \theta_1 +\sin \psi\,\cos \theta_2\,\cos \theta_1,
\nonumber\\[2ex]
&&\cos \psi\,\cos \theta_1-\sin \psi\,\cos \theta_2\,\sin \theta_1)\,,
\nonumber\\[2ex]
\sin^2 \frac{\psi}{2}&=&\frac{P_{12}\,P_{34}}{(P_{12}-P_{35})\,(P_{12}-P_{45})}
\,,
\end{eqnarray}
\begin{eqnarray}
\label{eqn3.16}
\sigma_{ab\rightarrow cd~{\rm H}}&=&K_{ab}\,Z_O^2\,\frac{G_{\rm H}^2}{2P_{12}}\,
\frac{1}{(4\pi)^n}\,\frac{1}{\Gamma(n-3)}\,\left (\frac{P_{12}}{\mu^2}
\right )^{1-n/2}\,\int_{m^2}^{P_{12}}dP_{35}
\nonumber\\[2ex]
&&\times\int_{P_{12}\,m^2
/P_{35}}^{P_{12}+m^2-P_{35}}dP_{45}\,\left (\frac{P_{35}\,P_{45}-P_{12}\,
m^2}{\mu^4}\right )^{n/2-2}
\nonumber\\[2ex]
&&\times\left (\frac{P_{12}+m^2-P_{35}-P_{45}}{\mu^2}\right )^{n/2-2}\,
\int_0^{\pi}d\theta_1\,\sin^{n-3}\theta_1
\nonumber\\[2ex]
&&\times\int_0^{\pi}d\theta_2\,\sin^{n-4}\theta_2\,
|M_{ab\rightarrow cd~{\rm B}}|^2\,.
\end{eqnarray}
For the next frame we choose the centre-of-mass frame of the two outgoing
particles indicated by the momenta $p_3$ and $p_4$.
\begin{eqnarray}
\label{eqn3.17}
p_1&=&\omega_1\,(1,0,\cdots ,0,0,1)\,,
\nonumber\\[2ex]
p_2&=&(\omega_2,0,\cdots ,0,|\vec p_5|\,\sin \psi,
|\vec p_5|\,\cos \psi-\omega_1)\,,
\nonumber\\[2ex]
-p_3&=&\frac{1}{2}\,\sqrt{P_{34}}\,(1,0,\cdots,\sin \theta_1\,\sin \theta_2,
\sin \theta_1\,\cos \theta_2,\cos \theta_1)\,,
\nonumber\\[2ex]
-p_4&=&\frac{1}{2}\,\sqrt{P_{34}}\,(1,0,\cdots,-\sin \theta_1\,\sin \theta_2,
-\sin \theta_1\,\cos \theta_2,-\cos \theta_1)\,,
\nonumber\\[2ex]
-p_5&=&(\omega_5,0,\cdots ,0,|\vec p_5|\,\sin \psi,|\vec p_5|\,\cos \psi)\,,
\nonumber\\[2ex]
\omega_1&=&\frac{P_{12}+P_{15}-m^2}{2\,\sqrt{P_{34}}}
\quad\,,\quad \omega_2=\frac{P_{12}+P_{25}-m^2}{2\,\sqrt{P_{34}}}
\nonumber\\[2ex]
&&\omega_5=-\frac{P_{15}+P_{25}}{2\,\sqrt{P_{34}}}\,,
\nonumber\\[2ex]
\cos \psi&=&\frac{(P_{34}-m^2)\,(P_{15}-m^2)-P_{12}\,(P_{25}
+m^2)}{(P_{12}+P_{15}-m^2)\sqrt{(P_{15}+P_{25})^2-4
\,m^2\, P_{34}}}\,,
\end{eqnarray}
\begin{eqnarray}
\label{eqn3.18}
&&\sigma_{ab\rightarrow cd~{\rm H}}=K_{ab}\,Z_O^2\,\frac{G_{\rm H}^2}{2P_{12}}\,
\frac{1}{(4\pi)^n}\,\frac{1}{\Gamma(n-3)}\,\left (\frac{P_{12}}{\mu^2}
\right )^{1-n/2}\,\int_{m^2-P_{12}}^0dP_{25}
\nonumber\\[2ex]
&&\times\int^{m^2+P_{12}\,m^2/(P_{25}-m^2)}
_{m^2-P_{12}-P_{25}}dP_{15}\,\left (\frac{(P_{15}-m^2)\,(P_{25}
-m^2)-P_{12}\,m^2}{\mu^4}\right )^{n/2-2}
\nonumber\\[2ex]
&&\times\left (\frac{P_{12}+P_{15}+P_{25}-m^2}{\mu^2}\right )^{n/2-2}\,
\int_0^{\pi}d\theta_1\,\sin^{n-3}\theta_1
\nonumber\\[2ex]
&&\times\int_0^{\pi}d\theta_2\,\sin^{n-4}\theta_2\,
|M_{ab\rightarrow cd~{\rm B}}|^2\,.
\end{eqnarray}
For the last frame we choose the centre-of-mass frame of one of the outgoing
partons and the (pseudo-)scalar Higgs boson indicated by the momenta $p_4$ and 
$p_5$ respectively
\begin{eqnarray}
\label{eqn3.19}
p_1&=&\omega_1\,(1,0,\cdots ,0,0,1)\,,
\nonumber\\[2ex]
p_2&=&(\omega_2,0,\cdots ,0,\omega_3\,\sin \psi,\omega_3\,\cos \psi-\omega_1)\,,
\nonumber\\[2ex]
-p_3&=&\omega_3\,(1,0,\cdots,0,\sin \psi, \cos \psi)\,,
\nonumber\\[2ex]
-p_4&=&\omega_4\,(1,0,\cdots,\sin \theta_1\,\sin \theta_2,
\sin \theta_1\,\cos \theta_2,\cos \theta_1)
\nonumber\\[2ex]
-p_5&=&(\omega_5,0,\cdots,-\omega_4\,\sin \theta_1\,\sin \theta_2,
-\omega_4\,\sin \theta_1\,\cos \theta_2,-\omega_4\,\cos \theta_1)\,,
\nonumber\\[2ex]
\omega_1&=&\frac{P_{12}+P_{13}}{2\,\sqrt{P_{45}}}
\quad\,,\quad \omega_2=\frac{P_{12}+P_{23}}{2\,\sqrt{P_{45}}}
\nonumber\\[2ex]
\omega_3&=&-\frac{P_{13}+P_{23}}{2\,\sqrt{P_{45}}}
\quad\,,\quad \omega_4=\frac{P_{45}-m^2}{2\,\sqrt{P_{45}}}
\nonumber\\[2ex]
\omega_5&=&\frac{P_{45}+m^2}{2\,\sqrt{P_{45}}}
\quad\,,\quad \cos \psi=\frac{P_{12}\,P_{23}-P_{45}\,P_{13}}{(P_{12}+P_{13})\,
(P_{13}+P_{23})}\,,`
\end{eqnarray}
\begin{eqnarray}
\label{eqn3.20}
&&\sigma_{ab\rightarrow cd~{\rm H}}=K_{ab}\,Z_O^2\,\frac{G_{\rm H}^2}{2P_{12}}\,
\frac{1}{(4\pi)^n}\,\frac{1}{\Gamma(n-3)}\,\left (
\frac{P_{12}}{\mu^2} \right )^{1-n/2}\,\int_{m^2-P_{12}}^0dP_{23}
\nonumber\\[2ex]
&&\times\int_{m^2-P_{12}-P_{23}}^0dP_{13}
\left (\frac{P_{13}\,P_{23}}{\mu^4}\right )^{n/2-2}
\left (\frac{P_{12}+P_{13}+P_{23}-m^2}{\mu^2}\right )^{n-3}\,
\nonumber\\[2ex]
&&\times\left (\frac{P_{12}+P_{13}+P_{23}}{\mu^2}\right )^{1-n/2}\,
\int_0^{\pi}d\theta_1\,\sin^{n-3}\theta_1
\int_0^{\pi}d\theta_2\,\sin^{n-4}\theta_2
\nonumber\\[2ex]
&&\times |M_{ab\rightarrow cd~{\rm B}}|^2\,.
\end{eqnarray}
The integration over the angular independent invariants $P_{ij}$
can be performed by rescaling them with respect to $P_{12}$
(see Appendix E in \cite{mmn2}). The integrals are such that they
can be performed in an algebraic way by a program
based on FORM \cite{form} which has been also used
to compute the NNLO coefficient functions of the Drell-Yan process
in \cite{mmn1}, \cite{mmn2}, \cite{ham} (for more details see \cite{rasm2}). 
The partonic cross sections Eq. (\ref{eqn2.12}) can be written as follows
\begin{eqnarray}
\label{eqn3.21}
\sigma_{ab,{\rm H}}\left (z,\frac{m^2}{\mu^2}\right )&=&
\frac{\pi\,G_{\rm H}^2}{8\,(N^2-1)}\,
\frac{1}{1+\varepsilon/2}\,Z_O^2\,\hat W_{ab,{\rm H}}\left (z,\frac{m^2}{\mu^2}
\right )
\nonumber\\[2ex]
\sigma_{ab,{\rm A}}\left (z,\frac{m^2}{\mu^2}\right )&=&
\frac{\pi\,G_{\rm A}^2}{8\,(N^2-1)}\,
\frac{1+\varepsilon}{1+\varepsilon/2}\,Z_{11}^2\,\hat W_{ab,{\rm A}}
\left (z,\frac{m^2}{\mu^2}\right )
\end{eqnarray}
\begin{eqnarray}
\label{eqn3.22}
z^{-1}\,\hat W_{gg,{\rm B}}&=&\delta(1-z)+\hat a_s\,S_{\varepsilon}\,
\left (\frac{m^2}{\mu^2}
\right )^{\varepsilon/2}\,\Bigg [\frac{2}{\varepsilon}\,\Big (P_{gg}^{(0)}
-2\,\beta_0)+w_{gg,{\rm B}}^{(1)}
+\varepsilon\,{\bar w}_{gg,{\rm B}}^{(1)}\Bigg ]
\nonumber\\[2ex]
&&+\hat a_s^2\,S_{\varepsilon}^2\,
\left (\frac{m^2}{\mu^2}\right )^{\varepsilon}
\Bigg [\frac{1}{\varepsilon^2}\,\Big (2\,P_{gg}^{(0)} \otimes P_{gg}^{(0)}
+P_{gq}^{(0)}\otimes P_{qg}^{(0)}-10\,\beta_0\,P_{gg}^{(0)}
\nonumber\\[2ex]
&&+12\,\beta_0^2\Big ) +\frac{1}{\varepsilon}\,\Big (P_{gg}^{(1)}
-6\,\beta_0\,w_{gg,{\rm B}}^{(1)}+
2\,P_{gg}^{(0)}\otimes w_{gg,{\rm B}}^{(1)} 
\nonumber\\[2ex]
&&+2\,w_{qg,{\rm B}}^{(1)}\otimes P_{qg}^{(0)}+z_{\rm B}\Big )
-6\,\beta_0\,\bar w_{gg,{\rm B}}^{(1)}+2\,P_{gg}^{(0)}\otimes \bar 
w_{gg,{\rm B}}^{(1)} 
\nonumber\\[2ex]
&&+2\,\bar w_{qg,{\rm B}}^{(1)}\otimes P_{qg}^{(0)}+w_{gg,{\rm B}}^{(2)}\Bigg ]
\nonumber\\[2ex]
&&\mbox{with}\quad z_{\rm H}=-4\,\beta_1 \qquad z_{\rm A}=-2\,\beta_1\,,
\end{eqnarray}
The coefficients $z_{\rm B}$, multiplying the single pole terms in the 
equations above, originate from 
the operators $O$ in Eq. (\ref{eqn2.1}) and $O_1$ in Eq. (\ref{eqn2.2}) and 
they are removed via the operator renormalization constants $Z_O$ in 
Eq. (\ref{eqn2.21}) and $Z_{11}$ in Eq. (\ref{eqn2.24}).
\begin{eqnarray}
\label{eqn3.23}
z^{-1}\,\hat W_{gq,{\rm B}}&=&\hat a_s\,S_{\varepsilon}\,
\left (\frac{m^2}{\mu^2}
\right )^{\varepsilon/2}\,\Bigg [\frac{2}{\varepsilon}\,P_{gq}^{(0)}
+w_{gq,{\rm B}}^{(1)}
+\varepsilon\,{\bar w}_{gq,{\rm B}}^{(1)}\Bigg ]+\hat a_s^2\,S_{\varepsilon}^2\,
\left (\frac{m^2}{\mu^2}\right )^{\varepsilon}
\nonumber\\[2ex]
&& \times\Bigg [ \frac{1}{\varepsilon^2}\,\Big (
+\frac{3}{2}\,P_{gg}^{(0)}\otimes
P_{gq}^{(0)}+\frac{1}{2}\,P_{qq}^{(0)}\otimes P_{gq}^{(0)}-5\,\beta_0\,
P_{gq}^{(0)} \Big )
\nonumber\\[2ex]
&&+\frac{1}{\varepsilon}\,\Big (\frac{1}{2}\,P_{gq}^{(1)}
-6\,\beta_0\,w_{gq,{\rm B}}^{(1)}+\frac{1}{2}\,P_{qg}^{(0)}\otimes 
w_{q\bar q,{\rm B}}^{(1)}
+P_{gq}^{(0)}\otimes w_{gg,{\rm B}}^{(1)}
\nonumber\\[2ex]
&&+(P_{gg}^{(0)}+P_{qq}^{(0)})
\otimes w_{gq,{\rm B}}^{(1)}\Big  ) 
-6\,\beta_0\,\bar w_{gq,{\rm B}}^{(1)}+\frac{1}{2}\,P_{qg}^{(0)}\otimes 
\bar w_{q\bar q,{\rm B}}^{(1)}
\nonumber\\[2ex]
&&+P_{gq}^{(0)}\otimes \bar w_{gg,{\rm B}}^{(1)}+(P_{gg}^{(0)}+P_{qq}^{(0)})
\otimes \bar w_{gq,{\rm B}}^{(1)} +w_{gq,{\rm B}}^{(2)}\Bigg ]\,,
\end{eqnarray}
\begin{eqnarray}
\label{eqn3.24}
z^{-1}\,\hat W_{q\bar q,{\rm B}}&=&\hat a_s\,S_{\varepsilon}\,
\left (\frac{m^2}{\mu^2}
\right )^{\varepsilon/2}\,\Bigg [w_{q\bar q,{\rm B}}^{(1)}+\varepsilon\,
{\bar w}_{q\bar q,{\rm B}}^{(1)} \Bigg ]+\hat a_s^2\,S_{\varepsilon}^2\,
\left (\frac{m^2}{\mu^2}\right )^{\varepsilon}\,\Bigg [
\nonumber\\[2ex]
&&\frac{1}{\varepsilon^2}\,P_{gq}^{(0)}\otimes P_{gq}^{(0)}
+\frac{1}{\varepsilon}\,\Big (-6\,\beta_0\,w_{q\bar q}^{(1)}+2\,P_{gq}^{(0)}
\otimes w_{gq,{\rm B}}^{(1)}
\nonumber\\[2ex]
&&+2\,P_{qq}^{(0)}\otimes w_{q\bar q,{\rm B}}^{(1)}\Big )
-6\,\beta_0\, \bar w_{q\bar q,{\rm B}}^{(1)}
+2\,P_{gq}^{(0)}\otimes \bar w_{gq,{\rm B}}^{(1)}
\nonumber\\[2ex]
&&+2\,P_{qq}^{(0)}\otimes \bar w_{q\bar q,{\rm B}}^{(1)}
+w_{q\bar q,{\rm B}}^{(2)} \Bigg ]\,,
\end{eqnarray}
\begin{eqnarray}
\label{eqn3.25}
z^{-1}\,\hat W_{q_1q_2,{\rm B}}&=&z^{-1}\,\hat W_{q_1\bar q_2,{\rm B}}\,,
\nonumber\\[2ex]
&=&\hat a_s^2\,S_{\varepsilon}^2\,
\left (\frac{m^2}{\mu^2}\right )^{\varepsilon}\,\Bigg [
\frac{1}{\varepsilon^2}\,P_{gq}^{(0)}\otimes P_{gq}^{(0)}
+\frac{2}{\varepsilon}\,P_{gq}^{(0)}\otimes w_{gq,{\rm B}}^{(1)}
\nonumber\\[2ex]
&&+2\,P_{gq}^{(0)}\otimes \bar w_{gq,{\rm B}}^{(1)}+w_{q_1q_2,{\rm B}}^{(2)} 
\Bigg ]\,,
\end{eqnarray}
where $\otimes$ denotes the convolution symbol defined by
\begin{eqnarray}
\label{eqn3.26}
f\otimes g(z)=\int_0^1dz_1\,\int_0^1dz_2\,\delta(z-z_1\,z_2)\,f(z_1)\,g(z_2)
\,.
\end{eqnarray}
For identical quark-quark scattering we have the same formula as in  
Eq. (\ref{eqn3.25}) except that 
$w_{qq,{\rm B}}^{(2)}\not =w_{q_1q_2,{\rm B}}^{(2)}$. 
The expressions
above follow from the renormalization group equations. They are constructed
in such a way that become finite after coupling constant renormalization,
operator renormalization
and mass factorization are carried out. The splitting functions $P_{ab}(z)$
and the coefficients $w_{ab}(z)$ with $z=m^2/s$ also occur in the coefficient 
functions given below except for the NLO terms $\bar w^{(1)}_{ab}(z)$, which
are proportional to $\varepsilon$. They are given by
\begin{eqnarray}
\label{eqn3.27}
\bar w_{gg,{\rm H}}^{(1)}&=&C_A\,\Bigg [8\,{\cal D}_2(z)-6\,\zeta(2)\,
{\cal D}_0(z) +\left (\frac{8}{z}-16+8\,z-8\,z^2\right )\,\Big (\ln^2 (1-z)
\nonumber\\[2ex]
&&-\frac{3}{4}\,
\zeta(2)\Big )+\left (\frac{8}{1-z}+\frac{8}{z}-16+8\,z-8\,z^2\right )\,
\Big (\frac{1}{4}\,\ln^2 z
\nonumber\\[2ex]
&&-\ln z\,\ln (1-z)\Big )-\frac{22}{3}\,\frac{(1-z)^3}{z}\,\Big (
\ln (1-z)-\frac{1}{2}\,\ln z\Big )-\frac{55}{3}+\frac{67}{9z}
\nonumber\\[2ex]
&&+\frac{55}{3}\,z -\frac{67}{9}\,z^2+2\,\delta(1-z)\Bigg ]\,,
\nonumber\\[2ex]
\bar w_{gq,{\rm H}}^{(1)}&=&C_F\,\Bigg [\left (\frac{4}{z}-4+2\,z\right )\,
\Big ( \ln^2 (1-z)-\ln z\,\ln (1-z)+\frac{1}{4}\ln^2 z
\nonumber\\[2ex]
&&-\frac{3}{4}\,\zeta(2)\Big )
+\left (-\frac{3}{z}+6-z\right )\,\Big (\ln (1-z)-\frac{1}{2}\,\ln z\Big )
+\frac{7}{2z}-5+\frac{3}{2}\,z\Bigg ]\,,
\nonumber\\[2ex]
\bar w_{q\bar q,{\rm H}}^{(1)}&=&C_F^2\,\Bigg [\frac{8}{3}\,
\frac{(1-z)^3}{z}\,\Big (\ln (1-z)
-\frac{1}{2}\,\ln z+\frac{1}{6} \Big )\Bigg ]\,,   
\end{eqnarray}
where ${\cal D}_i$ denotes the distribution
\begin{eqnarray}
\label{eqn3.28}
{\cal D}_i=\left (\frac{\ln^i(1-z)}{1-z}\right )_+\,.
\end{eqnarray}
The difference between scalar and pseudo-scalar Higgs production is given
by
\begin{eqnarray}
\label{eqn3.29}
\bar w_{gg,{\rm A-H}}^{(1)}&=&C_A\,\Bigg [-8\,(1-z)-14\,\delta(1-z)\Bigg ]\,,
\nonumber\\[2ex]
\bar w_{gq,{\rm A-H}}^{(1)}&=&0\,,
\nonumber\\[2ex]
\bar w_{q\bar q,{\rm A-H}}^{(1)}&=&0\,.
\end{eqnarray}
To render the partonic cross sections finite one has first to
perform coupling constant renormalization. This is done by replacing
the bare coupling constant by the renormalized one see Eq. (\ref{eqn2.25}).
Then one has to carry out operator renormalization which is achieved by 
multiplying
the expressions in Eqs. (\ref{eqn3.22})-(\ref{eqn3.25}) with
the operator renormalization constants in Eqs. (\ref{eqn2.21}),
(\ref{eqn2.24}). The remaining divergences, which are of collinear origin,
are removed by mass factorization 
\begin{eqnarray}
\label{eqn3.30}
z^{-1}\,Z_O^2\,\hat W_{ab,{\rm B}}\left (\frac{1}{\varepsilon},\frac{m^2}{\mu^2}
\right )= \sum_{c,d=q,\bar q,g}\Gamma_{ca}\left (\frac{1}{\varepsilon}\right )
\otimes \Gamma_{db}\left (\frac{1}{\varepsilon}\right )\otimes z^{-1}\,
\Delta_{cd,{\rm B}} \left ( \frac{m^2}{\mu^2}\right )\,,
\nonumber\\[2ex]
\end{eqnarray}
where $\Gamma_{ca}(z)$ denote the kernels containing the splitting
functions which multiply the collinear divergences represented by
the pole terms $1/\varepsilon^k$. Note that in the case of
pseudo-scalar Higgs production the $Z_O$ (see Eq. (\ref{eqn2.21}))
in the above equation has to be
replaced by $Z_{11}$ in Eq. (\ref{eqn2.24}).
For the four different subprocesses the mass 
factorization relations in Eq. (\ref{eqn3.30}) become equal to
\begin{eqnarray}
\label{eqn3.31}
z^{-1}\,Z_O^2\hat W_{gg,{\rm B}}&=&\Gamma_{gg}\otimes \Gamma_{gg}\otimes 
z^{-1}\,\Delta_{gg,{\rm B}}+4\,\Gamma_{gg}\otimes \Gamma_{qg}\otimes 
z^{-1}\,\Delta_{gq,{\rm B}}\,,
\nonumber\\[2ex]
\\[2ex]
\label{eqn3.32}
z^{-1}\,Z_O^2\,\hat W_{gq,{\rm B}}&=&\Gamma_{qq}\otimes \Gamma_{gg}
\otimes z^{-1}\,
\Delta_{gq,{\rm B}}+\Gamma_{gq} \otimes \Gamma_{gg}\otimes z^{-1}\,
\Delta_{gg,{\rm B}}
\nonumber\\[2ex]
&&+\Gamma_{qq}\otimes\Gamma_{qg}\otimes z^{-1}\,\Delta_{q\bar q,{\rm B}}\,,
\\[2ex]
\label{eqn3.33}
z^{-1}\,Z_O^2\,\hat W_{q\bar q,{\rm B}}&=&\Gamma_{gq}\otimes \Gamma_{gq}\otimes 
z^{-1}\, \Delta_{gg,{\rm B}}+\Gamma_{qq}\otimes\Gamma_{qq}\otimes z^{-1}\,
\Delta_{q\bar q,{\rm B}}
\nonumber\\[2ex]
&&+2\,\Gamma_{gq}\otimes \Gamma_{qq}\otimes z^{-1}\,\Delta_{gq,{\rm B}}\,,
\\[2ex]
\label{eqn3.34}
z^{-1}\,Z_O^2\,\hat W_{q_1q_2,{\rm B}}&=&\Gamma_{gq}\otimes \Gamma_{gq}\otimes 
z^{-1}\,\Delta_{gg,{\rm B}} +2\,\Gamma_{gq}\otimes \Gamma_{qq}\otimes 
z^{-1}\,\Delta_{gq,{\rm B}}
\nonumber\\[2ex]
&&+\Gamma_{qq} \otimes \Gamma_{qq}\otimes z^{-1}\,\Delta_{q_1q_2,{\rm B}}\,,
\end{eqnarray}
where we have identified
\begin{eqnarray}
\label{eqn3.35}
&&\Gamma_{qg}=\Gamma_{\bar qg}\quad , \quad \Gamma_{gq}=\Gamma_{g\bar q}\,,
\nonumber\\[2ex]
&& \hat W_{gq,{\rm B}}=\hat W_{g\bar q,{\rm B}}=\hat W_{qg,{\rm B}}
=\hat W_{\bar qg,{\rm B}}\,,
\nonumber\\[2ex]
&&\Delta_{gq,{\rm B}}=\Delta_{g\bar q,{\rm B}}=\Delta_{qg,{\rm B}}
=\Delta_{\bar qg,{\rm B}} \,.
\end{eqnarray}
Since we need the finite expressions up to order $\alpha_s^2$ it is 
sufficient to expand the kernels $\Gamma_{ca}$ up to the following order
in the renormalized coupling constant
\begin{eqnarray}
\label{eqn3.36}
\Gamma_{qq}&=&\delta(1-z)+a_s(\mu^2)\,S_{\varepsilon}\,\Bigg 
[\frac{1}{\varepsilon}\, P_{qq}^{(0)}\Bigg ]\,,
\\[2ex]
\label{eqn3.37}
\Gamma_{qg}&=&a_s(\mu^2)\,S_{\varepsilon}\,\Bigg [\frac{1}{2\varepsilon}\,
P_{qg}^{(0)}\Bigg ]\,,
\\[2ex]
\label{eqn3.38}
\Gamma_{qq}^{\rm PS}&=&a_s^2(\mu^2)\,S_{\varepsilon}^2\,\Bigg [
\frac{1}{4\varepsilon^2}\,P_{qg}^{(0)}\otimes P_{gq}^{(0)}+
\frac{1}{4\varepsilon^2}\,P_{qq}^{(1),{\rm PS}}\Bigg ]\,,
\\[2ex]
\label{eqn3.39}
\Gamma_{gq}&=&a_s(\mu^2)\,S_{\varepsilon}\,\Bigg [\frac{1}{\varepsilon}\,
P_{gq}^{(0)}\Bigg ]+a_s^2(\mu^2)\,S_{\varepsilon}^2\,\Bigg [
\frac{1}{\varepsilon^2}\,\Big (
\frac{1}{2}\,P_{gq}^{(0)}\otimes P_{gg}^{(0)} 
\nonumber\\[2ex]
&&+\frac{1}{2}\,P_{qq}^{(0)}\otimes P_{gq}^{(0)}
+\beta_0\,P_{gq}^{(0)}\Big )+\frac{1}{2\varepsilon}\,P_{gq}^{(1)} \Bigg ]\,,
\\[2ex]
\label{eqn3.40}
\Gamma_{gg}&=&\delta(1-z)+a_s(\mu^2)\,S_{\varepsilon}\,
\Bigg [\frac{1}{\varepsilon}\,
P_{gg}^{(0)}\Bigg ]+a_s^2(\mu^2)\,S_{\varepsilon}^2\,\Bigg 
[\frac{1}{\varepsilon^2}
\,\Big (\frac{1}{2}\,P_{gg}^{(0)}\otimes P_{gg}^{(0)}
\nonumber\\[2ex]
&&+\frac{1}{2}\,P_{gq}^{(0)}\otimes P_{qg}^{(0)}
+\beta_0\,P_{gg}^{(0)}\Big )+\frac{1}{2\varepsilon}\,P_{gg}^{(1)}\Bigg ]\,,
\end{eqnarray}
(PS denotes pure-singlet).
After renormalization and mass factorization the coefficient functions have the 
following algebraic form
\begin{eqnarray}
\label{eqn3.41}
z^{-1}\,\Delta_{gg,{\rm B}}&=&\delta(1-z)+a_s(\mu^2)\,\Bigg [P_{gg}^{(0)}\,
\ln \frac{m^2}
{\mu^2}+w_{gg,{\rm B}}^{(1)}\Bigg ]+a_s^2(\mu^2)\,\Bigg [
\nonumber\\[2ex]
&&\left \{\frac{1}{2}\,P_{gg}^{(0)} \otimes P_{gg}^{(0)}+\frac{1}{4}\,
P_{gq}^{(0)}\otimes P_{qg}^{(0)}-\frac{5}{2}\,
\beta_0\,P_{gg}^{(0)}+3\,\beta_0^2\right \}\,\ln^2 \frac{m^2}{\mu^2}
\nonumber\\[2ex]
&&+\Big \{P_{gg}^{(1)}-3\,\beta_0\,w_{gg,{\rm B}}^{(1)}+P_{gg}^{(0)}\otimes 
w_{gg,{\rm B}}^{(1)}+w_{gq,{\rm B}}^{(1)} \otimes P_{qg}^{(0)}
\nonumber\\[2ex]
&&+z_{\rm B}\Big \} \,\ln \frac{m^2}{\mu^2}+w_{gg,{\rm B}}^{(2)}
\Bigg ]\,,
\\[2ex]
\label{eqn3.42}
z^{-1}\,\Delta_{gq,{\rm B}}&=&a_s(\mu^2)\,\Bigg [\frac{1}{2}\,P_{gq}^{(0)}\,
\ln \frac{m^2}
{\mu^2}+w_{gq,{\rm B}}^{(1)}\Bigg ]+a_s^2(\mu^2)\,\Bigg [\left \{\frac{3}{8}\,
P_{gg}^{(0)}
\otimes P_{gq}^{(0)}\right.
\nonumber\\[2ex]
&&\left. +\frac{1}{8}\,P_{qq}^{(0)}\otimes P_{gq}^{(0)}-\frac{5}{4}\,
\beta_0\,P_{gq}^{(0)}\right \}\,\ln^2 \frac{m^2}{\mu^2}+\left \{
\frac{1}{2}\,P_{gq}^{(1)}-3\,\beta_0\,w_{gq,{\rm B}}^{(1)}\right. 
\nonumber\\[2ex]
&&\left.+\frac{1}{4}\,P_{qg}^{(0)}\otimes w_{q\bar q,{\rm B}}^{(1)}
+\frac{1}{2}\,P_{gq}^{(0)}\otimes w_{gg,{\rm B}}^{(1)} \right.
\nonumber\\[2ex]
&&\left.+\frac{1}{2}\,\Big(P_{gg}^{(0)}+P_{qq}^{(0)}\Big )\otimes 
w_{gq,{\rm B}}^{(1)}\right \}\, \ln \frac{m^2}{\mu^2}
+w_{gq,{\rm B}}^{(2)}\Bigg ]\,,
\\[2ex]
\label{eqn3.43}
z^{-1}\,\Delta_{q\bar q,{\rm B}}&=&a_s(\mu^2)\,\Bigg [
w_{q\bar q,{\rm B}}^{(1)}\Bigg ]
+a_s^2(\mu^2)\,
\Bigg [\frac{1}{4}\,P_{gq}^{(0)}\otimes P_{gq}^{(0)}\,\ln^2 \frac{m^2}
{\mu^2}
\nonumber\\[2ex]
&& +\left \{-3\,\beta_0\,w_{q\bar q,{\rm B}}^{(1)}+P_{gq}^{(0)}\otimes 
w_{gq,{\rm B}}^{(1)}+P_{qq}^{(0)}\otimes w_{q\bar q,{\rm B}}^{(1)}\right \}\,
\ln \frac{m^2}{\mu^2}
\nonumber\\[2ex]
&& +w_{q\bar q,{\rm B}}^{(2)} \Bigg ]\,,
\\[2ex]
\label{eqn3.44}
z^{-1}\,\Delta_{q_1q_2,{\rm B}}&=&z^{-1}\,\Delta_{q_1\bar q_2,{\rm B}}\,,
\nonumber\\[2ex]
&=&a_s^2(\mu^2)\,\Bigg [\frac{1}{4}\,P_{gq}^{(0)}\otimes 
P_{gq}^{(0)}\,\ln^2 \frac{m^2}{\mu^2} + P_{gq}^{(0)}\otimes w_{gq,{\rm B}}^{(1)}
\,\ln \frac{m^2}{\mu^2} 
\nonumber\\[2ex]
&&+w_{q_1q_2,{\rm B}}^{(2)} \Bigg ]\,.
\end{eqnarray}
In Appendix A and B we give the explicit expressions for the
coefficient functions so that one can determine the coefficients 
$P_{ab}^{(k)}$ and $w_{ab,{\rm B}}^{(k)}$. Our results which are expressed in 
the colour factors $C_A$, $C_F$ and $T_f$ agree with those published
in \cite{haki3}, \cite{anme1}, \cite{anme2} for $N=3$.
In the representation of the coefficient functions above we have
put the renormalization scale $\mu_r$ equal to the mass 
factorization scale $\mu$. If one wants to distinguish between both scales
one can make the simple substitution 
\begin{eqnarray}
\label{eqn3.45}
\alpha_s(\mu^2)=\alpha_s(\mu_r^2)\left [1 + \frac{\alpha_s(\mu_r^2)}{4\pi}
\beta_0\,\ln \frac{\mu_r^2}{\mu^2}\right ]\,.
\end{eqnarray}

\mysection{Total cross sections for the process\\
 $p + p\rightarrow H +'X'$}
In this section we will present total cross sections (see Eq. (\ref{eqn2.1}))
for Higgs-boson production in proton-proton collisions at the LHC and
in proton-anti-proton collisions at the TEVATRON. 
The cross section can be written in two ways (for the definitions see 
\cite{cafl1}, \cite{cafl2})
\begin{eqnarray}
\label{eqn4.1}
\sigma_{\rm tot}=\frac{\pi\,G_{\rm B}^2}{8\,(N^2-1)}\,\sum_{a,b=q,\bar q,g}
\int_x^1\,dy\, \tilde \Phi_{ab}(y,\mu^2)\,
\Delta_{ab}\left (\frac{x}{y},\frac{m^2}{\mu^2}\right )\,,
\end{eqnarray}
where $x=m^2/S$ and $\tilde \Phi_{ab}$ is the momentum fraction luminosity 
defined by
\begin{eqnarray}
\label{eqn4.2}
\tilde \Phi_{ab}(y,\mu^2)=\int_y^1\frac{du}{u}\,f_a(u,\mu^2)\,f_b\left (
\frac{y}{u},\mu^2\right )\,.
\end{eqnarray}
However Eq. (\ref{eqn4.1}) can be also cast in the form
\begin{eqnarray}
\label{eqn4.3}
&&\sigma_{\rm tot}=\frac{\pi\,G_{\rm B}^2}{8\,(N^2-1)}\,\sum_{a,b=q,\bar q,g}
x\,\int_x^1\,dy\, \Phi_{ab}(y,\mu^2)\,\left (\frac{y}{x}\,
\Delta_{ab}\left (\frac{x}{y},\frac{m^2}{\mu^2}\right )\right )\,,
\nonumber\\[2ex]
&&\mbox{with}\quad \Phi_{ab}(y,\mu^2)=y^{-1}\,\tilde \Phi_{ab}(y,\mu^2)\,,
\end{eqnarray}
where $\Phi_{ab}$ denotes the parton luminosity.
For the exact cross section the expressions in Eqs. (\ref{eqn4.1})
and (\ref{eqn4.3}) lead to the same results but if one wants to study
the various approximations in the literature like the soft-plus-virtual
(SV) gluon approximation then it makes a difference if $\Delta_{ab}(z)$
or $z^{-1}\,\Delta_{ab}(z)$ will be expanded around $z=1$. This difference
will show up in the less singular terms as we will show at the
end of this section.
In the subsequent part of this section we study the dependence of the cross 
section on the input parameters like the QCD scale $\Lambda$,
the renormalization/factorization scale $\mu$, the mass $m$ of the Higgs boson
and the input parton densities. We are also interested in the theoretical
uncertainty of the NNLO cross section. One uncertainty is due to the
corrections which show up beyond NNLO. They can be guessed from the rate
of convergence represented by the $K$-factor and the variation 
with respect to the scale $\mu$. We will show that the coefficient functions
$\Delta_{ab}(z)$ in Eq. (\ref{eqn4.1}) are dominated by  
the logarithmic terms of the type $\ln^i(1-z)$. A determination of these
logarithms is of importance to estimate the higher order corrections.
Another uncertainty is caused by the input parton 
densities. Parton density sets differ from each other in several aspects. 
They are based on fitting data from different experiments.
Furthermore the shapes of the densities at the specific input scale
chosen for $\mu$ differ from each other. Finally these sets are based on 
different theoretical assumptions, for instance in the treatment of heavy
flavours. In some sets the heavy flavour is treated as a massless particle
which is described by a parton density similar to the treatment of the
light quarks. In other sets the mass of the heavy flavour is considered to be
on the same order of magnitude as the other large scales. Then the
heavy flavour production is described by exact perturbation theory at a 
certain order so that threshold effects are fully taken into account. 
Although the coefficient functions
in the effective Lagrangian approach are computed exactly in NNLO this is not
the case for the parton densities because the exact three-loop splitting
functions (anomalous dimensions) are not known yet. Until now only
a finite number of moments are available (see \cite{larin2}) which are used in 
\cite{nevo} together with other constraints to approximate the splitting 
functions.
These approximations are very reliable as long as $y>10^{-4}$ in Eq. 
(\ref{eqn4.1}). The approximated splitting functions were used in \cite{mrst02}
to obtain NNLO parton density sets. One of them called MRST02 (see Table 
\ref{table1}) will be used in our paper. For the approximations in LO
and NLO we use the sets in \cite{mrst02} and \cite{mrst01} respectively. 
For the LO, NLO and NNLO plots we employ the one-, two-, and three-loop
asymptotic forms of the running coupling constant as given in Eq. (3) of
\cite{chkn}. In order to make a comparison with other parametrizations
for the parton densities we also choose the sets made by CTEQ \cite{lai} and 
GRV \cite{grv}. However these sets do not contain NNLO versions so that we
can present the NLO results only.
All sets are listed in Table \ref{table1} 
\footnote{Notice that according to the prescription of the CTEQ group 
\cite{lai} also the LO corrected quantities have to be computed with 
the NLO running coupling constant.}
together with the 
corresponding QCD scale $\Lambda_{n_f}$ determined for
$n_f=5$. The same number of flavours is also chosen for the coefficient 
functions.
\begin{table}
\begin{center}
\begin{tabular}{|c|c|c|}\hline
MRST02 (LO, lo2002.dat) & $\Lambda_5^{\rm LO}=167~{\rm MeV}$  &
$\alpha_s^{\rm LO}(M_Z)=0.130$      \\
MRST01 (NLO, alf119.dat)    & $\Lambda_5^{\rm NLO}=239~{\rm MeV}$ &
$\alpha_s^{\rm NLO}(M_Z)=0.119$       \\
MRST02 (NNLO, vnvalf1155.dat) & $\Lambda_5^{\rm NNLO}=176~{\rm MeV}$ &
$\alpha_s^{\rm NNLO}(M_Z)=0.115$ \\
CTEQ6 (LO, cteq6l.tbl)     & $\Lambda_5^{\rm NLO}=226~{\rm MeV}$   &
$\alpha_s^{\rm NLO}(M_Z)=0.118$       \\
CTEQ6 (NLO, cteq6m.tbl)    & $\Lambda_5^{\rm NLO}=226~{\rm MeV}$  &
$\alpha_s^{\rm NLO}(M_Z)=0.118$    \\
GRV98 (LO, grv98lo.grid) & $\Lambda_5^{\rm LO}=132~{\rm MeV}$   &
$\alpha_s^{\rm LO}(M_Z)=0.125$   \\
GRV98 (NLO, grvnlm.grid) & $\Lambda_5^{\rm NLO}=174~{\rm MeV}$  &
$\alpha_s^{\rm NLO}(M_Z)=0.114$       \\
\hline
\end{tabular}
\end{center}
\caption{Various parton density sets with their values for the QCD scale
$\Lambda$ and $\alpha_s(M_Z)$.}
\label{table1}
\end{table}
Notice that the GRV sets do not contain densities for charm and bottom quarks.
For simplicity the factorization scale is set equal to the renormalization
scale $\mu_r$. For our plots we take $\mu=m$ and use the 
MRST sets for the LO, NLO and NNLO computations (see Table \ref{table1}) 
unless mentioned otherwise.
Here we want to emphasize that the magnitude of the $\sigma_{\rm tot}$ is
extremely sensitive to the renormalization scale because the effective 
coupling constant $G_{\rm B}\sim \alpha_s(\mu_r)$, which implies that
$\sigma^{\rm LO}\sim \alpha_s^2$. The sensitivity to the factorization scale
is much smaller.

For the computation of the effective coupling constant $G_{\rm B}$ in Eq. 
(\ref{eqn2.3}) we choose the top quark mass $m_t=173.4~{\rm GeV/c^2}$ and the
Fermi constant $G_F=1.16639~{\rm GeV}^{-2}=4541.68~{\rm pb}$. 
In this paper we will only study scalar Higgs production  
and omit pseudo-scalar Higgs production.
The cross sections of the latter are about $9/4~\cot^2 \beta$ larger
than those for the standard model Higgs so that all our conclusions also
apply to pseudo-scalar Higgs production. 
We give results for both proton-proton collisions at the centre-of-mass 
energy $\sqrt{S}=14~{\rm TeV}$ and proton-anti-proton collisions
at $\sqrt{S}=2~{\rm TeV}$. 

In Fig. 1 we have plotted the contributions coming from the various 
subprocesses up to NLO. From this figure we infer that $\sigma_{\rm tot}$
is completely dominated by the $gg$ reaction whereas the other contributions
are down by two ($gq+g\bar q$-subprocess) or even by 
three ($q\bar q$-subprocess) 
orders of magnitude. This picture is different from the behaviour of the
transverse momentum $p_T$- and rapidity $y$-distributions in \cite{rasm1} where 
the $gq+g\bar q$-reaction amounted to about $1/3$ of that for 
the $gg$-reaction when $p_T>30~{\rm GeV/c}$. 
This is because $\sigma_{\rm tot}$ receives all its
contribution from the small $p_T$-region ($p_T<30~{\rm GeV/c}$) where the $gg$ 
subprocess overwhelms all other reactions. This picture is not changed when
we study the cross section in NNLO (see Fig. 2) where new subprocesses
contribute given by the $qq$, $q\bar q$ and $\bar q\bar q$ reactions. Notice 
that the incoming (anti-) quarks can be identical and non-identical. The 
relative order of magnitude is the same as in NLO and the 
new reactions are down by three orders of magnitude. 
In Fig. 3 we have plotted $\sigma_{\rm tot}$ 
in LO, NLO and NNLO as a function of $m$. 
The cross section decreases when $m$ increases due to a reduction in
the available phase space. In NNLO the curve shows a steeper decrease 
than in NLO and LO.
Furthermore the growth of the cross section slows down while going to higher
orders. To show more clearly how the cross sections change with respect
to $m$ we have put them in Table \ref{table2}. We have also
shown them for the TEVATRON (proton-anti-proton collisions 
at $\sqrt{S}=2~{\rm TeV}$) in Table \ref{table3}
where they are appreciably smaller than in the case of the LHC.

The NNLO cross sections in Tables \ref{table2}, \ref{table3} have some
theoretical errors which can be estimated in various ways.
First one can study their variation with respect to the scale $\mu$. This
can be achieved by computing the ratio
\begin{table}
\begin{center}
\begin{tabular}{|c|c|c|c|}\hline
   mass &  LO  &  NLO &  NNLO \\
\hline
  100 &  30.35  & 52.75 &  60.84 \\
  110 &  25.53  & 44.75 &  51.80 \\
  120 &  21.77  & 38.43 &  44.62 \\
  130 &  18.77  & 33.37 &  38.85 \\
  140 &  16.36  & 29.27 &  34.17 \\
  150 &  14.38  & 25.88 &  30.29 \\
  160 &  12.74  & 23.06 &  27.07 \\
  170 &  11.37  & 20.69 &  24.36 \\
  180 &  10.23  & 18.69 &  22.09 \\
  190 &   9.254 & 17.00 &  20.10 \\
  200 &   8.425 & 15.53 &  18.43 \\
  210 &   7.714 & 14.28 &  16.97 \\
  220 &   7.102 & 13.20 &  15.71 \\
  230 &   6.573 & 12.26 &  14.62 \\
  240 &   6.115 & 11.45 &  13.69 \\
  250 &   5.718 & 10.74 &  12.86 \\
  260 &   5.375 & 10.13 &  12.16 \\
  270 &   5.080 &  9.604 & 11.55 \\
  280 &   4.827 &  9.152 & 11.03 \\
  290 &   4.615 &  8.772 & 10.59 \\
  300 &   4.441 &  8.468 & 10.24 \\
\hline
\end{tabular}
\end{center}
\caption{Total cross sections in pb for Higgs masses between
100 GeV and 300 GeV at the LHC. The LO, NLO and NNLO results
are generated with the MRST parton densities listed in Table 1. }
\label{table2}
\end{table}
\begin{table}
\begin{center}
\begin{tabular}{|c|c|c|c|}\hline
   mass &  LO  &  NLO &  NNLO \\
\hline
  100  &  0.6116  &  1.351  &  1.781  \\
  110  &  0.4608  &  1.025  &  1.363  \\
  120  &  0.3530  &  0.791  &  1.060  \\
  130  &  0.2745  &  0.619  &  0.835  \\
  140  &  0.2161  &  0.491  &  0.666  \\
  150  &  0.1721  &  0.393  &  0.538  \\
  160  &  0.1385  &  0.318  &  0.438  \\
  170  &  0.1124  &  0.259  &  0.359  \\
  180  &  0.0920  &  0.213  &  0.293  \\
  190  &  0.0758  &  0.177  &  0.248  \\
  200  &  0.0629  &  0.147  &  0.207  \\
  210  &  0.0526  &  0.124  &  0.176  \\
  220  &  0.0442  &  0.105  &  0.149  \\
  230  &  0.0374  &  0.089  &  0.127  \\
  240  &  0.0318  &  0.076  &  0.109  \\
  250  &  0.0270  &  0.065  &  0.095  \\
  260  &  0.0234  &  0.056  &  0.082  \\
  270  &  0.0202  &  0.049  &  0.072  \\
  280  &  0.0176  &  0.043  &  0.063  \\
  290  &  0.0154  &  0.038  &  0.055  \\
  300  &  0.0136  &  0.033  &  0.049  \\
\hline
\end{tabular}
\end{center}
\caption{Total cross sections in pb for Higgs masses between
100 GeV and 300 GeV at the TEVATRON. The LO, NLO and NNLO results
are generated with the MRST parton densities listed in Table 1. }
\label{table3}
\end{table}
\begin{eqnarray}
\label{eqn4.4}
N\left (\frac{\mu}{\mu_0}\right )=\frac{\sigma_{\rm tot}(\mu)}
{\sigma_{\rm tot}(\mu_0)}\,,
\end{eqnarray}
where $\mu_0=m$ and $\mu$ is varied in the range $0.1<\mu/\mu_0<10$.
In Fig. 4 a plot of $N$ is shown for $m=100~{\rm GeV/c^2}$. Here one observes
a clear improvement while going from LO to NNLO. In particular the curve
for NNLO is flatter than that for NLO. The improvement becomes even better
when we choose a larger Higgs mass see e.g. Fig. 5 ($m=200~{\rm GeV/c^2}$).
However for still larger masses there is hardly any improvement anymore
(see Fig. 6 where $m=300~{\rm GeV/c^2}$). This is in contrast to 
the NNLO corrected cross section for vector boson production in \cite{ham}
which is insensitive to the choice of scale chosen in the range above.

A second way to study the reliability of our prediction is to study the 
rate of convergence of the perturbation series, which is 
represented by the $K$-factor. We choose the following definitions
\begin{eqnarray}
\label{eqn4.5}
K^{\rm NLO}=\frac{\sigma_{\rm tot}^{\rm NLO}}{\sigma_{\rm tot}^{\rm LO}} 
\quad\,,\quad
K^{\rm NNLO}=\frac{\sigma_{\rm tot}^{\rm NNLO}}{\sigma_{\rm tot}^{\rm NLO}}\,. 
\end{eqnarray}
Notice that in the above expression the definition of $K^{\rm NNLO}$
differs from the usual one which is given by $K^{\rm NNLO}=
\sigma_{\rm tot}^{\rm NNLO}/\sigma_{\rm tot}^{\rm LO}$. We have chosen
this definition because the rate of convergence can be shown in a better way.
In Fig. 7 one observes that both $K$-factors vary slowly as 
$m$ increases. Moreover there is considerable improvement in the rate of
convergence if one goes from NLO to NNLO. At $m=100~{\rm GeV/c^2}$
we have $K^{\rm NLO}\sim 1.74$ whereas $K^{\rm NNLO}\sim 1.15$. Still
the corrections for Higgs boson production are larger than those
obtained from $Z$-boson production at the LHC where one gets
$K^{\rm NLO}\sim 1.22$ and $K^{\rm NNLO}\sim 0.95$ (see \cite{ham}). 
In Fig. 8
we compare the $K$-factor in NLO obtained from the MRST with the results
computed by using the parton density sets given by the CTEQ \cite{lai} and 
GRV \cite{grv} collaborations (see Table \ref{table1}). From this figure we 
infer that the $K$-factor computed using GRV98 and CTEQ6 is larger than the one 
obtained from MRST01(NLO) at $m=100~{\rm GeV/c^2}$. However at large $m$ there 
is a cross over and we observe $K^{\rm CTEQ}>K^{\rm MRST}>K^{\rm GRV98}$ .
The large result shown by the CTEQ6-set is due to the behaviour of their 
gluon density in LO. In NLO this density behaves in the same way
as those in the MRST and GRV sets.  We also investigated older sets
provided by the CTEQ-collaboration, namely CTEQ5 and CTEQ6. It turns out that
all these LO CTEQ sets lead to the same large value for $K^{\rm NLO}$ as shown
by Fig. 8. A study of $K^{\rm NLO}$, using the most recent parton densities,
was also made in \cite{cafl2} but now for the TEVATRON.

After the K-factor we now make a comparison between
$\sigma_{\rm tot}$ computed from the various sets which leads to
the third uncertainty in the theoretical prediction. For that purpose we plot
the ratios
\begin{eqnarray}
\label{eqn4.6}
R^{\rm CTEQ}=\frac{\sigma_{\rm tot}^{\rm CTEQ}}{\sigma_{\rm tot}^{\rm MRST}}
\quad\,,\quad
R^{\rm GRV}=\frac{\sigma_{\rm tot}^{\rm GRV}}{\sigma_{\rm tot}^{\rm MRST}}\,,
\end{eqnarray}
which are shown in NLO in Fig. 9.
In this figure we see that the CTEQ6 set leads to the same cross section as the
one obtained by MRST01. The cross section given by the GRV98 is $20\%$ 
above the MRST01 result and the ratio decreases with increasing $m$ to
$5\%$ above at $m=300~{\rm GeV/c^2}$. 
The reason that
the CTEQ6 result is closer to the MRST sets can be attributed to the fact that
their parton density sets are more recent and are constructed to fit
the same experiments. Notice that a study of the ratios in Eq. (\ref{eqn4.6})
was also made for the TEVATRON in \cite{cafl2} using the most recent parton 
densities. 

The factorization scale dependence in Figs. 4-6 and
the $K$-factor in Fig. 7 of the NNLO cross section can be used to give an error 
estimate
of the latter for the LHC. Since $(\sigma^{\rm NNLO}-\sigma^{\rm NLO})/
\sigma^{\rm NLO}=K^{\rm NNLO}-1$ the error on $\sigma^{\rm NNLO}$ is
equal to $15\%$, $19\%$ and $21\%$ for $m=100~{\rm GeV/c^2}$,
$m=200~{\rm GeV/c^2}$ and $m=300~{\rm GeV/c^2}$ respectively (see Fig. 7). 
This is corroborated by the scale variation if we choose the range
$0.25<\mu/\mu_0<4.0$. For $\mu/\mu_0=0.25$ the errors above become
$18\%$, $15\%$ and $14\%$ respectively. In the case $\mu/\mu_0=0.25$ 
they are given by $16\%$ for all $m$ given above. 
Notice that the error due the $K$-factor increases when $m$
gets larger whereas the opposite happens if the error estimate is derived
from the mass factorization/renormalisation scale.
On top of this we have to add the error due to the chosen parton density.
For this we choose the NLO plots in Fig. 9 
where the cross section predicted by CTEQ6 hardly 
differs from the one given by MRST01 but the GRV98 plot deviates
from the latter by $20\%$ for small $m$ to $5\%$ for large $m$.

An important feature of the Higgs cross section, which is
very often emphasised in the literature, is that the integral in Eq.
(\ref{eqn4.1}) is dominated by $y\sim x=m^2/S$. Since $x$ is small
the $gg$ part of $\sigma_{\rm tot}$ dominates the contributions coming from 
the other partonic subprocesses because of the steep rise of the gluon-gluon 
luminosity in Eq. (\ref{eqn4.2}) at small $y$.
The importance of the small $y$-region is revealed
when we impose an upper cut on the integral in Eq. (\ref{eqn4.1}). Hence
we compute
\begin{eqnarray}
\label{eqn4.7}
\sigma_{\rm tot}(x_{\rm max})&=&\frac{\pi\,G_{\rm B}^2}{8\,(N^2-1)}\,\Bigg [
\tilde \Phi_{gg}(x,\mu^2)+\sum_{a,b=q,g}\sum_{i=1}^{\infty}a_s^i(\mu^2)\,
\int_x^{x_{\rm max}} dy\, \tilde \Phi_{ab}(y,\mu^2)
\nonumber\\[2ex] 
&&\times \Delta_{ab}^{(i)}\left (\frac{x}{y},\frac{m^2}{\mu^2}\right )
\Bigg ]\,,
\end{eqnarray}
and plot the ratio
\begin{eqnarray}
\label{eqn4.8}
R(x_{\rm max})=\frac{\sigma_{\rm tot}(x_{\rm max})}{\sigma_{\rm tot}(1)}
\quad\,,\quad \sigma_{\rm tot}(1)=\sigma_{\rm tot}^{\rm EXACT}\,,
\end{eqnarray}
which is shown for NLO and NNLO in Fig. 10a for the choice $x_{\rm max}=5~x$.
The figure reveals that more than 95 $\%$ of the cross section 
comes from the integration region $x\le y \le 5~x$ where $x<5.1\times 10^{-5}$.
When $m$ gets larger $R(x_{\rm max})$ will increase because
the available phase space for Higgs boson production decreases. The latter
also happens when the centre-of-mass energy decreases like in the 
case for the TEVATRON where $\sqrt{S}=2~{\rm TeV}$ (Here $x<2.5\times 10^{-3}$)
see Fig. 10b.
(Note Figs. 10a and 10b have different scales.) 
In the latter case more than 99 $\%$ of the cross section receives
its contribution from $x\le y \le 5~x$ which becomes 100 $\%$ when
$m=300~{\rm GeV/c^2}$.

Finally we study the validity of the soft-plus-virtual gluon approximation
including subleading terms. The result of this approximation depends
on the definitions for the total cross section given in Eq. (\ref{eqn4.1})
and Eq. (\ref{eqn4.3}) as has been pointed out
in \cite{cafl2}. If we follow the definition in \cite{kls}, \cite{haki1},
where one choses the expression in Eq. (\ref{eqn4.1}), the sof-plus-virtual
coefficient function becomes
\begin{eqnarray}
\label{eqn4.9}
\Delta_{ab}^{\rm SV}\left(z,\frac{m^2}{\mu^2}\right )
&=&\delta(1-z)+\sum_{i=1}^{\infty}a_s^i(\mu^2)\,\Bigg [\sum_{j=0}^{2i-1}
a_{i,j}\,{\cal D}_j+b_i\,\delta(1-z)\Bigg ]\,.
\end{eqnarray}
where the distributions ${\cal D}_i(z)$ are defined in Eq. (\ref{eqn3.28}).
Furthermore the logarithmic terms $\ln^j (1-z)$
including the constants ($j=0$) are
important. The latter are denoted by
\begin{eqnarray}
\label{eqn4.10}
\Delta_{ab}^{\rm L}\left(z,\frac{m^2}{\mu^2}\right )=\sum_{i=1}^{\infty}
a_s^i(\mu^2)\,\sum_{j=0}^{2i-1}c_{i,j}\,\ln^j(1-z)\,.
\end{eqnarray}
Notice that the coefficients $a_{i,j},b_i,c_{i,j}$ also contain terms in  
$\ln m^2/\mu^2$. The soft-plus-virtual gluon contributions only emerge from 
the $gg$-subprocess and they can be found in Eqs. (\ref{eqnA.2}) and
(\ref{eqnA.6}). This subprocess also leads to the logarithms of the type 
$c_{i,j}\,\ln^j(1-z)$. The latter also show up in the 
$gq+g\bar q$-channel but they are absent in the $q\bar q$ and $qq$
subprocesses provided one neglects the terms $(1-z)^k$ for $k>0$. 
The coefficients
$c_{i,j}$ can be found in Eqs. (\ref{eqnA.26})-(\ref{eqnA.32}) (see also 
\cite{haki2}) (scalar Higgs) and Eqs. (\ref{eqnB.21}), (\ref{eqnB.22}) 
(pseudo-scalar Higgs). If we use the definition of the cross section
in Eq. (\ref{eqn4.3}) advocated in \cite{cafl1}, \cite{cafl2}
the expansion of $z^{-1}\,\Delta_{ab}^{\rm SV}$
will not differ from the one shown in Eq. (\ref{eqn4.9}) but
$\Delta_{ab}^{\rm L}$ will change due to contribution of
subleading logarithms and it becomes equal to
\begin{eqnarray}
\label{eqn4.11}
\Delta_{ab}^{\rm L}\left(z,\frac{m^2}{\mu^2}\right )=\sum_{i=1}^{\infty}
a_s^i(\mu^2)\,\sum_{j=0}^{2i-1}\Big (a_{i,j}+c_{i,j}\Big )\,\ln^j(1-z)\,.
\end{eqnarray}
Because of the different luminosities $\tilde \Phi(z)$ and $\Phi(z)$ in Eqs.
(\ref{eqn4.1}) and (\ref{eqn4.3}) respectively the value of
the soft-plus-virtual cross section $\sigma_{\rm tot}^{\rm SV}$ will change.
This will be only partially compensated by $\sigma_{\rm tot}^{\rm L}$ coming
from Eqs. (\ref{eqn4.10}) and (\ref{eqn4.11}). In fact the two results for
$\sigma_{\rm tot}^{\rm SVL}$ coming from definitions (\ref{eqn4.1}) 
and (\ref{eqn4.3}) will differ by terms in $(1-z)^k\,\ln^j(1-z)$ for $k\ge 1$ 
convoluted by one of the luminosities. In order to study the effect of the
two definitions on the approximation we will plot
\begin{eqnarray}
\label{eqn4.12}
R^{\rm SV}=\frac{\sigma_{\rm tot}^{\rm SV}}{\sigma_{\rm tot}}\,,\qquad
R^{\rm SVL}=\frac{\sigma_{\rm tot}^{\rm SVL}}{\sigma_{\rm tot}}\,.
\end{eqnarray}
The same study is made in \cite{kls}, \cite{cafl1}, \cite{haki1}
and \cite{cafl2}. However in these papers the complete result
for the NNLO coefficient function was not known yet. Therefore one
could only study the leading logarithmic terms in Eqs. (\ref{eqn4.10}),
(\ref{eqn4.11}) which are of collinear origin.
In Fig. 11a and Fig. 11b we have shown the ratios above for the LHC in NLO 
and NNLO respectively.
The figures reveal that $\sigma_{\rm tot}^{\rm SV}$ using Eq. (\ref{eqn4.3})
gives a much better approximation than Eq. (\ref{eqn4.1}). This also
holds if we include the logarithms in Eqs. (\ref{eqn4.10}) and
(\ref{eqn4.11}) denoted by $\sigma_{\rm tot}^{\rm SVL}$. The latter 
overestimates the exact cross section a little bit. Further we observe
that both SV approximations become worse in NNLO with respect to NLO.
In Figs. 12a,b we have performed the same analysis but now for the TEVATRON.
The features are the same as in Figs. 11a,b for LHC except that for the 
TEVATRON the SV approximations derived from Eq. (\ref{eqn4.1})
and Eq. (\ref{eqn4.3}) become better. 
This is no surprise because for the TEVATRON the Higgs mass is
much closer to the boundary of phase space.
This also explains why at larger masses the SV approximation improves.
From this study we infer that the soft-plus-virtual approximation works much 
better for Eq. (\ref{eqn4.3}) than for Eq. (\ref{eqn4.1}). The reason is that 
the cross section
is dominated by the small $x$-region (see Eq. (\ref{eqn4.7}) and Fig. 10).
By choosing the gluon luminosity in Eq. (\ref{eqn4.3}) this region will become
even more important than the gluon luminosity chosen in Eq. (\ref{eqn4.2}) 
which is used in Eq. (\ref{eqn4.7}). For NLO this was already shown in 
\cite{cafl2} but it is now also confirmed in NNLO. Further this approximation
becomes worse in NNLO which holds for both definitions.
In spite of this drawback one should still choose Eq. (\ref{eqn4.3}) to 
perform the 
soft-plus-virtual gluon resummation (see \cite{cafl3}) in order to get a 
better estimate of the cross section.
We also noticed that the contribution of logarithmic terms in
Eqs. (\ref{eqn4.10}), Eqs. (\ref{eqn4.11}) is substantial.
The determination of the coefficient
of the leading term, which is of collinear origin, is done in \cite{kls},
\cite{cafl1}. A systematic approach to determine $c_{i,j}$ can be found
in \cite{soak}.

Summarising our results we have recalculated the NNLO corrections
to the total cross section for (pseudo-) scalar Higgs production
using a different method than those presented in \cite{haki3} and
\cite{anme2}. In particular our approach differs from the one given in
\cite{haki3} since we directly evaluated the multi-particle phase
space integrals without fitting the coefficient functions to an
expansion in the terms $(1-z)^k\,\ln^l(1-z)$. Further we presented
the radiative parts of the coefficient functions for general colour factors 
and checked that
for $N=3$ our answers are in agreement with the literature. 
It turns out that the total cross section is almost completely
determined by the $gg$-subprocess which is in contrast
to the differential cross section where also the $gq+g\bar q$-channel
gives a considerable contribution. We studied
the $K$-factors and the dependence of the cross sections on the
chosen mass factorization/renormalization scale and the adopted
parton density set from which one can infer an error estimate of the
cross section. If we exclude the GRV98 parton density the error
can be mainly attributed to the missing higher order contributions
to the coefficient functions which we estimate to lie between $14\%$ and 
$21\%$. Finally we studied the validity of the soft-plus-virtual gluon
approximation. Depending on the definition of the total cross section
(Eq. (\ref{eqn4.1}) versus Eq. (\ref{eqn4.3})) this approximation
is excellent in NLO but it becomes less good in NNLO but Eq.
(\ref{eqn4.3}) is the best way to resum the soft-plus-virtual
corrections.
\\[2mm]
Acknowledgement: V. Ravindran would like to thank C. Anastasiou and
K. Melnikov for providing him with the unpublished results.
In particular he could check that we had the same partonic cross section 
for $g + g \rightarrow g + g + {\rm H}({\rm A})$ prior to renormalization
and mass factorization were carried out.


\appendix
\mysection*{Appendix A}
\setcounter{section}{1}
In this Appendix we present the coefficient functions for scalar Higgs
boson production. In the case of the $gg$ subprocess one can split
the corresponding coefficient function into
\begin{eqnarray}
\label{eqnA.1}
\Delta_{gg,{\rm B}}\left (z,\frac{m^2}{\mu^2}\right )
=\Delta_{gg,{\rm B}}^{\rm S+V}\left (z,\frac{m^2}{\mu^2}\right )
+\Delta_{gg,{\rm B}}^{\rm H}\left (z,\frac{m^2}{\mu^2}\right )\,,
\end{eqnarray}
where the superscripts $S+V$ and $H$ denote the soft-plus-virtual and
hard gluon parts respectively. The former contains the distributions
$\delta(1-z)$ and ${\cal D}_i(z)$ where the latter is defined in 
Eq.(\ref{eqn3.28}). We find
\begin{eqnarray}
\label{eqnA.2}
\Delta_{gg,{\rm H}}^{(1),{\rm S+V}}=C_A\,\Bigg [8\,{\cal D}_0(z)\,
\ln \left(\frac{m^2}
{\mu^2}\right)+16\,{\cal D}_1(z)+8\,\zeta(2)\,\delta(1-z) \Bigg ]\,,
\end{eqnarray}
\begin{eqnarray}
\label{eqnA.3}
\Delta_{gg,{\rm H}}^{(1)}&=&C_A\Bigg [\Bigg \{-16\,z+8\,z^2-8\,z^3\Bigg \}\,
\ln \left(\frac{m^2}{\mu^2}\right) + \Big (-32\,z+16\,z^2
-16\,z^3\Big )
\nonumber\\[2ex]
&&\times \Big (\ln (1-z)-\frac{1}{2}\,\ln z\Big )-\frac{8}{1-z}\,\ln z
-\frac{22}{3}\,(1-z)^3 \Bigg ]\,,
\\[2ex]
\label{eqnA.4}
\Delta_{gq,{\rm H}}^{(1)}&=&C_F\,\Bigg [\Bigg \{4-4\,z+2\,z^2\Bigg \}\,
\ln \left(\frac{m^2}{\mu^2}\right) + \Big (8-8\,z+4\,z^2\Big )\,
\Big (\ln (1-z)
\nonumber\\[2ex]
&&-\frac{1}{2}\,\ln z\Big )-3+6\,z-z^2\Bigg ]\,,
\\[2ex]
\label{eqnA.5}
\Delta_{q\bar q,{\rm H}}^{(1)}&=&C_F^2\,\Bigg [\frac{8}{3}\,(1-z)^3\Bigg ]\,,
\end{eqnarray}
\begin{eqnarray}
\label{eqnA.6}
\Delta_{gg,{\rm H}}^{(2),{\rm S+V}}&=&C_A^2\,\Bigg [\Bigg \{64\,{\cal D}_1(z)
- \frac{44}{3}\,{\cal D}_0(z)-32\,\zeta(2)\,\delta(1-z)\Bigg \}
\,\ln^2 \left(\frac{m^2}{\mu^2}\right)
\nonumber\\[2ex]
&&+\Bigg \{ 192\,{\cal D}_2(z)- \frac{176}{3}\,{\cal D}_1(z)
+\Big (\frac{536}{9}- 80\,\zeta(2)\Big )\,{\cal D}_0(z)
\nonumber\\[2ex]
&&+\delta(1-z)\,\Big (-24-\frac{88}{3}\,\zeta(2)
+152\,\zeta(3)\Big )\Bigg \}\,\ln \left(\frac{m^2}{\mu^2}\right)
\nonumber\\[2ex]
&&+128\,{\cal D}_3(z)-\frac{176}{3}\,{\cal D}_2(z)
+\Big (\frac{1072}{9} - 160\,\zeta(2)\Big )\,{\cal D}_1(z)
\nonumber\\[2ex]
&&+\Big ( - \frac{1616}{27}+\frac{176}{3}\,\zeta(2)+312\,\zeta(3)\Big )\, 
{\cal D}_0(z) 
\nonumber\\[2ex]
&&+\delta(1-z)\,\Big (93+\frac{536}{9}\,\zeta(2)-\frac{220}{3}\,\zeta(3)
-\frac{4}{5}\zeta^2(2)\Big )\Bigg ]
\nonumber\\[2ex]
&& +n_f\,T_f\,C_A\,\Bigg [\Bigg \{\frac{16}{3}\,{\cal D}_0(z)\Bigg \}
\ln^2 \left(\frac{m^2}{\mu^2}\right)+\Bigg \{\frac{64}{3}\,{\cal D}_1(z)
- \frac{160}{9}\,{\cal D}_0(z)
\nonumber\\[2ex]
&& +\delta(1-z)\Big (16+\frac{32}{3}\,\zeta(2)\Big )\Bigg \}
\,\ln \left(\frac{m^2}{\mu^2}\right)+\frac{64}{3}\,{\cal D}_2(z)
-\frac{320}{9}\,{\cal D}_1(z)
\nonumber\\[2ex]
&&+\Big (\frac{448}{27}
-\frac{64}{3}\,\zeta(2)\Big )\,{\cal D}_0(z)+\delta(1-z)\,\Big (
- \frac{2024}{27}-\frac{160}{9}\,\zeta(2)
\nonumber\\[2ex]
&&+\frac{80}{9}\, \zeta(3)\Big )\Bigg ]
\nonumber\\[2ex]
&& + n_f\,T_f\,C_F\,\Bigg [8\,\delta(1-z)\,\ln \left(\frac{m^2}{\mu^2}
\right) + 4\,\delta(1-z)\Bigg ]\,.
\end{eqnarray}
Notice that the colour decomposition of the $\delta(1-z)$ term 
into $n_f\,T_f\,C_A$ and $n_f\,T_f\,C_F$ is arbitrary due to our ignorance
of the same parts appearing in the two-loop virtual corrections
computed in \cite{harland}. They are only correct when $N=3$.
\begin{eqnarray}
\label{eqnA.7}
\Delta_{gg,{\rm H}}^{(2)}=C_A^2\,\Delta_{gg,{\rm H}}^{(2),{\rm C_A^2}}
+n_f\,T_f\,C_A\,\Delta_{gg,{\rm H}}^{(2),{\rm C_AT_f}}+n_f\,T_f\,C_F\,
\Delta_{gg,{\rm H}}^{(2),{\rm C_FT_f}}\,,
\end{eqnarray}
\begin{eqnarray}
\label{eqnA.8}
\Delta_{gg,{\rm H}}^{(2),{\rm C_A^2}}&=&\Bigg [\Bigg \{ \Big (-128\,z +64\,z^2
-64\,z^3 \Big )\,\ln(1-z)+\Big (- 96\,z^2 + 32\,z^3 
\nonumber\\[2ex]
&& - \frac{32}{1-z}\Big )\,
\ln z - \frac{352}{3} + \frac{376}{3}\,z - \frac{332}{3}\,z^2 
+ 132\,z^3\Bigg \}\,\ln^2 \left(\frac{m^2}{\mu^2}\right)
\nonumber\\[2ex]
&& +\Bigg \{\Big ( 64\,z + 32\,z^2 + 32\,z^3 + \frac{32}{1+z} \Big )\,
\Big ({\rm Li}_2(-z)+\ln z\,\ln(1+z)\Big )
\nonumber\\[2ex]
&&-\Big (256\,z+256\,z^2\Big )\,{\rm Li}_2(1-z)+\Big (-384\,z+192\,z^2
-192\,z^3 \Big )
\nonumber\\[2ex]
&& \times \ln^2(1-z)+\Big (192\,z-480\,z^2+224\,z^3-\frac{224}{1-z}
\Big  )
\nonumber\\[2ex]
&&\times \ln z\,\ln(1-z)
+\Big (128\,z^2-48\,z^3+\frac{40}{1-z}-\frac{8}{1+z} \Big )\,\ln^2 z
\nonumber\\[2ex]
&&+\Big (-528+\frac{2032}{3}\,z-\frac{1856}{3}\,z^2+\frac{1760}{3}\,z^3
\Big )\,\ln(1-z)+\Big (176 
\nonumber\\[2ex]
&&-\frac{904}{3}\,z+\frac{1328}{3}\,z^2-\frac{1496}{3}\,z^3+\frac{88}{3}\,
\frac{1}{1-z} \Big )\,\ln z +\Big (192\,z-64\,z^2
\nonumber\\[2ex]
&&+96\,z^3+\frac{16}{1+z}\Big )\,\zeta(2)+\frac{1262}{3}-\frac{3562}{9}\,z
+\frac{3026}{9}\,z^2-\frac{4322}{9}\,z^3 \Big \}
\nonumber\\[2ex]
&&\times \ln \left(\frac{m^2}{\mu^2}\right) 
\nonumber\\[2ex]
&& +\Big (128\,z+64\,z^2+64\,z^3+\frac{64}{1+z} \Big)\,\Big (
\ln z\,\ln(1-z)\,\ln(1+z)
\nonumber\\[2ex]
&&+\ln(1-z)\,{\rm Li}_2(-z)
+{\rm Li}_3\left (-\frac{1-z}{1+z}\right ) 
-{\rm Li}_3\left (\frac{1-z}{1+z}\right )\Big )
\nonumber\\[2ex]
&&+\Big (-24+8\,z-32\,z^2-32\,z^3+\frac{16}{1+z} \Big )\,\Big (
2\,{\rm S}_{1,2}(-z)
\nonumber\\[2ex]
&&+\zeta(2)\,\ln(1+z)+2\,\ln(1+z)\,{\rm Li}_2(-z)+\ln z\,\ln^2(1+z)\Big )
\nonumber\\[2ex]
&&+\Big (-8+24\,z -64\,z^3 +\frac{16}{1+z} \Big )\,{\rm Li}_3(-z)
+\Big (-16-208\,z
\nonumber\\[2ex]
&&-624\,z^2+112\,z^3-\frac{144}{1-z}
-\frac{64}{1+z} \Big )\,{\rm S}_{1,2}(1-z) +\Big (-16
\nonumber\\[2ex]
&&+672\,z+536\,z^2+88\,z^3-\frac{8}{1-z}+\frac{64}{1+z} \Big )\,
{\rm Li}_3(1-z) +\Big ( 24   
\nonumber\\[2ex]
&& - 104\,z- 16\,z^2 + 16\,z^3 - \frac{64}{1+z}\Big ) \ln z\,{\rm Li}_2(-z)
+\Big ( 8 +8\,z   
\nonumber\\[2ex]
&&- 88\,z^2+ 8\,z^3- \frac{56}{1-z} - \frac{48}{1+z}\Big )
\,\ln z\,{\rm Li}_2(1-z)
\nonumber\\[2ex]
&&-512\,z\,\Big (1+z\Big )\,\ln(1- z)\,
{\rm Li}_2(1-z) +\Big ( 240\,z -504\,z^2
\nonumber\\[2ex]
&&+248\,z^3-\frac{248}{1-z}\Big )\,\ln z\,\ln^2(1-z)
+\Big (- 96\,z+304\,z^2 -144\,z^3
\nonumber\\[2ex]
&&+\frac{128}{1-z}-\frac{16}{1+z}
\Big )\,\ln^2 z\,\ln(1-z)+\Big (20- 92\,z-16\,z^2
-16\,z^3
\nonumber\\[2ex]
&&-\frac{56}{1+z}\Big )\ln^2 z\,\ln(1+z)+\Big (-256\,z+128\,z^2
-128\,z^3 \Big )
\nonumber\\[2ex]
&&\times \ln^3(1-z) +\Big (\frac{8}{3}+\frac{16}{3}\,z-\frac{136}{3}\,z^2
+24\,z^3 -\frac{56}{3}\,\frac{1}{1-z}
\nonumber\\[2ex]
&&+\frac{16}{3}\,\frac{1}{1+z}\Big)\,\ln^3 z+\Big ( 384\,z-128\,z^2+192\,z^3
+\frac{32}{1+z}\Big ) 
\nonumber\\[2ex]
&&\times \zeta(2)\,\ln(1-z)+\Big (8-72\,z+392\,z^2-168\,z^3+\frac{144}{1-z}
\nonumber\\[2ex]
&&-\frac{24}{1+z} \Big )\,\zeta(2)\,\ln z+\Big (-\frac{ 112}{3}-32\,z
+56\,z^2+\frac{88}{3}\,z^3\Big )\,\Big ({\rm Li}_2(-z)
\nonumber\\[2ex]
&&+\ln z\,\ln(1+z)\Big ) +\Big (-\frac{1112}{3} +\frac{464}{3}\,z
+\frac{548}{3}\,z^2-\frac{572}{3}\, z^3
\nonumber\\[2ex]
&&-\frac{44}{3}\,\frac{1}{1-z} \Big )\,{\rm Li}_2(1-z)+
\Big (\frac{1232}{3}-\frac{2144}{3}\,z+\frac{2848}{3}\,z^2
-\frac{2992}{3}\,z^3
\nonumber\\[2ex]
&&+\frac{176}{3}\,\frac{1}{1-z}\Big )\,\ln z\,\ln(1-z)
+\Big (-528+\frac{2032}{3}\,z-\frac{1856}{3}\,z^2
\nonumber\\[2ex]
&&+\frac{1760}{3}\,z^3 \Big )
\,\ln^2(1-z)+\Big (-\frac{286}{3}+164\,z-302\,z^2+\frac{748}{3}\,z^3
\nonumber\\[2ex]
&&-\frac{22}{3}\,\frac{1}{1-z}\Big )\,\ln^2 z
+\Big (\frac{8108}{9}-\frac{8636}{9}\,z +\frac{7576}{9}\,z^2
-1020\,z^3 \Big )
\nonumber\\[2ex]
&&\times \ln(1-z)+\Big (-\frac{3602}{9} +\frac{2188}{3}\,z
-\frac{1582}{3}\,z^2+\frac{8108}{9}\,z^3
\nonumber\\[2ex]
&&-\frac{536}{9}\,\frac{1}{1-z}\Big )
\,\ln z +\Big (-608\,z+320\,z^2
-352\,z^3+\frac{8}{1+z} )\,\zeta(3)
\nonumber\\[2ex]
&&+\Big (\frac{1220}{3}-\frac{1156}{3}\,z
+\frac{1016}{3}\,z^2-\frac{1408}{3}\, z^3 )\,\zeta(2)
-\frac{16705}{27}
\nonumber\\[2ex]
&&+\frac{18523}{27}\,z-\frac{19751}{27}\,z^2 +\frac{7055}{9}\,z^3 \Bigg ]\,,
\end{eqnarray}
\begin{eqnarray}
\label{eqnA.9}
\Delta_{gg,{\rm H}}^{(2),{\rm C_AT_f}}&=&\Bigg [\Bigg \{ -\frac{32}{3}\,z
+\frac{16}{3}\,z^2-\frac{16}{3}\,z^3\Bigg \}\,\ln^2\left (\frac{m^2}{\mu^2}
\right )
\nonumber\\[2ex]
&&+\Bigg \{\Big (-\frac{128}{3}\,z+\frac{64}{3}\,z^2-\frac{64}{3}\,z^3\Big )\,
\ln (1-z)+\Big (\frac{32}{3}\,z-\frac{64}{3}\,z^2
\nonumber\\[2ex]
&&+\frac{32}{3}\,z^3-\frac{32}{3}\,
\frac{1}{1-z}\Big )\,\ln z-\frac{296}{9}+\frac{728}{9}\,z-\frac{568}{9}\,z^2
\nonumber\\[2ex]
&&+\frac{152}{3}\,z^3\Bigg \}\,\ln\left (\frac{m^2}{\mu^2}\right )
\nonumber\\[2ex]
&&+\Big (-16+16\,z-8\,z^2\Big )\,\Big ({\rm S}_{1,2}(1-z)+{\rm Li}_3(1-z)\Big )
+\Big (\frac{16}{3}
\nonumber\\[2ex]
&&-\frac{160}{3}\,z+\frac{8}{3}\,z^2-\frac{32}{3}\,z^3+\frac{16}{3}
\,\frac{1}{1-z}\Big )\,{\rm Li}_2(1-z)+\Big (-\frac{128}{3}\,z
\nonumber\\[2ex]
&&+\frac{64}{3}\,z^2
-\frac{64}{3}\,z^3\Big )\,\Big (\ln^2 (1-z)-\zeta(2)\Big )
+\Big (\frac{64}{3}\,z-\frac{128}{3}\,z^2
\nonumber\\[2ex]
&&+\frac{64}{3}\,z^3-\frac{64}{3}\,\frac{1}{1-z}\Big )\,\ln z\,\ln (1-z)
+\Big (\frac{8}{3}-8\,z+20\,z^2 
\nonumber\\[2ex]
&&-\frac{16}{3}\,z^3+\frac{8}{3}\,\frac{1}{1-z}\Big )
\,\ln^2 z+ \Big (-\frac{592}{9}+\frac{1456}{9}\,z-\frac{1160}{9}\,z^2
\nonumber\\[2ex]
&&+\frac{304}{3}\,z^3\Big )\,\ln (1-z)+\Big (\frac{296}{9}-\frac{208}{3}\,z
+\frac{268}{3}\,z^2-64\,z^3
\nonumber\\[2ex]
&&+\frac{160}{9}\,\frac{1}{1-z}\Big )\,\ln z+\frac{3142}{27}
-\frac{6188}{27}\,z+\frac{5218}{27}\,z^2
\nonumber\\[2ex]
&& -\frac{3068}{27}\,z^3\Bigg ]\,,
\end{eqnarray}
\begin{eqnarray}
\label{eqnA.10}
\Delta_{gg,{\rm H}}^{(2),{\rm C_FT_f}}&=&\Bigg [\Bigg \{ 16\,z\,\Big (1+z\Big )
\,\ln z+\frac{32}{3}+8\,z-8\,z^2-\frac{32}{3}\,z^3\Bigg \}\,
\ln^2\left (\frac{m^2}{\mu^2}\right )
\nonumber\\[2ex]
&&+\Bigg \{32\,z\,\Big (1+z\Big ) \,\Big (2\,{\rm Li}_2(1-z)+2\,\ln z \ln (1-z)
-\ln^2 z\Big )
\nonumber\\[2ex]
&& +\Big (\frac{128}{3}+32\,z-32\,z^2-\frac{128}{3}\,z^3\Big )
\ln(1-z)+\Big (-\frac{64}{3}
\nonumber\\[2ex]
&&-64\,z-48\,z^2+\frac{128}{3}\,z^3\Big )\,\ln z
-\frac{256}{9}-\frac{448}{3}\,z+\frac{352}{3}\,z^2
\nonumber\\[2ex]
&&+\frac{544}{9}\,z^3 \Bigg \}\,\ln \left (\frac{m^2}{\mu^2}\right )
\nonumber\\[2ex]
&&+\Big (32+32\,z+80\,z^2\Big )\,{\rm S}_{1,2}(1-z)
+\Big (32-160\,z-112\,z^2\Big )
\nonumber\\[2ex]
&&\times {\rm Li}_3(1-z)+(\frac{64}{3}-48\,z
-160\,z^2+\frac{64}{3}\,z^3\Big )\,{\rm Li}_2(1-z)
\nonumber\\[2ex]
&&+z\,\Big (1+z \Big )\,\Big (128\,\ln(1-z)\,{\rm Li}_2(1-z)
-32\,\ln z\,{\rm Li}_2(1-z)
\nonumber\\[2ex]
&&-64\,\zeta(2)\,\ln z+64\,\ln z \,\ln^2(1-z)
-64\,\ln^2 z \,\ln(1-z) 
\nonumber\\[2ex]
&&+\frac{40}{3}\,\ln^3 z\Big )+\Big (\frac{128}{3}
+32\,z-32\,z^2-\frac{128}{3}\,z^3\Big )\,\Big (\ln^2 (1-z)
\nonumber\\[2ex]
&&-\zeta(2)\Big )+\Big (-\frac{128}{3}-128\,z-96\,z^2+\frac{256}{3}\,z^3\Big )
\,\ln z\,\ln(1-z)
\nonumber\\[2ex]
&&+\Big (\frac{32}{3}+52\,z+28\,z^2-32\,z^3\Big )\,\ln^2 z+\Big (
-\frac{512}{9}-\frac{896}{3}\,z
\nonumber\\[2ex]
&&+\frac{704}{3}\,z^2+\frac{1088}{9}\,z^3\Big )
\,\ln (1-z)+\Big ( \frac{256}{9}+192\,z-128\,z^2
\nonumber\\[2ex]
&&-\frac{1088}{9}\,z^3\Big )\,\ln z-\frac{608}{27}+\frac{4144}{9}\,z
-\frac{3280}{9}\,z^2-\frac{1984}{27}\,z^3 \Bigg ]\,,
\end{eqnarray}
\begin{eqnarray}
\label{eqnA.11}
\Delta_{gq,{\rm H}}^{(2)}=C_F^2\,\Delta_{gq,{\rm H}}^{(2),{\rm C_F^2}}
+C_A\,C_F\,\Delta_{gq,{\rm H}}^{(2),{\rm C_AC_F}}+n_f\,T_f\,C_F\,
\Delta_{gq,{\rm H}}^{(2),{\rm C_FT_f}}\,,
\end{eqnarray}
\begin{eqnarray}
\label{eqnA.12}
\Delta_{gq,{\rm H}}^{(2),{\rm C_F^2}}&=&\Bigg [\Bigg \{ \Big (8-8\,z+4\,z^2
\Big )\,\ln(1-z)+\Big (4\,z-2\,z^2\Big )\,\ln z+4\,z-z^2\Bigg \}
\nonumber\\[2ex]
&&\times \ln^2 \left (\frac{m^2}{\mu^2}\right )
\nonumber\\[2ex]
&&+ \Bigg \{16\,{\rm Li}_2(1-z)+\Big (8-8\,z+4\,z^2\Big )\,\Big (3\,\ln^2(1-z)
-4\,\zeta(2)\Big )
\nonumber\\[2ex]
&&+\Big (-16+32\,z-16\,z^2\Big )\,\ln z\,\ln(1-z)+\Big (-8\,z
+4\,z^2\Big )\,\ln^2 z
\nonumber\\[2ex]
&&+\Big (-36+64\,z-28\,z^2\Big )\,\ln(1-z)
+\Big (-8\,z+38\,z^2+\frac{16}{3}\,z^3\Big )
\nonumber\\[2ex]
&&\times \ln z+\frac{106}{9}-12\,z-4\,z^2
-\frac{124}{9}\,z^3\Bigg \}\,\ln \left (\frac{m^2}{\mu^2}\right )
\nonumber\\[2ex]
&&+\Big (-32+48\,z-24\,z^2\Big )\,\Big ({\rm S}_{1,2}(1-z)-\zeta(2)\,\ln z
\Big ) 
\nonumber\\[2ex]
&&+\Big (-16-16\,z+8\,z^2\Big )\,\Big ({\rm Li}_3(1-z) -\ln(1-z)
{\rm Li}_2(1-z)\Big )
\nonumber\\[2ex]
&&+\Big (4-4\,z+2\,z^2\Big )\,\Big ( \frac{13}{3}\,\ln^3(1-z)
-16\,\zeta(2)\,\ln(1-z)+8\,\zeta(3)\Big )
\nonumber\\[2ex]
&&+\Big (-16+8\,z-4\,z^2\Big )\, \ln z\,{\rm Li}_2(1-z)+\Big (8-24\,z
+12\,z^2\Big )
\nonumber\\[2ex]
&&\times \ln^2 z\,\ln(1-z) +\Big (-24+40\,z -20\,z^2\Big )\,\ln z\,\ln^2(1-z)
\nonumber\\[2ex]
&&+\Big (\frac{10}{3}\,z-\frac{5}{3}\,z^2\Big )\,\ln^3 z+\Big (\frac{32}{3}
+16\,z+16\,z^2+\frac{32}{3}\,z^3\Big )
\nonumber\\[2ex]
&&\times \Big ( {\rm Li}_2(-z)+\ln z\,\ln(1+z)\Big )
+\Big (-\frac{92}{3}+56\,z+42\,z^2
\nonumber\\[2ex]
&&+\frac{16}{3}\, z^3\Big )\,{\rm Li}_2(1-z)+\Big (-60+94\,z-45\,z^2\Big )\,
\ln^2(1-z)
\nonumber\\[2ex]
&&+\Big (36-56\,z+88\,z^2+\frac{32}{3}\,z^3\Big )\,\ln z\,\ln(1-z)
+\Big (-14\,z
\nonumber\\[2ex]
&&-\frac{63}{2}\,z^2-\frac{40}{3}\,z^3\Big )\,\ln^2 z
+\Big ( \frac{232}{3}-104\,z+60\,z^2
+\frac{16}{3}\,z^3\Big )\,\zeta(2)
\nonumber\\[2ex]
&&+\Big (\frac{878}{9}-144\,z+48\,z^2-\frac{248}{9}\,z^3\Big )\,\ln(1-z)
+\Big (-\frac{106}{9}
\nonumber\\[2ex]
&&+69\,z-\frac{214}{3}\,z^2+\frac{112}{9}\,z^3\Big )\,\ln z
-\frac{1393}{54}-\frac{130}{9}\,z +\frac{17}{18}\,z^2 
\nonumber\\[2ex]
&&+\frac{1304}{27}\,z^3 \Bigg ]\,,
\end{eqnarray}
\begin{eqnarray}
\label{eqnA.13}
\Delta_{gq,{\rm H}}^{(2),{\rm C_AC_F}}&=&\Bigg [\Bigg \{ \Big (24-24\,z
+12\,z^2\Big )\,\ln(1-z)+\Big (-24-24\,z-24\,z^2 \Big )\,\ln z
\nonumber\\[2ex]
&&-\frac{230}{3}+\frac{188}{3}\,z-\frac{4}{3}\,z^2+8\,z^3\Bigg \}\,
\ln^2 \left (\frac{m^2}{\mu^2}\right )
\nonumber\\[2ex]
&&+ \Bigg \{\Big (-48-144\,z-72\,z^2\Big )\,{\rm Li}_2(1-z)+
\Big (16+16\,z+8\,z^2\Big )
\nonumber\\[2ex]
&&\times \Big ({\rm Li}_2(-z)+\ln z\,\ln(1+z)\Big )
+\Big (72-72\,z+36\,z^2\Big ) 
\nonumber\\[2ex]
&&\times \ln^2(1-z)+\Big (-128-64\,z-112\,z^2\Big )\,\ln z\,\ln(1-z) 
\nonumber\\[2ex]
&&+\Big (24+32\,z+28\,z^2\Big )\, \ln^2 z+16\,z\,\zeta(2)
+\Big (-\frac{868}{3}+\frac{736}{3}\,z
\nonumber\\[2ex]
&&+\frac{40}{3}\,z^2+32\,z^3\Big )\,\ln(1-z)+\Big (124-\frac{448}{3}\,z
+\frac{8}{3}\,z^2-32\,z^3\Big )
\nonumber\\[2ex]
&&\times \ln z+\frac{2422}{9}-\frac{1724}{9}\,z-\frac{362}{9}\,z^2
-\frac{32}{9}\,z^3\Bigg \}\,\ln \left (\frac{m^2}{\mu^2}\right )
\nonumber\\[2ex]
&&+\Big (-136-248\,z-148\,z^2\Big )\,{\rm S}_{1,2}(1-z)+\Big (104+344\,z
\nonumber\\[2ex]
&&+148\,z^2 \Big )\,{\rm Li}_3(1-z)+\Big (8+8\,z+4\,z^2\Big )\,
\Big (4\,{\rm Li}_3\left (-\frac{1-z}{1+z} \right )
\nonumber\\[2ex]
&&-4\,{\rm Li}_3\left (\frac{1-z}{1+z}\right )+2\,{\rm Li}_3(-z)-4\,
\ln z\,{\rm Li}_2(-z)
\nonumber\\[2ex]
&&+4\,\ln (1-z)\,{\rm Li}_2(-z) +4\,\ln z\,\ln (1-z)\,\ln(1+z)
\nonumber\\[2ex]
&&-3\,\ln^2 z\,\ln(1+z)\Big )+\Big (-80-304\,z-136\,z^2\Big )\,\ln(1-z)
\nonumber\\[2ex]
&&\times {\rm Li}_2(1-z)+\Big (-48-32\,z-32\,z^2\Big )\,\ln z\,
{\rm Li}_2(1-z)+\Big (\frac{140}{3}
\nonumber\\[2ex]
&&-\frac{140}{3}\,z+\frac{70}{3}\,z^2\Big )
\,\ln^3(1-z) +\Big (-8-12\,z-10\,z^2\Big )\,\ln^3 z
\nonumber\\[2ex]
&&+\Big (64+48\,z+64\,z^2\Big )\, \ln^2 z\,\ln(1-z) + \Big (-132-60\,z
\nonumber\\[2ex]
&&-114\,z^2\Big )\, \ln z\,\ln^2(1-z)+32\,z\,\zeta(2)\,\ln(1-z)+\Big (
64+112\,z
\nonumber\\[2ex]
&&+80\,z^2\Big )\,\zeta(2)\,\ln z+\Big (136-112\,z+68\,z^2\Big )\,\zeta(3)
+\Big (-\frac{538}{3}
\nonumber\\[2ex]
&&+40\,z+34\,z^2 -\frac{40}{3}\,z^3\Big )
\,{\rm Li}_2(1-z)+\Big (-\frac{88}{3}-40\,z-12\,z^2
\nonumber\\[2ex]
&&-\frac{16}{3}\,z^3 \Big )\,\Big ({\rm Li}_2(-z)
+\ln z\,\ln (1+z)\Big )+\Big (-\frac{784}{3}+\frac{634}{3}\,z
\nonumber\\[2ex]
&&+\frac{97}{3}\,z^2+32\,z^3\Big )
\,\ln^2(1-z)+\Big (\frac{692}{3}-\frac{832}{3}\,z-\frac{40}{3}\,z^2
\nonumber\\[2ex]
&&-64\,z^3\Big )\,\ln z\,\ln(1-z)+\Big (-62+\frac{344}{3}\,z
-\frac{10}{3}\,z^2 +\frac{68}{3}\,z^3\Big )
\nonumber\\[2ex]
&&\times \ln^2 z+\Big (\frac{574}{3}-128\,z-58\,z^2-\frac{104}{3}\,z^3\Big )
\,\zeta(2)+\Big (\frac{4342}{9}
\nonumber\\[2ex]
&&-\frac{904}{3}\,z-116\,z^2-\frac{64}{9}\, z^3\Big )\,\ln(1-z)
+\Big (-\frac{2402}{9}+\frac{2660}{9}\,z
\nonumber\\[2ex]
&&+\frac{1079}{9}\,z^2 +\frac{224}{9}\,z^3\Big )\,\ln z
-\frac{21539}{54}+\frac{9962}{27}\,z
+\frac{1171}{54}\,z^2
\nonumber\\[2ex]
&&-\frac{238}{27}\,z^3\Bigg ]\,,
\end{eqnarray}
\begin{eqnarray}
\label{eqnA.14}
\Delta_{gq,{\rm H}}^{(2),{\rm C_FT_f}}&=&\Bigg [\Bigg \{ \frac{16}{3}
-\frac{16}{3}\,z+\frac{8}{3}\,z^2 \Bigg \}\,\ln^2 \left (\frac{m^2}{\mu^2}
\right )
\nonumber\\[2ex]
&&+\Bigg \{\Big ( \frac{32}{3}-\frac{32}{3}\,z+\frac{16}{3}\,z^2\Big )\,
\Big (\ln (1-z)-\ln z)\Big )-\frac{232}{9}+\frac{304}{9}\,z
\nonumber\\[2ex]
&&-\frac{152}{9}\,z^2 \Bigg \}\,\ln \left (\frac{m^2}{\mu^2}\right )
\nonumber\\[2ex]
&&+\Big ( \frac{8}{3}-\frac{8}{3}\,z+\frac{4}{3}\,z^2 \Big )\,\Big (
\ln^2(1-z)+2\,\ln^2 z-4\,\ln z\,\ln(1-z)\Bigg )
\nonumber\\[2ex]
&&+\Big (-\frac{104}{3}
+\frac{128}{3}\,z-24\,z^2\Big )\,\ln(1-z)+\Big (\frac{232}{9}-\frac{304}{9}\,z
\nonumber\\[2ex]
&&+\frac{152}{9}\,z^2\Big )\,\ln z +\frac{1060}{27}-\frac{1672}{27}\,z
+\frac{716}{27}\,z^2\Bigg ]\,,
\end{eqnarray}
\begin{eqnarray}
\label{eqnA.15}
\Delta_{q_1q_2,{\rm H}}^{(2)}=\Delta_{q_1\bar q_2,{\rm H}}^{(2)}=
C_F^2\,\Delta_{q_1q_2,{\rm H}}^{(2),{\rm C_F^2}}\,,
\end{eqnarray}
\begin{eqnarray}
\label{eqnA.16}
\Delta_{q_1q_2,{\rm H}}^{(2),{\rm C_F^2}}&=&\Bigg [\Bigg \{ \Big (-16-16\,z
-4\,z^2 \Big )\,\ln z -24+16\,z+8\,z^2 \Bigg \}\,\ln^2 \left (\frac{m^2}{\mu^2}
\right )
\nonumber\\[2ex]
&&+\Bigg \{\Big (16+16\,z+4\,z^2\Big )\,\Big (-4\,{\rm Li}_2(1-z)
-4\,\ln z\,\ln(1-z)
\nonumber\\[2ex]
&&+\ln^2 z\Big )+\Big (-96+64\,z+32\,z^2\Big )\,\ln(1-z)+\Big (48-32\,z
\nonumber\\[2ex]
&&-20\,z^2\Big )\,\ln z+102-72\,z-30\,z^2\Bigg \}
\ln \left (\frac{m^2}{\mu^2}\right )
\nonumber\\[2ex]
&&+ \Big (16+16\,z+4\,z^2\Big )\,\Big (-8\,{\rm S}_{1,2}(1-z)+8\,{\rm Li}_3(1-z)
-\frac{1}{3}\,\ln^3 z
\nonumber\\[2ex]
&&-8\,\ln(1-z)\,{\rm Li}_2(1-z)-2\,\ln z\,{\rm Li}_2(1-z)
-4\,\ln z\,\ln^2(1-z) 
\nonumber\\[2ex]
&&+2\,\ln^2 z\,\ln(1-z)+4\,\zeta(2)\,\ln z\Big )+\Big (-48+32\,z+8\,z^2\Big )
\nonumber\\[2ex]
&&\times {\rm Li}_2(1-z) +\Big (-96+64\,z+32\,z^2\Big )\,\Big (\ln^2(1-z)
-\zeta(2) \Big )
\nonumber\\[2ex]
&& +\Big (96-64\,z-40\,z^2\Big )\,\ln z\,\ln(1-z)
+\Big (-24 +32\,z
\nonumber\\[2ex]
&&+8\,z^2)\,\ln^2 z+\Big (204-144\,z-60\,z^2 \Big )\,\ln(1-z)
+\Big (-118
\nonumber\\[2ex]
&&+88\,z+58\,z^2 \Big )\,\ln z-210+188\,z+22\,z^2\Bigg ]\,,
\end{eqnarray}
\begin{eqnarray}
\label{eqnA.17}
\Delta_{qq,{\rm H}}^{(2)}=C_A\,C_F^2\,\Delta_{qq,{\rm H}}^{(2),{\rm C_AC_F^2}}
+C_F^3\,\Delta_{qq,{\rm H}}^{(2),{\rm C_F^3}}+C_F^2\,
\Delta_{qq,{\rm H}}^{(2),{\rm C_F^2}}\,,
\end{eqnarray}
\begin{eqnarray}
\label{eqnA.18}
\Delta_{qq,{\rm H}}^{(2),{\rm C_AC_F^2}}&=&\Bigg [\Bigg (-16+16\,z-8\,z^2\Big )
\,{\rm S}_{1,2}(1-z)+\Big (8\,z+12\,z^2\Big )\,\ln^2 z
\nonumber\\[2ex]
&&+\Big (-16\,z-36\,z^2 \Big )\,\ln z-10-32\,z+42\,z^2\Bigg ]\,,
\end{eqnarray}
\begin{eqnarray}
\label{eqnA.19}
\Delta_{qq,{\rm H}}^{(2),{\rm C_F^3}}&=&\Bigg [\Bigg (32-32\,z+16\,z^2\Big )
\,{\rm S}_{1,2}(1-z)+\Big (-16\,z-24\,z^2\Big )\,\ln^2 z
\nonumber\\[2ex]
&&+\Big (32\,z+72\,z^2 \Big )\,\ln z+20+64\,z-84\,z^2\Bigg ]\,,
\\[2ex]
\label{eqnA.20}
\Delta_{qq,{\rm H}}^{(2),{\rm C_F^2}}&=&
\Delta_{q_1q_2,{\rm H}}^{(2),{\rm C_F^2}}\,,
\\[2ex]
\label{eqnA.21}
\Delta_{q\bar q,{\rm H}}^{(2)}&=&C_A\,C_F^2\,
\Delta_{q\bar q,{\rm H}}^{(2),{\rm C_AC_F^2}}
+C_F^3\,\Delta_{q\bar q,{\rm H}}^{(2),{\rm C_F^3}}+C_F^2\,
\Delta_{q\bar q,{\rm H}}^{(2),{\rm C_F^2}} 
\nonumber\\[2ex]
&&\qquad +n_f\,T_f\,C_F^2\, \Delta_{q\bar q,{\rm H}}^{(2),{\rm C_F^2T_f}}\,,
\end{eqnarray}
\begin{eqnarray}
\label{eqnA.22}
\Delta_{q\bar q,{\rm H}}^{(2),{\rm C_AC_F^2}}&=&\Bigg [\Bigg \{-\frac{88}{3}
+88\,z-88\,z^2+\frac{88}{3}\,z^3\Bigg \}\,\ln \left (\frac{m^2}{\mu^2}
\right )
\nonumber\\[2ex]
&& +\Big (8+8\,z+4\,z^2\Big )\,\Big (-4\,{\rm S}_{1,2}(-z)-6\,{\rm Li}_3(-z)
\nonumber\\[2ex]
&&-4\,\ln(1+z)\,{\rm Li}_2(-z)+6\,\ln z\,{\rm Li}_2(-z)-2\,\ln z\,\ln^2(1+z)
\nonumber\\[2ex]
&&+3\,\ln^2 z\,\ln(1+z)-2\,\zeta(2)\,\ln(1+z)-4\,\zeta(3)\Big )
+\Big (\frac{16}{3} -16\,z
\nonumber\\[2ex]
&&+16\,z^2-\frac{16}{3}\,z^3\Big )\,\Big (5\,{\rm Li}_2(1-z)+2\,\ln^2(1-z)
-8\,\ln(1-z)\Big )
\nonumber\\[2ex]
&&+\Big (32\,z+16\,z^2+\frac{32}{3}\,z^3\Big )\,\Big (
{\rm Li}_2(-z)+\ln z\,\ln(1+z)\Big )
\nonumber\\[2ex]
&&+\Big (-\frac{176}{3}+192\,z-168\,z^2 +64\,z^3\Big )\,\zeta(2)
+\Big (-40\,z-\frac{32}{3}\,z^3\Big )
\nonumber\\[2ex]
&&\times \ln^2 z+\Big (\frac{88}{3}
-\frac{296}{3}\,z+\frac{500}{3}\,z^2-\frac{128}{3}\,z^3\Big )\,\ln z
+130-316\,z
\nonumber\\[2ex]
&&+\frac{890}{3}\,z^2-\frac{332}{3}\,z^3\Bigg ]\,,
\end{eqnarray}
\begin{eqnarray}
\label{eqnA.23}
\Delta_{q\bar q,{\rm H}}^{(2),{\rm C_F^3}}&=&\Bigg [\Bigg \{\Big (\frac{64}{3}
-64\,z+64\,z^2-\frac{64}{3}\,z^3\Big )\,\ln(1-z)+\Big (32\,z-32\,z^2
\nonumber\\[2ex]
&&+\frac{64}{3}\,z^3\Big )\,\ln z+\frac{64}{3}\,z-\frac{64}{3}\,z^2\Bigg \}
\,\ln \left (\frac{m^2}{\mu^2}\right )
\nonumber\\[2ex]
&& +\Big (16+16\,z+8\,z^2\Big )\,\Big (4\,{\rm S}_{1,2}(-z)+6\,{\rm Li}_3(-z)
+4\,\zeta(3)
\nonumber\\[2ex]
&&+4\,\ln(1+z)\,{\rm Li}_2(-z)-6\,\ln z\,{\rm Li}_2(-z)+2\,\ln z\,\ln^2(1+z)
\nonumber\\[2ex]
&&-3\,\ln^2 z\,\ln(1+z)+2\,\zeta(2)\,\ln(1+z)\Big )+\Big (-32
+128\,z
\nonumber\\[2ex]
&&-128\,z^2+\frac{160}{3}\,z^3\Big )\,{\rm Li}_2(1-z)
+\Big (-64\,z-32\,z^2\Big )\,\Big ({\rm Li}_2(-z)
\nonumber\\[2ex]
&&+\ln z\ln(1+z)\Big )
+\Big (\frac{32}{3} -32\,z+32\,z^2-\frac{32}{3}\,z^3\Big )
\nonumber\\[2ex]
&&\times\ln^2(1-z)+\Big (-\frac{64}{3}+96\,z-96\,z^2+\frac{128}{3}\,z^3\Big )
\nonumber\\[2ex]
&&\times\ln z\,\ln(1-z)+\Big (40\,z+48\,z^2-\frac{16}{3}\,z^3\Big )\,\ln^2 z
+\Big (\frac{224}{3}-256\,z
\nonumber\\[2ex]
&&+208\,z^2-\frac{224}{3}\,z^3\Big )\,\zeta(2)
+\Big (-32+\frac{352}{3}\,z-\frac{352}{3}\,z^2+32\,z^3\Big )
\nonumber\\[2ex]
&&\times \ln(1-z)+\Big (-\frac{256}{3}\,z-\frac{200}{3}\,z^2
-32\,z^3\Big )\,\ln z -\frac{164}{3}-64\,z
\nonumber\\[2ex]
&&+\frac{260}{3}\,z^2+32\,z^3\Bigg ]\,,
\\[2ex]
\label{eqnA.24}
\Delta_{q\bar q,{\rm H}}^{(2),{\rm C_F^2}}&=&
\Delta_{q_1q_2,{\rm H}}^{(2),{\rm C_F^2}}\,,
\end{eqnarray}
\begin{eqnarray}
\label{eqnA.25}
\Delta_{q\bar q,{\rm H}}^{(2),{\rm C_F^2T_f}}&=&\Bigg [\Bigg \{\frac{32}{3}
-32\,z+32\,z^2-\frac{32}{3}\,z^3\Bigg \}\,\ln \left (\frac{m^2}{\mu^2}\right )
\nonumber\\[2ex]
&&+\Big (\frac{64}{9}-\frac{64}{3}\,z+\frac{64}{3}\,z^2-\frac{64}{9}\,z^3\Big )
\,\ln(1-z)+\Big (-\frac{32}{3}+\frac{160}{3}\,z
\nonumber\\[2ex]
&&-\frac{128}{3}\,z^2
+\frac{128}{9}\,z^3\Big )\,\ln z -\frac{368}{27}+\frac{592}{9}\,z
-\frac{688}{9}\,z^2
\nonumber\\[2ex]
&&+\frac{656}{27}\,z^3\Bigg ]\,.
\end{eqnarray}
As discussed below Eq. (\ref{eqn4.10}) the corrections are dominated
by the terms $c_{i,j}\,\ln^j(1-z)$ in the coefficient functions.
They are obtained by taking the limit $z\rightarrow 1$ in the expressions
above. Most of them vanish. Those which survive are given by
\begin{eqnarray}
\label{eqnA.26}
\mathop{\mbox{lim}}\limits_{\vphantom{\frac{A}{A}} z \rightarrow 1}
\Delta_{gg,{\rm H}}^{(1),{\rm C_A}}&=&\Bigg [-16\,
\ln \left (\frac{m^2}{\mu^2}\right )-32\,\ln(1-z)+8\Bigg ]\,,
\\[2ex]
\label{eqnA.27}
\mathop{\mbox{lim}}\limits_{\vphantom{\frac{A}{A}} z \rightarrow 1}
\Delta_{gg,{\rm H}}^{(2),{\rm C_A^2}}&=&\Bigg [\Bigg \{-128\,\ln(1-z)
+\frac{184}{3}\Bigg \}\,\ln^2 \left (\frac{m^2}{\mu^2}\right )
+\Bigg \{-384\,\ln^2(1-z)
\nonumber\\[2ex]
&&+\frac{1024}{3}\,\ln(1-z)+160\,\zeta(2)
-\frac{1336}{9}\Bigg \}\,\ln \left (\frac{m^2}{\mu^2}\right )
\nonumber\\[2ex]
&&-256\,\ln^3(1-z)+\frac{1096}{3}\,\ln^2(1-z)+\Big (320\,\zeta(2)
\nonumber\\[2ex]
&&-\frac{2660}{9}\Big )\,\ln(1-z)-624\,\zeta(3)-\frac{784}{3}\,\zeta(2)
+\frac{4228}{27}\Bigg ]\,,
\end{eqnarray}
\begin{eqnarray}
\label{eqnA.28}
\mathop{\mbox{lim}}\limits_{\vphantom{\frac{A}{A}} z \rightarrow 1}
\Delta_{gg,{\rm H}}^{(2),{\rm C_AT_f}}&=&\Bigg [-\frac{32}{3}\,
\ln^2 \left (\frac{m^2}{\mu^2}\right )+\Bigg \{-\frac{128}{3}\,\ln(1-z)
+\frac{416}{9}\Bigg \}\,\ln \left (\frac{m^2}{\mu^2}\right )
\nonumber\\[2ex]
&&-\frac{128}{3}\,\ln^2(1-z)+\frac{808}{9}\,\ln(1-z)+\frac{128}{3}\,\zeta(2)
-\frac{1232}{27}\Bigg ]\,,
\nonumber\\[2ex]
\end{eqnarray}
\begin{eqnarray}
\label{eqnA.29}
\mathop{\mbox{lim}}\limits_{\vphantom{\frac{A}{A}} z \rightarrow 1}
\Delta_{gq,{\rm H}}^{(1),{\rm C_F}}&=&\Bigg [2\,
\ln \left (\frac{m^2}{\mu^2}\right )+4\,\ln(1-z)+2\Bigg ]\,,
\\[2ex]
\label{eqnA.30}
\mathop{\mbox{lim}}\limits_{\vphantom{\frac{A}{A}} z \rightarrow 1}
\Delta_{gq,{\rm H}}^{(2),{\rm C_F^2}}&=&\Bigg [\Bigg \{4\,\ln(1-z)
+3\Bigg \}\,\ln^2 \left (\frac{m^2}{\mu^2}\right )
+\Bigg \{12\,\ln^2(1-z)
\nonumber\\[2ex]
&&-16\,\zeta(2)-18\Bigg \}\,\ln \left (\frac{m^2}{\mu^2}
\right )+\frac{26}{3}\,\ln^3(1-z)-11\,\ln^2(1-z)
\nonumber\\[2ex]
&&+\Big (-32\,\zeta(2)
-26\Big )\, \ln(1-z)+16\,\zeta(3)+12\,\zeta(2)+9\Bigg ]\,,
\nonumber\\[2ex]
\\[2ex]
\label{eqnA.31}
\mathop{\mbox{lim}}\limits_{\vphantom{\frac{A}{A}} z \rightarrow 1}
\Delta_{gq,{\rm H}}^{(2),{\rm C_AC_F}}&=&\Bigg [\Bigg \{12\,\ln(1-z)
-\frac{22}{3}\Bigg \}\,\ln^2 \left (\frac{m^2}{\mu^2}\right )
+\Bigg \{36\,\ln^2(1-z)
\nonumber\\[2ex]
&&+\frac{4}{3}\,\ln(1-z)-4\,\zeta(2)+\frac{304}{9}\Bigg \}\,
\ln \left (\frac{m^2}{\mu^2}\right )+\frac{70}{3}\,\ln^3(1-z)
\nonumber\\[2ex]
&&+\frac{43}{3}\,\ln^2(1-z)
+\Big (-8\,\zeta(2)+58\Big )\, \ln(1-z)+62\,\zeta(3)
\nonumber\\[2ex]
&&+14\,\zeta(2)- \frac{460}{27}\Bigg ]\,,
\\[2ex]
\label{eqnA.32}
\mathop{\mbox{lim}}\limits_{\vphantom{\frac{A}{A}} z \rightarrow 1}
\Delta_{gq,{\rm H}}^{(2),{\rm C_FT_f}}&=&\Bigg [\frac{8}{3}\,
\ln^2 \left (\frac{m^2}{\mu^2}\right )+\Bigg \{\frac{16}{3}\,\ln(1-z)
-\frac{80}{9}\Bigg \}\,\ln \left (\frac{m^2}{\mu^2}\right )
\nonumber\\[2ex]
&&+\frac{4}{3}\,\ln^2(1-z)-16\,\ln(1-z)+\frac{104}{27}\Bigg ]\,.
\end{eqnarray}
The nonvanishing coefficients $c_{i,j}$ in Eq. (\ref{eqn4.9}) can be 
read off from the equations above.

\mysection*{Appendix B}
\setcounter{section}{2}
In this Appendix we present the coefficient functions for pseudo-scalar
Higgs boson production. To shorten the expressions we give the difference
between the coefficient functions originating from pseudo-scalar and
scalar production only, namely,
\begin{eqnarray}
\label{eqnB.1}
\Delta_{gg,{\rm A-H}}^{(1),{\rm S+V}}&=&C_A\,\Bigg [8\,\delta(1-z)\Bigg ]\,,
\\[2ex]
\label{eqnB.2}
\Delta_{gg,{\rm A-H}}^{(1)}&=&0\,,
\\[2ex]
\label{eqnB.3}
\Delta_{gq,{\rm A-H}}^{(1)}&=&0\,,
\\[2ex]
\label{eqnB.4}
\Delta_{q\bar q,{\rm A-H}}^{(1)}&=&0\,.
\end{eqnarray}
In the NNLO corrections below we indicate by $O_{12}$ the contribution
coming from the interferences between the operators $O_1$ and $O_2$ in 
Eqs. (\ref{eqn3.9}), Eqs. (\ref{eqn3.10}). In the coefficient functions
one has to put $O_{12}=1$, 
\begin{eqnarray}
\label{eqnB.5}
\Delta_{gg,{\rm A-H}}^{(2),{\rm S+V}}&=&C_A^2\,\Bigg [\Bigg \{64\,{\cal D}_0(z)
-\frac{20}{3}\,\delta(1-z)\Bigg \}\,\ln \left(\frac{m^2}{\mu^2}\right)
\nonumber\\[2ex]
&&+128\,{\cal D}_1(z) +\delta(1-z)\,\Big (\frac{215}{3}+64\,\zeta(2)\Big )
\Bigg ]
\nonumber\\[2ex]
&&+C_A\,T_f\,n_f\Bigg [\Bigg \{-\frac{8}{3}\,\delta(1-z)\Bigg \}\,
\ln \left(\frac{m^2}{\mu^2}\right)
\nonumber\\[2ex]
&&+\delta(1-z)\,\Big (-\frac{196}{9}+O_{12}\,\left (\frac{32}{3}\,
\ln \left(\frac{\mu_r^2}{m_t^2}\right) -\frac{16}{3}\right )\Bigg ]
\nonumber\\[2ex]
&&+C_F\,T_f\,n_f\Bigg [\Bigg \{-8\,\delta(1-z)\Bigg \}\,
\ln \left(\frac{m^2}{\mu^2}\right)-4\delta(1-z)\Bigg ]\,.
\end{eqnarray}
For the colour decomposition of the $n_f$-part of the expression above 
see the remark below Eq. (\ref{eqnA.6}). Also
\begin{eqnarray}
\label{eqnB.6}
\Delta_{gg,{\rm A-H}}^{(2),{\rm C_A^2}}&=&\Bigg [\Bigg \{-128\,z+64\,z^2
-64\,z^3\Bigg \}\,\ln \left(\frac{m^2}{\mu^2}\right)
\nonumber\\[2ex]
&&-16\,z\,\ln^2 z+\Big (-256\,z+128\,z^2-128\,z^3\Big )\,\ln(1-z)
+\Big (32
\nonumber\\[2ex]
&&+\frac{440}{3}\,z-64\,z^2+64\,z^3-\frac{64}{1-z}\Big )\,\ln z
+\frac{44}{3}+\frac{232}{3}\,z-\frac{452}{3}\,z^2
\nonumber\\[2ex]
&&+\frac{176}{3}\,z^3\Bigg ]\,,
\\[2ex]
\label{eqnB.7}
\Delta_{gg,{\rm A-H}}^{(2),{\rm C_AT_f}}&=&\Bigg [
\frac{32}{3}\,z\,\ln z+\frac{16}{3}+\frac{16}{3}\,z-\frac{32}{3}\,z^2\Bigg ]\,,
\\[2ex]
\label{eqnB.8}
\Delta_{gg,{\rm A-H}}^{(2),{\rm C_FT_f}}&=&\Bigg [
16\,z\,\ln^2 z -16+32\,z-16\,z^2\Bigg ]\,,
\\[2ex]
\label{eqnB.9}
\Delta_{gq,{\rm A-H}}^{(2),{\rm C_F^2}}&=&\Bigg [
8\,z\,\ln^2 z -24\,z\ln z - 4+16\,z-12\,z^2+O_{12}\,\Big (-12
\nonumber\\[2ex]
&&+12\,z^2\Big )\Bigg ]\,,
\\[2ex]
\label{eqnB.10}
\Delta_{gq,{\rm A-H}}^{(2),{\rm C_AC_F}}&=&\Bigg [\Bigg \{32-32\,z+16\,z^2
\Bigg \}\,\ln \left(\frac{m^2}{\mu^2}\right)
\nonumber\\[2ex]
&& -16\,z\,\ln^2 z +\Big (64-64\,z+32\,z^2\Big )\,\ln(1-z)
+\Big (80\,z 
\nonumber\\[2ex]
&&-16\,z^2\Big )\,\ln z+60-48\,z+4\,z^2 \Bigg ]\,,
\\[2ex]
\label{eqnB.11}
\Delta_{gq,{\rm A-H}}^{(2),{\rm C_FT_f}}&=&0\,,
\\[2ex]
\label{eqnB.12}
\Delta_{q_1q_2,{\rm A-H}}^{(2),{\rm C_F^2}}&=&\Bigg [-16\,z\,\ln^2 z+\Big (
32+48\,z\Big )\,\ln z+88-96\,z+8\,z^2\Bigg ]\,,
\\[2ex]
\label{eqnB.13}
\Delta_{qq,{\rm A-H}}^{(2),{\rm C_AC_F^2}}&=&\Bigg [-16\,z\,\ln^2 z+32\,z\,
\ln z+32-32\,z\Bigg ]\,,
\\[2ex]
\label{eqnB.14}
\Delta_{qq,{\rm A-H}}^{(2),{\rm C_F^3}}&=&\Bigg [32\,z\,\ln^2 z-64\,z\,
\ln z-64+64\,z\Bigg ]\,,
\\[2ex]
\label{eqnB.15}
\Delta_{qq,{\rm A-H}}^{(2),{\rm C_F^2}}&=&
\Delta_{q_1q_2,{\rm A-H}}^{(2),{\rm C_F^2}}\,,
\\[2ex]
\label{eqnB.16}
\Delta_{q_1\bar q_2,{\rm A-H}}^{(2),{\rm C_F^2}}&=&
\Delta_{q_1q_2,{\rm A-H}}^{(2),{\rm C_F^2}}\,,
\\[2ex]
\label{eqnB.17}
\Delta_{q\bar q,{\rm A-H}}^{(2),{\rm C_AC_F^2}}&=&\Bigg [16\,z\,\ln^2 z
+\frac{80}{3}\,z\, \ln z+\frac{8}{3}+\frac{56}{3}\,z^2-\frac{64}{3}\,z^3\Bigg ]
\,,
\\[2ex]
\label{eqnB.18}
\Delta_{q\bar q,{\rm A-H}}^{(2),{\rm C_F^3}}&=&\Bigg [-16\,z\,\ln^2 z
-32\,z\, \ln z+64-160\,z+96\,z^2
\nonumber\\[2ex]
&&+O_{12}\,\Big (-48+96\,z-48\,z^2\Big )\Bigg ]\,,
\\[2ex]
\label{eqnB.19}
\Delta_{q\bar q,{\rm A-H}}^{(2),{\rm C_F^2}}&=&
\Delta_{q_1q_2,{\rm A-H}}^{(2),{\rm C_F^2}}\,,
\\[2ex]
\label{eqnB.20}
\Delta_{q\bar q,{\rm A-H}}^{(2),{\rm C_F^2T_f}}&=&\Bigg [
-\frac{64}{3}\,z\, \ln z-\frac{32}{3}+\frac{32}{3}\,z^2\Bigg ]\,.
\end{eqnarray}
The nonvanishing coefficients $c_{i,j}$ in Eq. (\ref{eqn4.9}) 
appearing in the difference between the pseudo-scalar and scalar
coefficient functions are determined from the expressions above by
taking the limit $z\rightarrow 1$. Those which survive in this limit
are given by
\begin{eqnarray}
\label{eqnB.21}
\mathop{\mbox{lim}}\limits_{\vphantom{\frac{A}{A}} z \rightarrow 1}
\Delta_{gg,{\rm A-H}}^{(2),{\rm C_A^2}}&=&\Bigg [-128\,
\ln \left (\frac{m^2}{\mu^2}\right )-256\,\ln(1-z)+64\Bigg ]\,,
\\[2ex]
\label{eqnB.22}
\mathop{\mbox{lim}}\limits_{\vphantom{\frac{A}{A}} z \rightarrow 1}
\Delta_{gg,{\rm A-H}}^{(2),{\rm C_AC_F}}&=&\Bigg [16\,
\ln \left (\frac{m^2}{\mu^2}\right )+32\,\ln(1-z)+16\Bigg ]\,.
\end{eqnarray}

%


\centerline{\bf \large{Figure Captions}}
\begin{description}
\item[Fig. 1]
The total cross section $\sigma_{\rm tot}$ plotted as a function of the
Higgs mass at $\sqrt{S}=14~{\rm TeV}$ with $\mu=m$.
The NLO plots are presented for the total (solid line) and the
subprocesses $gg$ (long-dashed line),
$10\times{\rm abs}(gq + g \bar q)$ (dot-dashed line)
and  $100 \times (q\bar q)$ (short-dashed line). Also shown is the 
LO result (dotted line).
\item[Fig. 2]
The total cross section $\sigma_{\rm tot}$ plotted as a function of the
Higgs mass at $\sqrt{S}=14~{\rm TeV}$ with $\mu=m$. The NNLO
plots are presented for the total (solid line) and the 
subprocesses $gg$ (long-dashed line),
${\rm abs}(gq + g\bar q)$ (dot-dashed line), 
$100 \times (q\bar q)$ (dotted line)
and $100 \times (qq + \bar q \bar q)$ (short-dashed line).
\item[Fig. 3]
The total cross section $\sigma_{\rm tot}$ with all channels included
plotted as a function of the Higgs mass at $\sqrt{S}=14~{\rm TeV}$
with $\mu=m$; NNLO (solid line), NLO (dashed line) and LO (dotted line).
\item[Fig. 4]
The quantity $N(\mu/\mu_0)$ (see Eq. (\ref{eqn4.3}) at
$\sqrt{S}=14~{\rm TeV}$ plotted in the range
$0.1<\mu/\mu_0<10$ with $\mu_0=m$ and $m=100~{\rm GeV/c^2}$.
The results are shown for LO (dotted line), NLO (dashed line) and
NNLO (solid line).
\item[Fig. 5]
Same as Fig. 4 but now for $m=200~{\rm GeV/c^2}$.
\item[Fig. 6]
Same as Fig. 4 but now for $m=300~{\rm GeV/c^2}$.
\item[Fig. 7]
The $K$-factors in NLO and NNLO at $\sqrt{S}=14~{\rm TeV}$ as a function
of the Higgs mass using the MRST-sets;
$K^{{\rm NNLO}}$ (solid line), $K^{{\rm NLO}}$  (dot-dashed line).
\item[Fig. 8]
The $K$-factors in NLO at $\sqrt{S}=14~{\rm TeV}$ as a function
of the Higgs mass using the following parton density sets
MRST01 (solid line), GRV98  (dashed line) and CTEQ6 (dotted line).
\item[Fig. 9]
The ratios $R=\sigma_{\rm tot}/\sigma_{\rm tot}^{\rm MRST}$ (Eq.(\ref{eqn4.5})) 
in NLO at $\sqrt{S}=14~{\rm TeV}$ and $\mu=m$ as a function of the Higgs mass;
$R^{\rm GRV}$ (solid line),
$R^{\rm CTEQ}$ (dotted line).
\item[Fig. 10a]
The ratio $R(x_{\rm max})$ (see Eq. (\ref{eqn4.7})) for
proton-proton collisions (LHC) where $x_{\rm max}=5~x$ with
$x=m^2/S$. The CM energy and scale are given by $\sqrt{S}=14~{\rm TeV}$ and 
$\mu=m$ respectively; NNLO (solid line), NLO (dashed line).
\item[Fig. 10b]
Same as in Fig. 10a but for proton-anti-proton collisions (TEVATRON)
at $\sqrt{S}=2~{\rm TeV}$ and $\mu=m$.
\item[Fig. 11a]
The ratios in Eq. (\ref{eqn4.12}) in NLO for proton-proton collisions (LHC)
as a function of the Higgs mass at $\sqrt{S}=14~{\rm TeV}$ and $\mu=m$;
$R^{\rm SV}$(Eq. (\ref{eqn4.1})) (dot-dashed line), 
$R^{\rm SVL}$(Eq. (\ref{eqn4.1})) (dashed line),
$R^{\rm SV}$(Eq. (\ref{eqn4.3})) (dotted line), 
$R^{\rm SVL}$(Eq. (\ref{eqn4.3})) (solid line).
\item[Fig. 11b]
The ratios in Eq. (\ref{eqn4.12}) in NNLO for proton-proton collisions (LHC)
as a function of the Higgs mass at $\sqrt{S}=14~{\rm TeV}$ and $\mu=m$;
$R^{\rm SV}$(Eq. (\ref{eqn4.1})) (dot-dashed line),     
$R^{\rm SVL}$(Eq. (\ref{eqn4.1})) (dashed line),
$R^{\rm SV}$(Eq. (\ref{eqn4.3})) (dotted line),         
$R^{\rm SVL}$(Eq. (\ref{eqn4.3})) (solid line).
\item[Fig. 12a]
Same as in Fig. 11a but for proton-anti-proton collisions (TEVATRON)
at $\sqrt{S}=2~{\rm TeV}$ and $\mu=m$.
\item[Fig. 12b]
Same as in Fig. 11b but now for proton-anti-proton collisions (TEVATRON)
at $\sqrt{S}=2~{\rm TeV}$ and $\mu=m$.
\end{description}

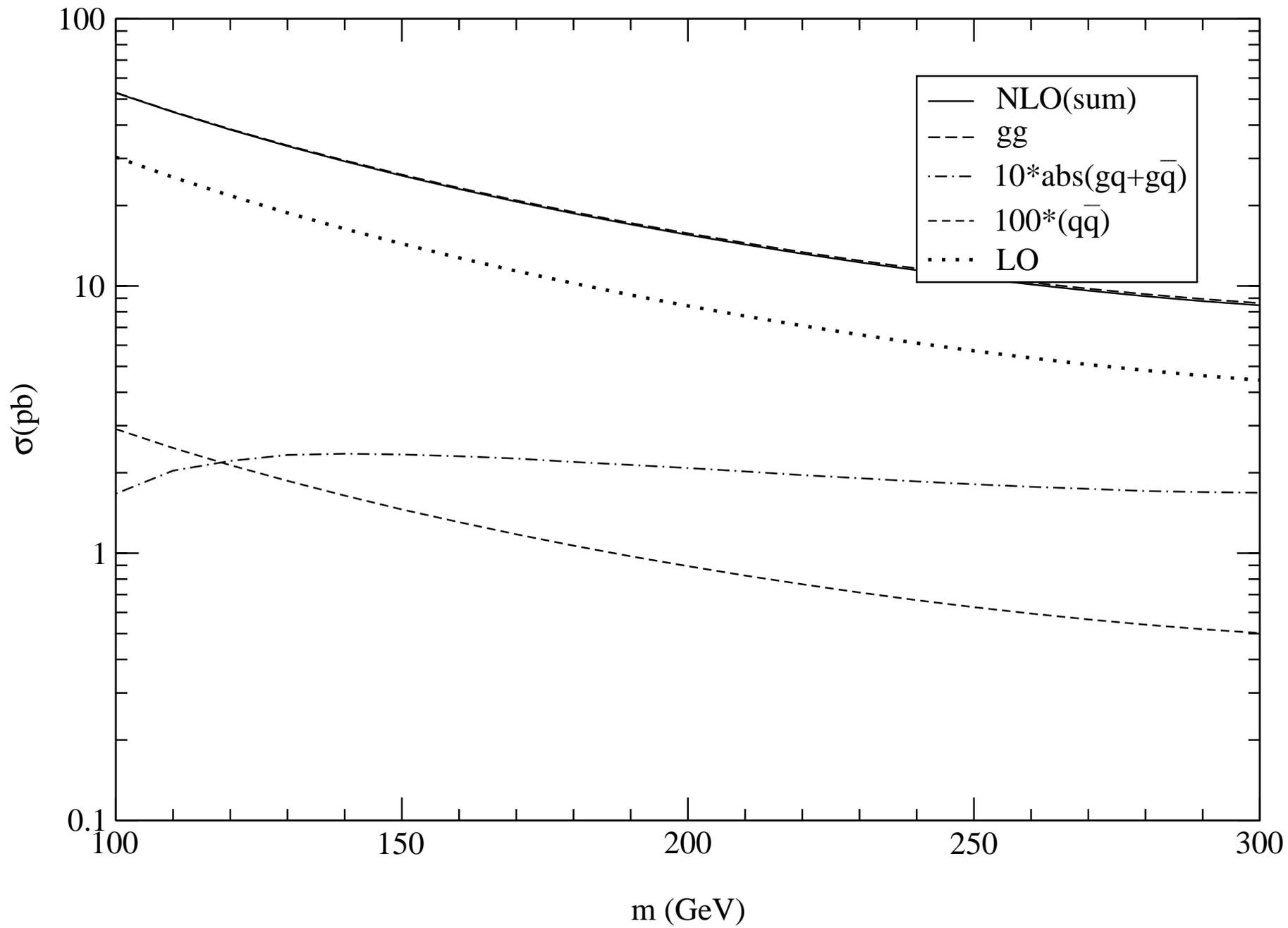
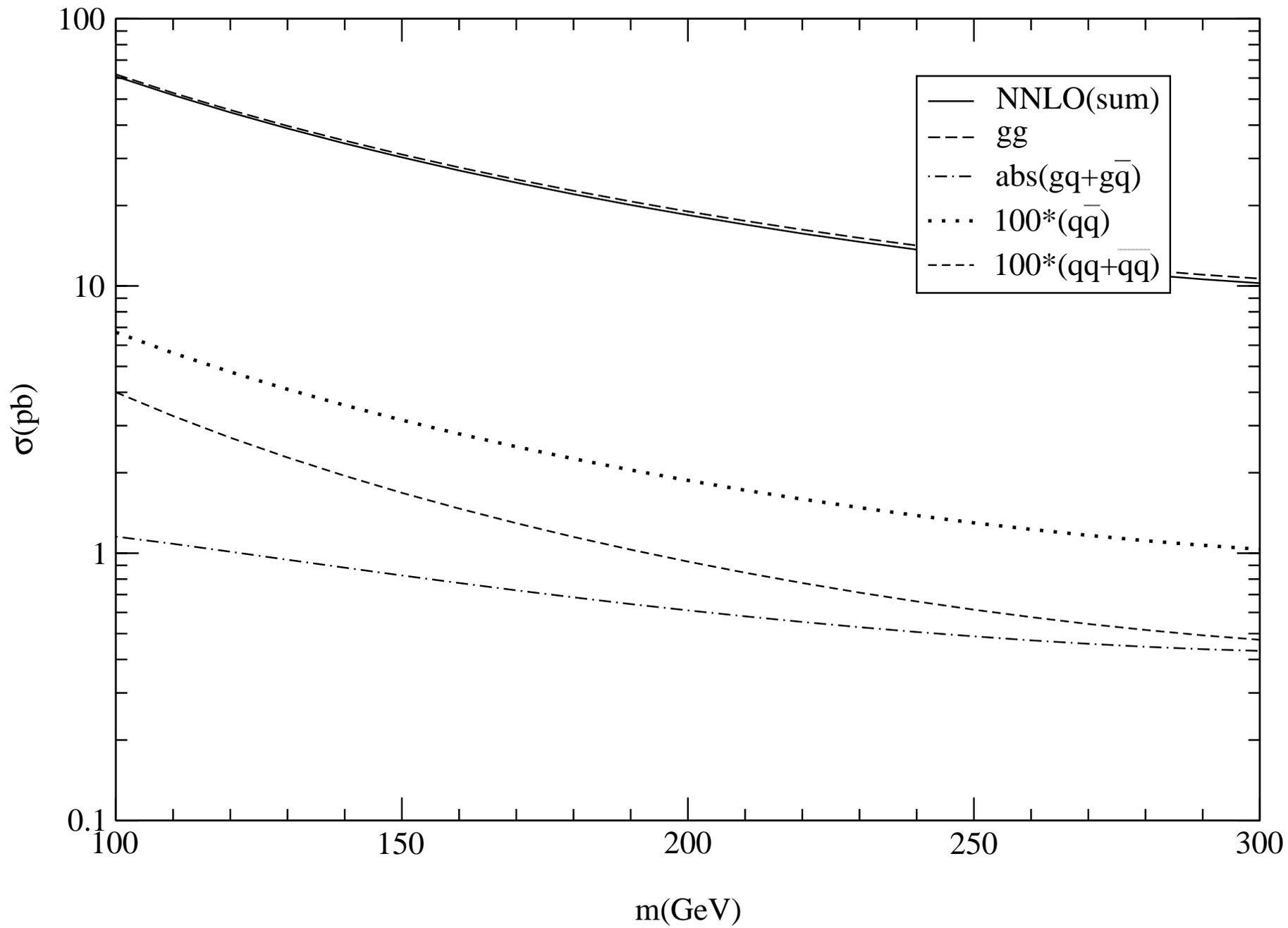
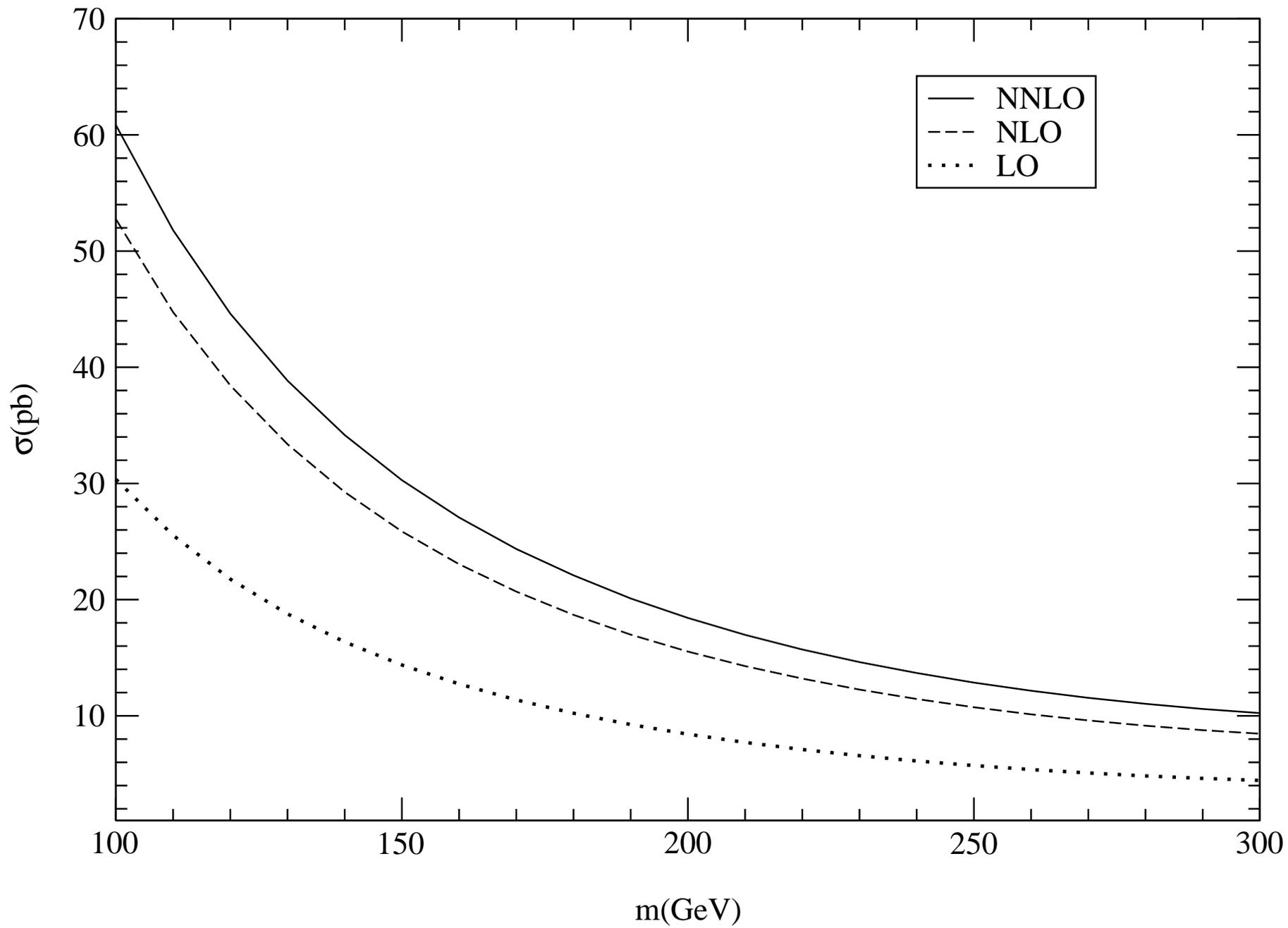
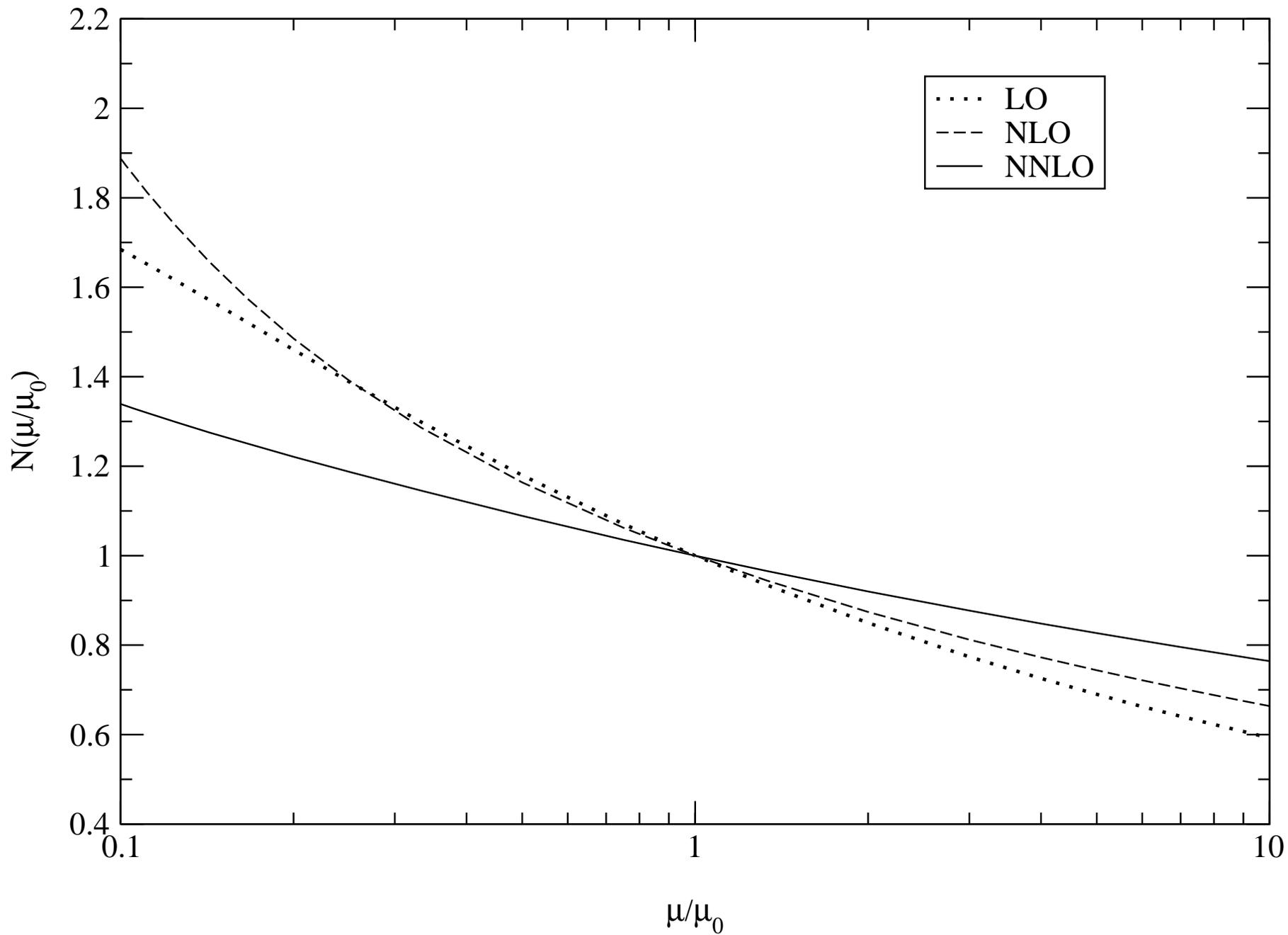
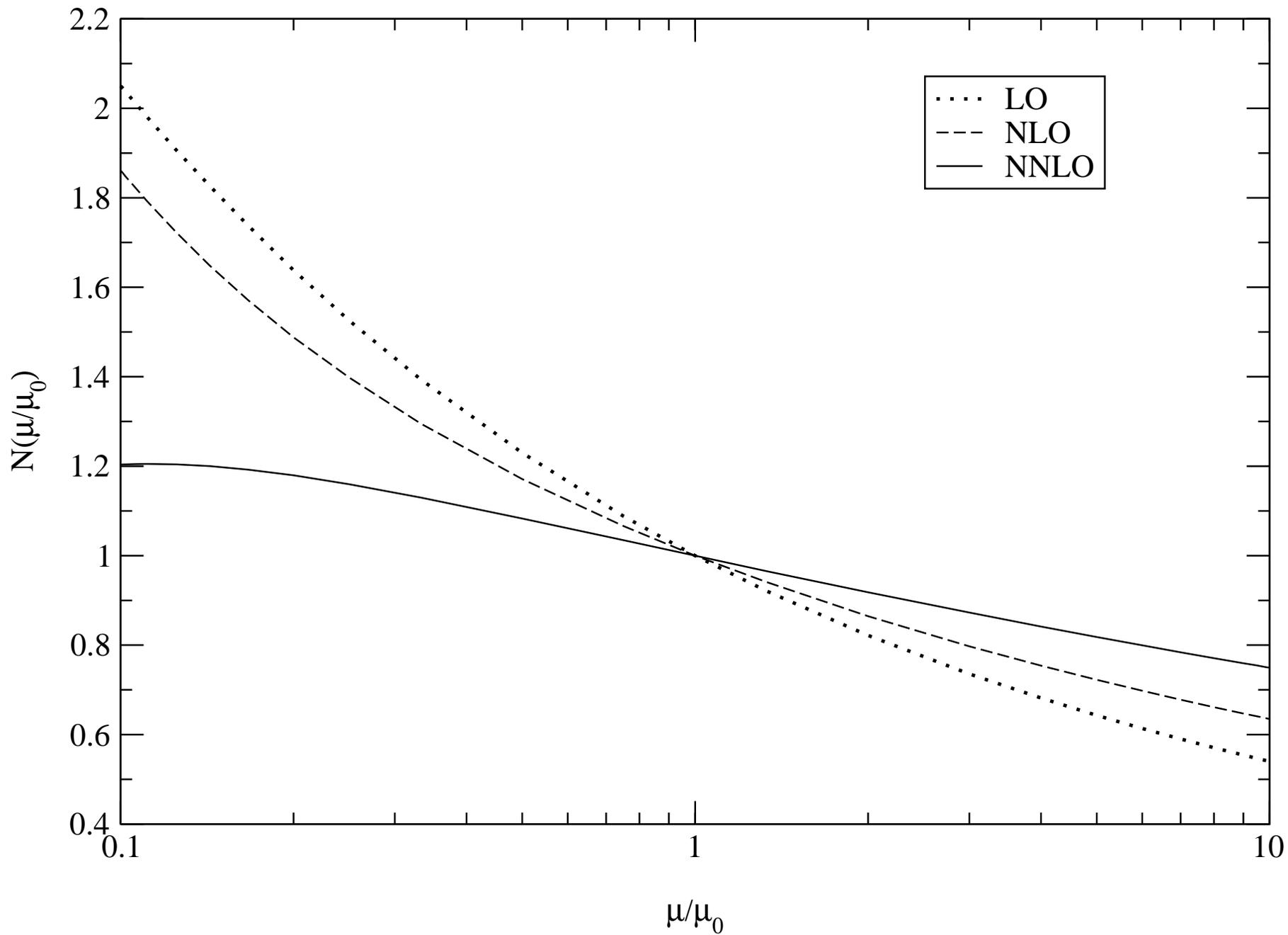
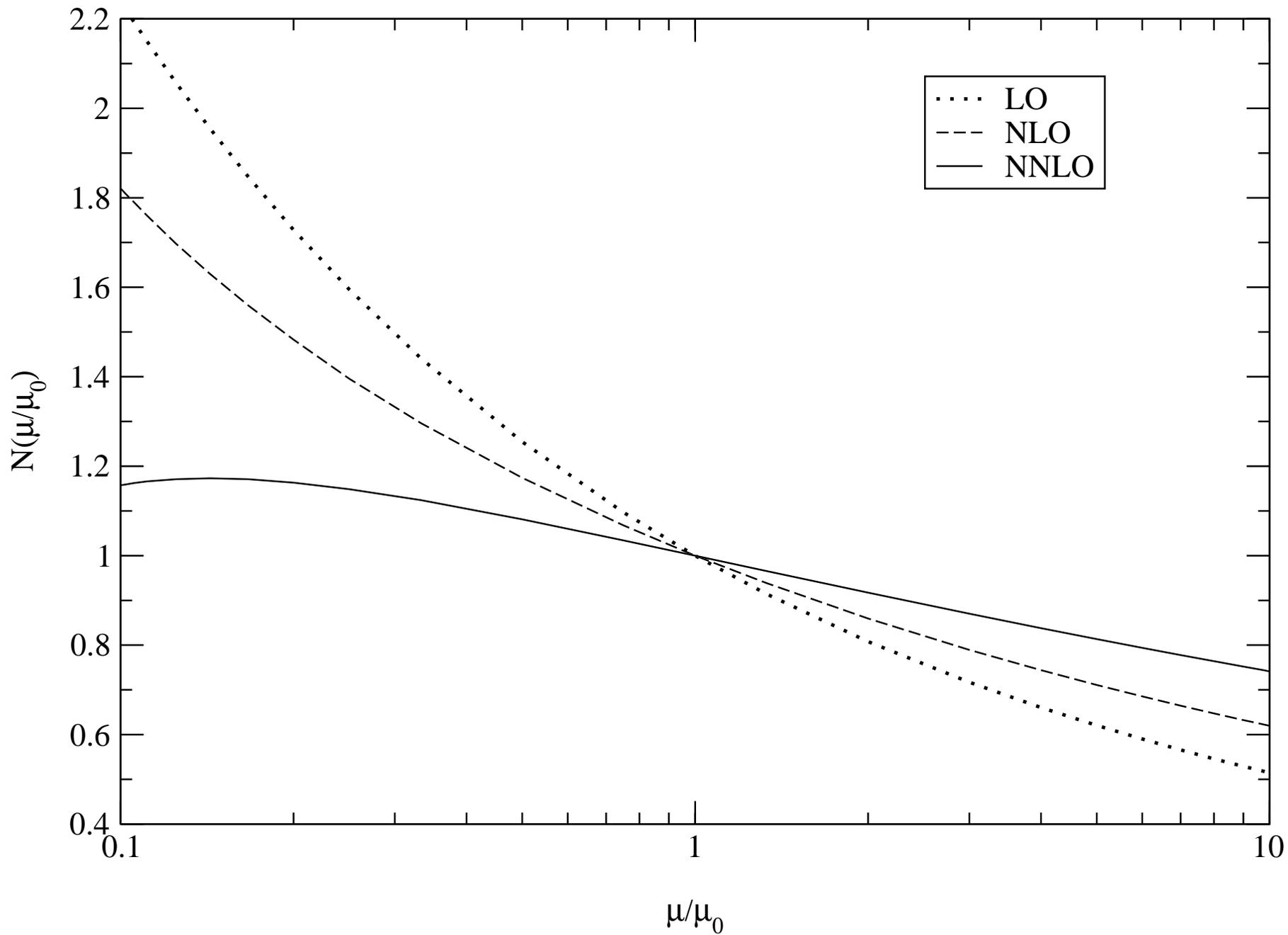
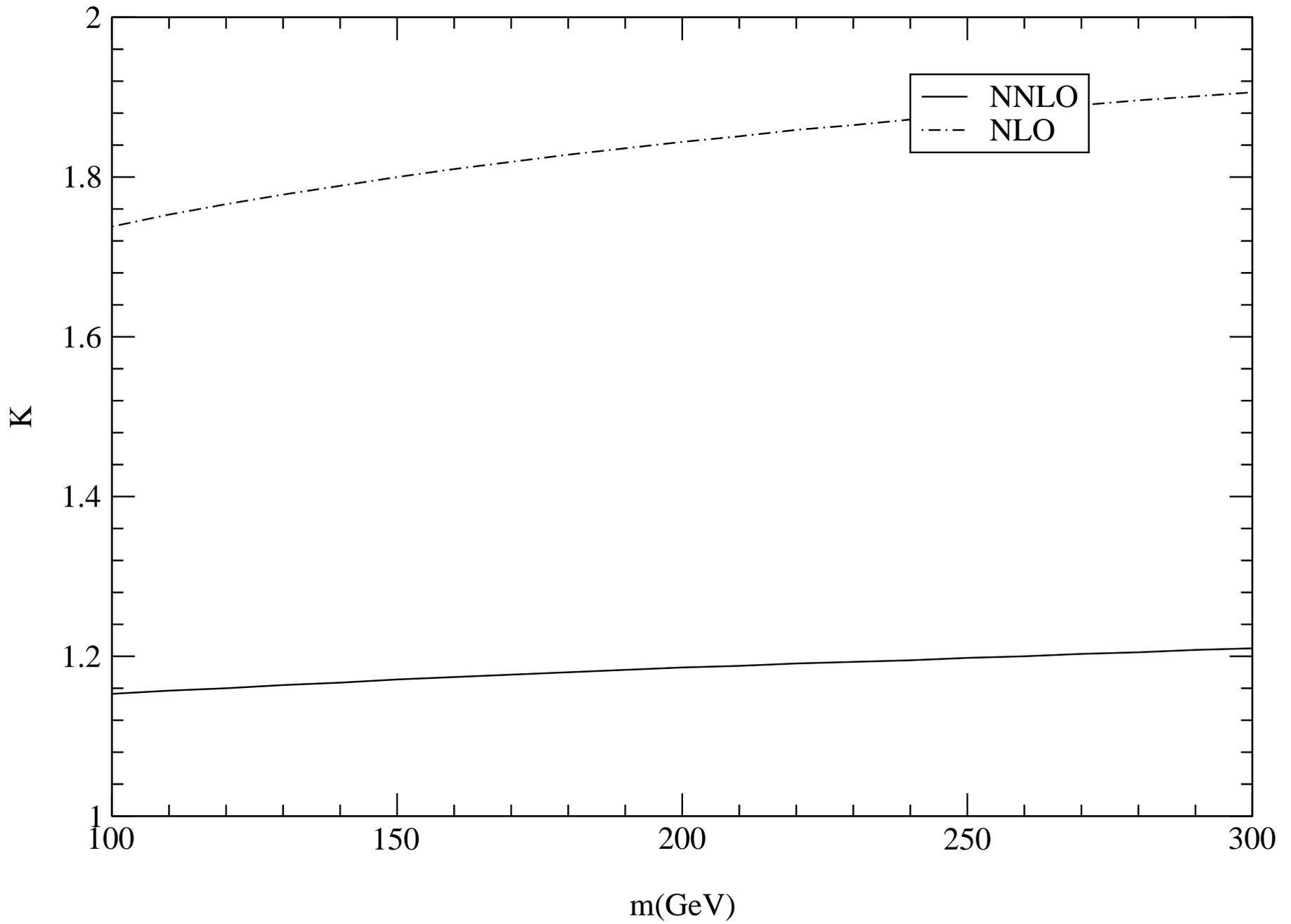
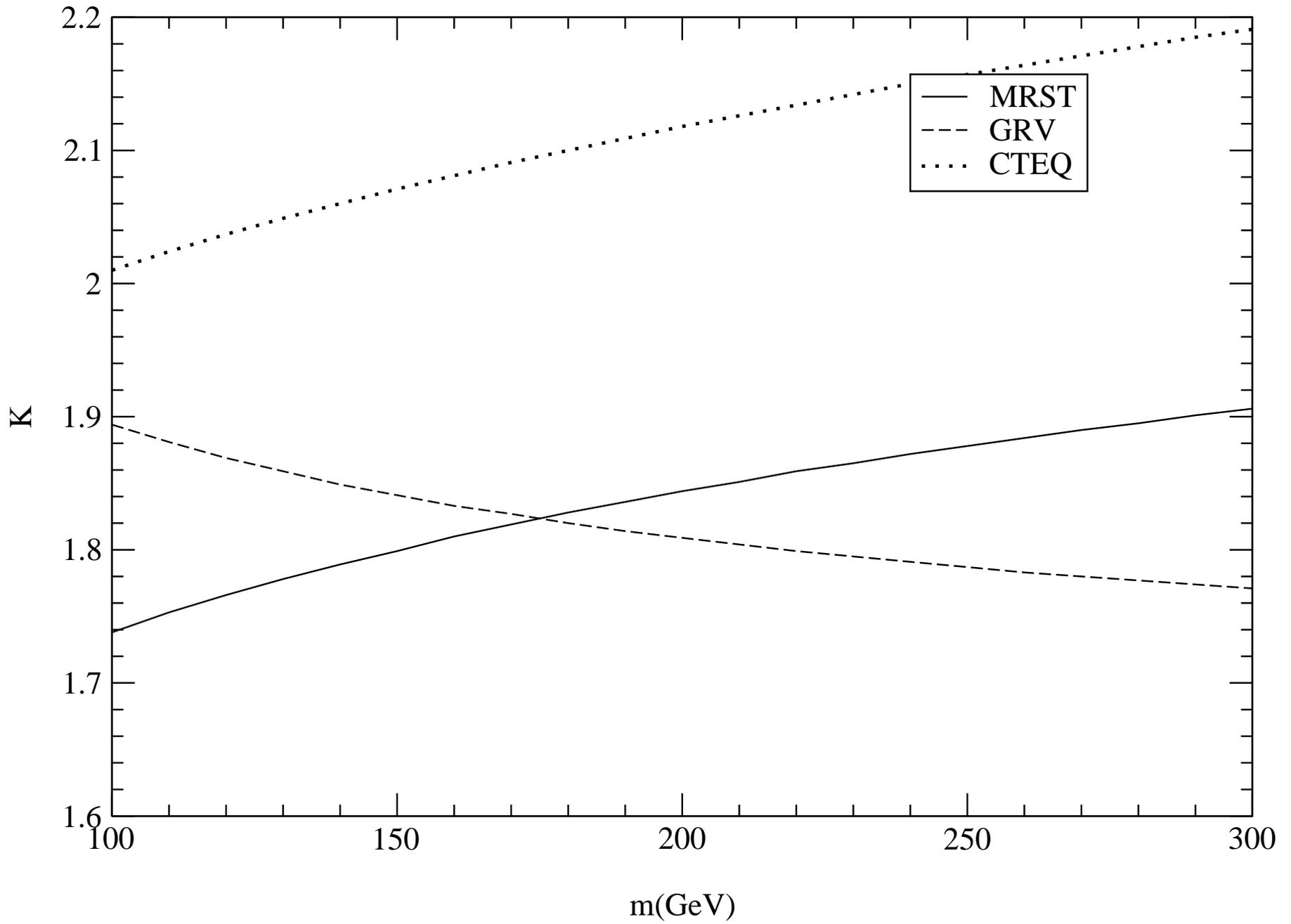
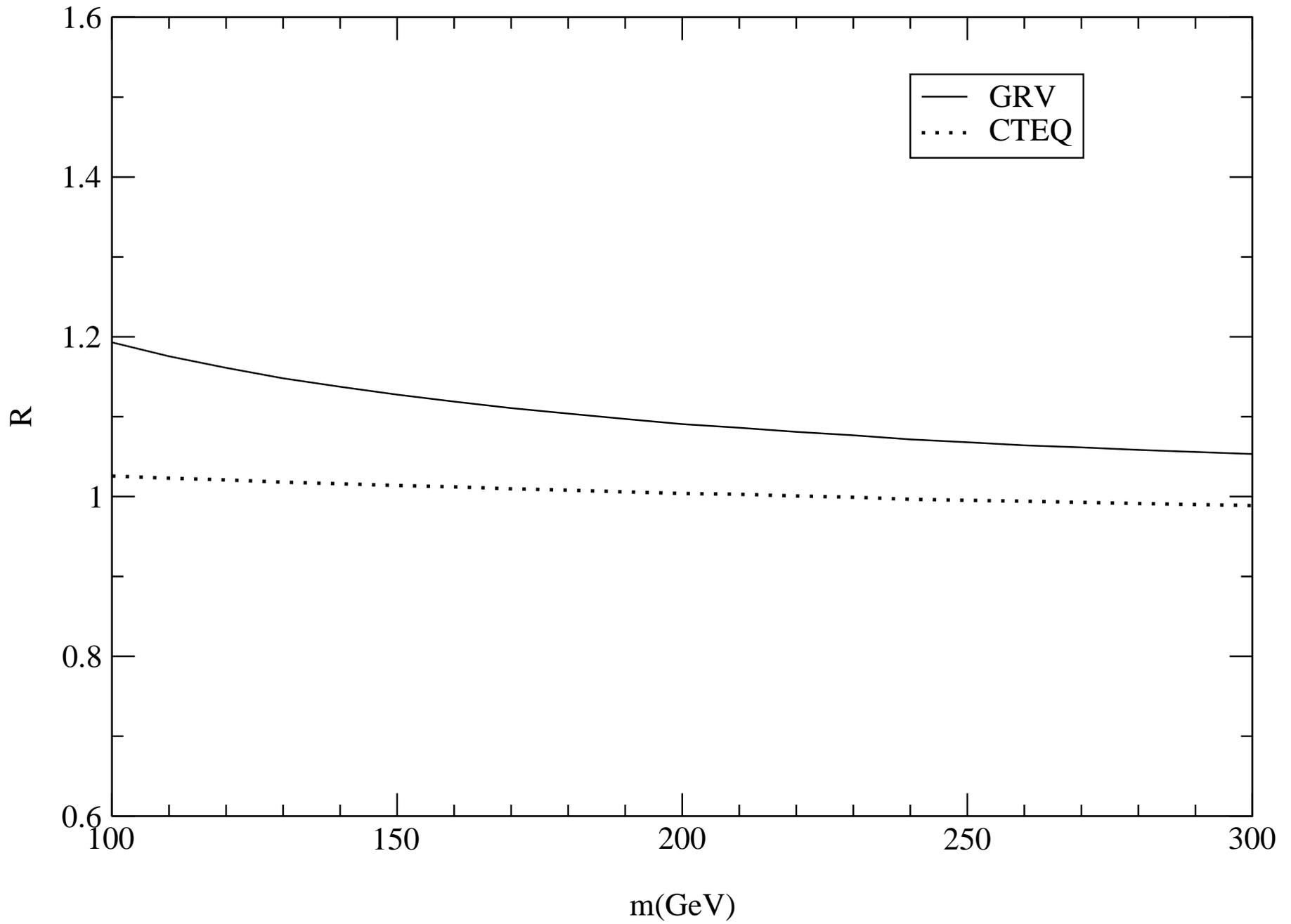
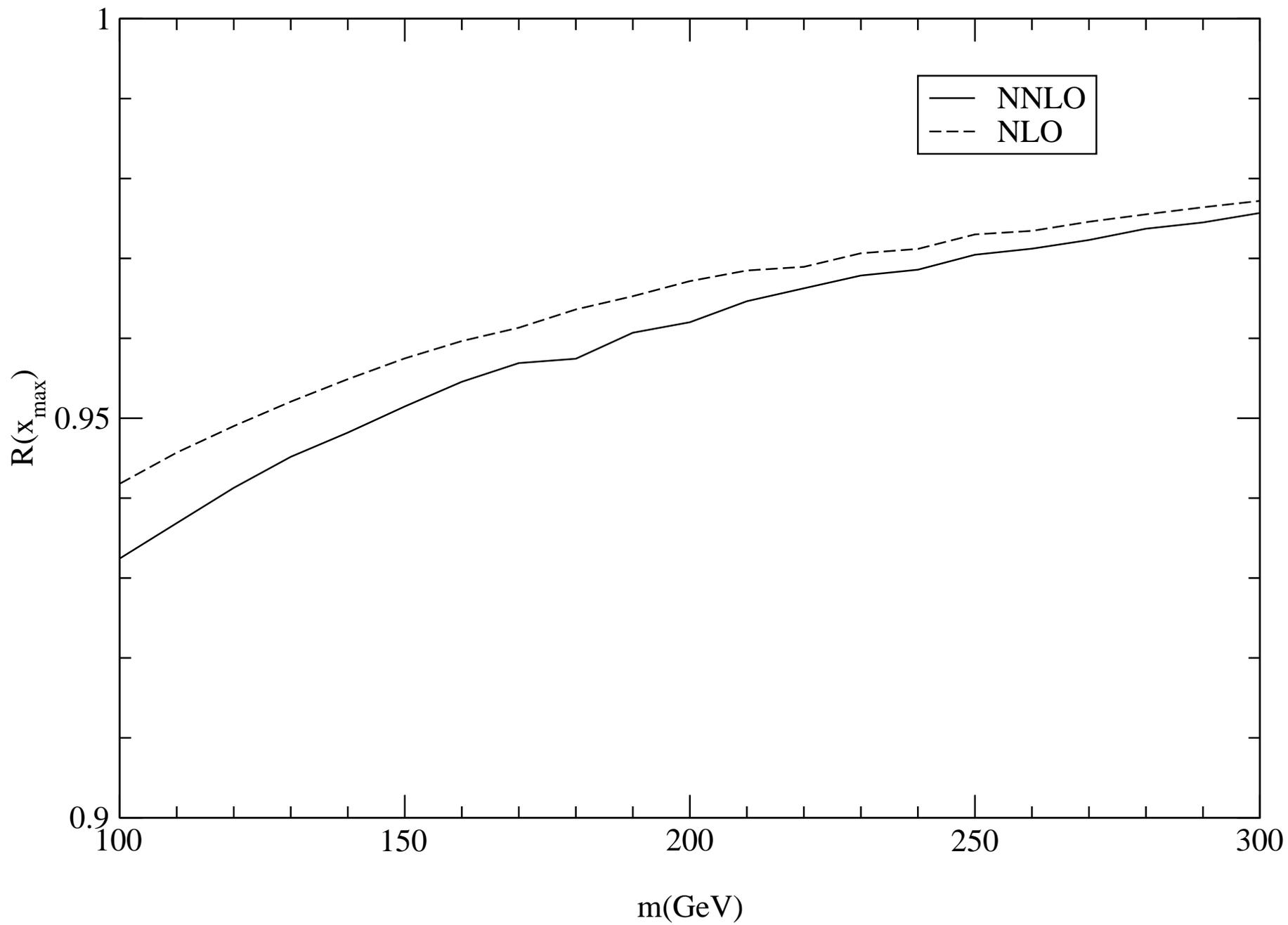
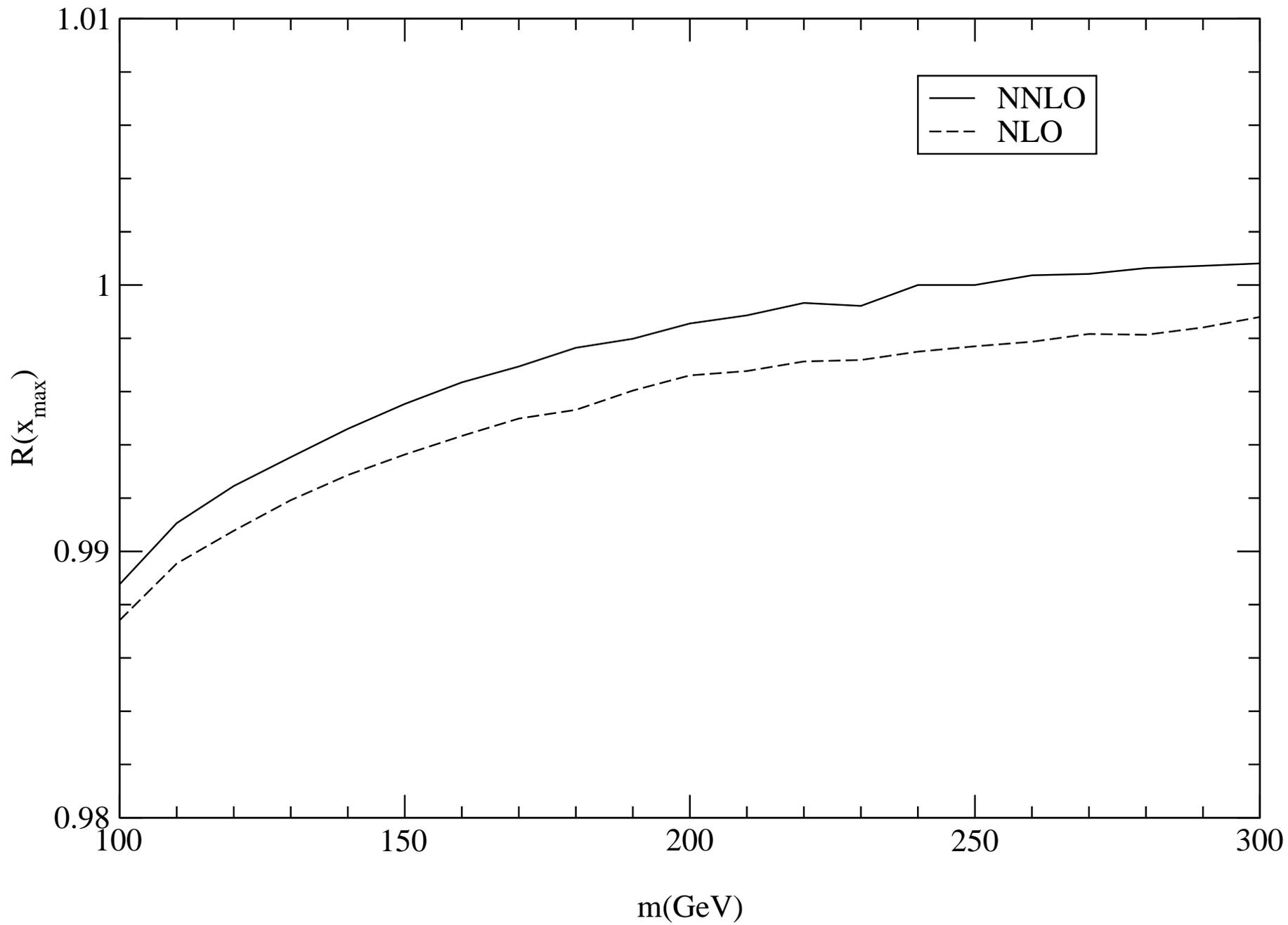
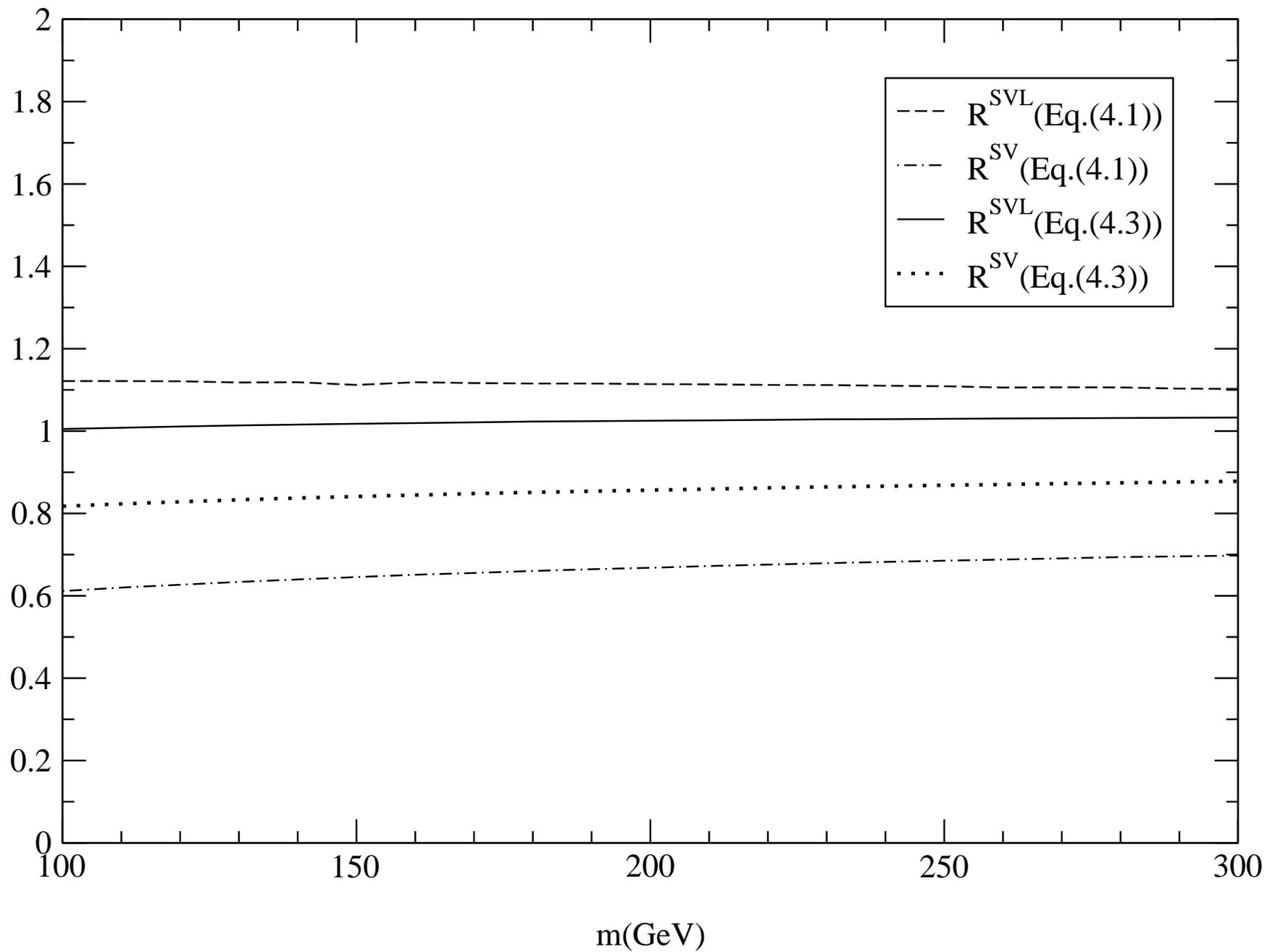
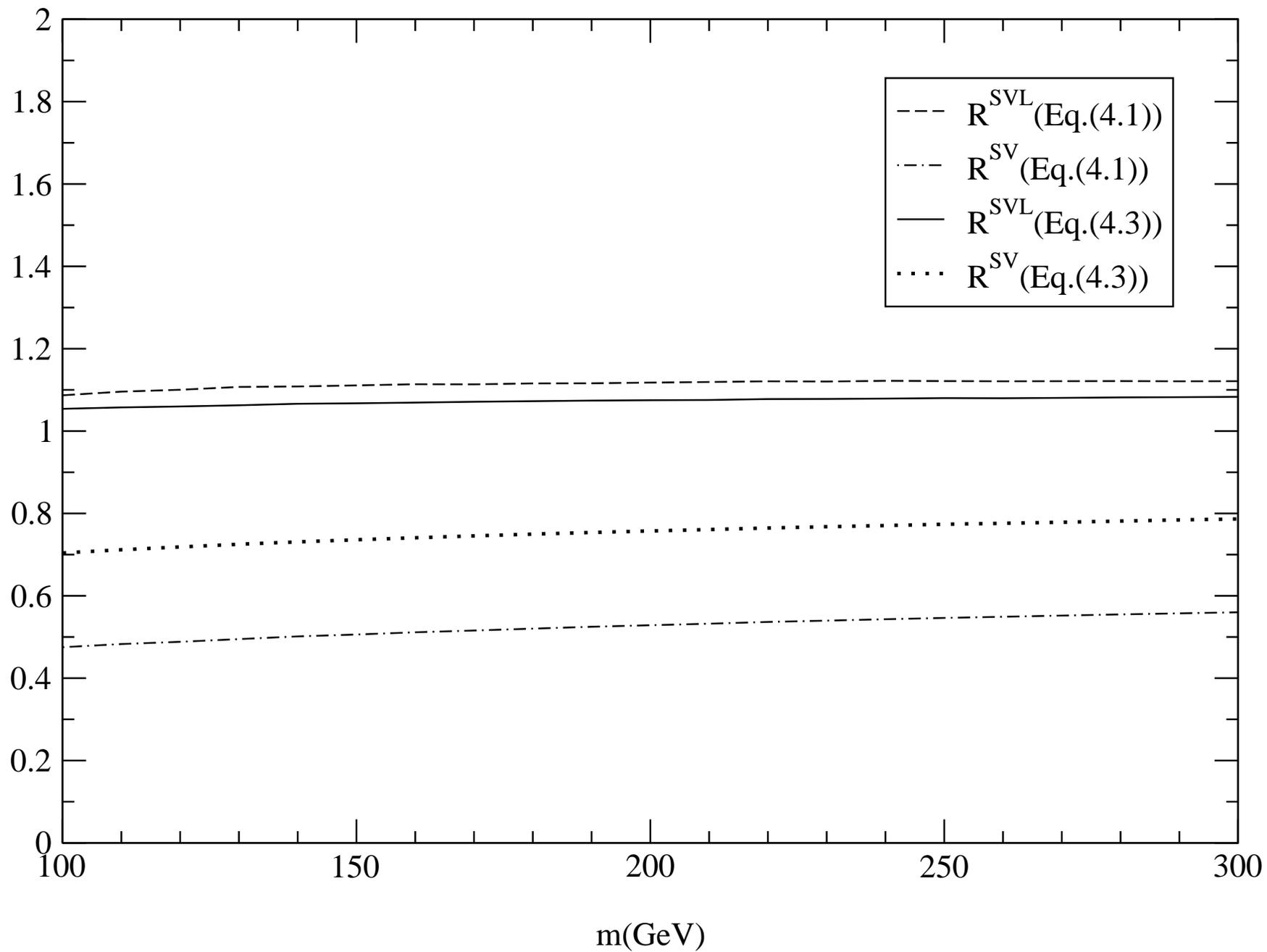
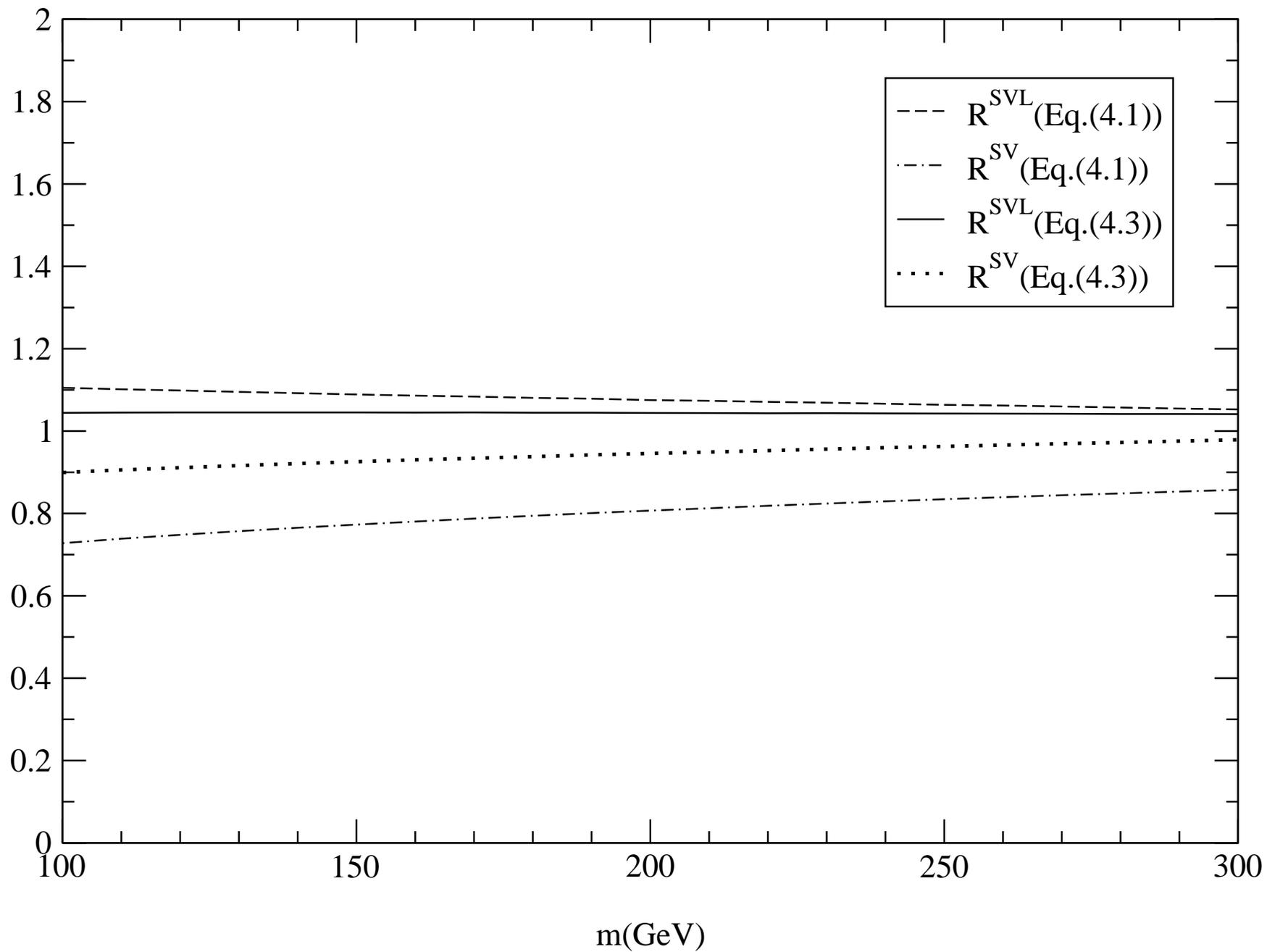
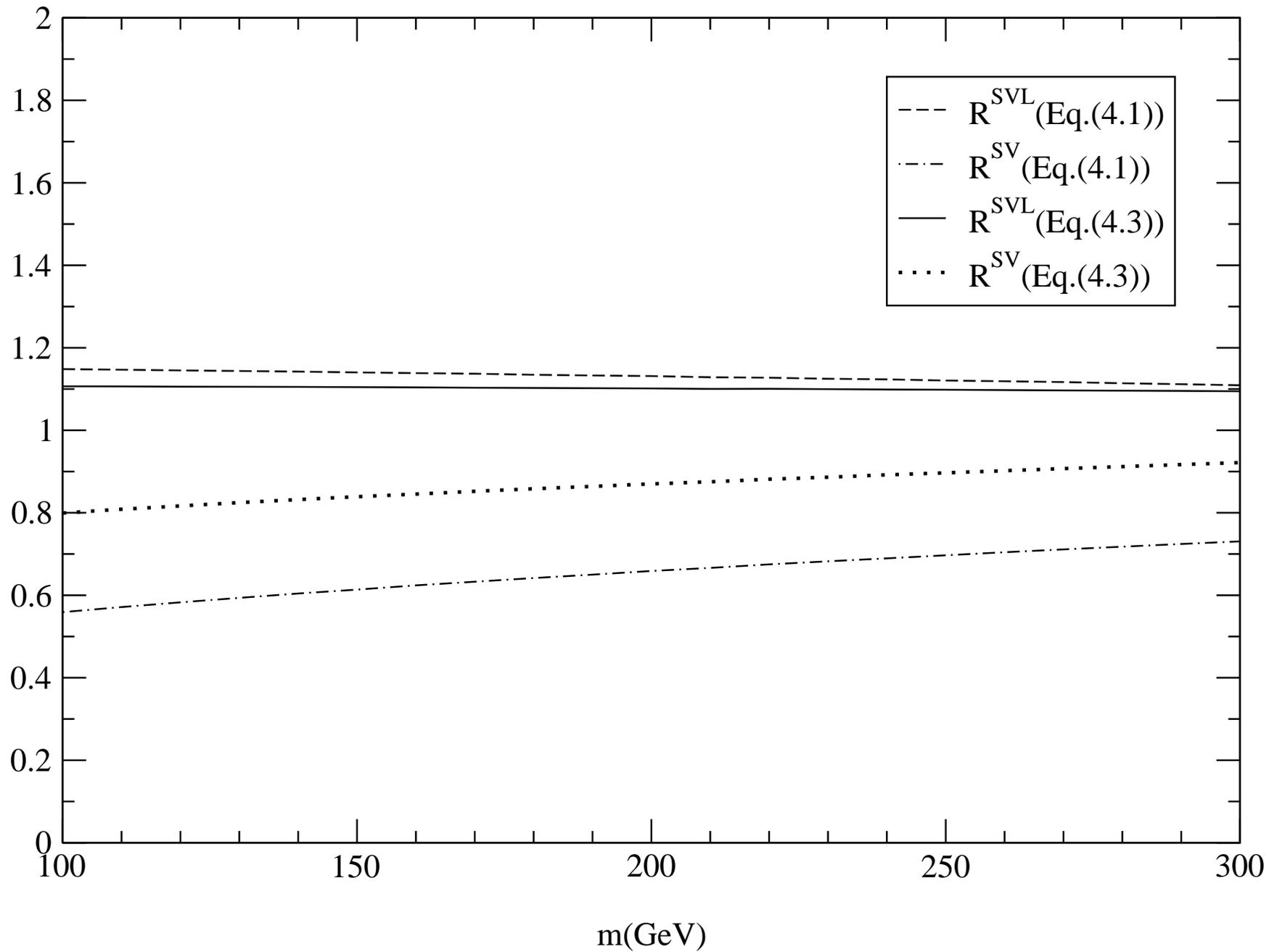
\end{document}